\documentclass[12pt]{article}
\pdfoutput=1
\usepackage{amsmath,amssymb,amsthm,bm,graphicx,float,array,multirow,multicol,rotfloat,caption,subcaption,enumerate,geometry,mathdots,adjustbox,booktabs,parskip,mathtools,tikz,tikz-cd,pdflscape,csquotes,lscape,rotating,empheq,mathrsfs}
\usepackage[para]{threeparttable}
\usepackage[all]{xy}
\usepackage[shortlabels]{enumitem}
\usepackage[normalem]{ulem}
\usepackage[vcentermath]{youngtab}
\usepackage[boxsize=.8em]{ytableau}
\usepackage[numbers,sort&compress]{natbib}
\usepackage[hidelinks]{hyperref}
\usepackage{cleveref}
\numberwithin{equation}{section}
\numberwithin{figure}{section}
\allowdisplaybreaks
\makeatletter
\usetikzlibrary{arrows,arrows.meta,shapes,shapes.misc,patterns,decorations.pathmorphing,decorations.pathreplacing,positioning,chains,fit}
\pgfdeclarepatternformonly{coarsedots}{\pgfqpoint{-1pt}{-1pt}}{\pgfqpoint{5pt}{5pt}}{\pgfqpoint{12pt}{12pt}}{
    \pgfpathcircle{\pgfqpoint{.5pt}{.5pt}}{.5pt}
    \pgfusepath{fill}}
\tikzset{gauge/.style={rounded rectangle, draw=black!100, thick, minimum size=5mm},  gaugeD/.style={rounded rectangle, draw=black!100,double,thick,minimum size=5mm},  empty/.style={rounded rectangle, draw=white!100, thick, minimum size=5mm}, flavor/.style={rectangle, draw=black!100, thick, minimum size=5mm},flavorD/.style={rectangle, draw=black!100, double,thick, minimum size=5mm}}
\hypersetup{colorlinks=true}
\hypersetup{linkcolor=black}
\hypersetup{citecolor=black}
\hypersetup{urlcolor=black}
\theoremstyle{plain}
\newtheorem*{thm*}{Theorem}
\newtheorem{thm}{Theorem}[section]

\newtheorem{conj}[thm]{Conjecture}

\newtheorem*{conj:3dmirror}{Conjecture \ref{conj:3dmirror}}

\def\CS{\mathcal{S}}
\def\CN{\mathcal{N}}
\def\CB{\mathcal{B}}
\def\CD{\mathcal{D}}

\newcommand\scalemath[2]{\scalebox{#1}{\mbox{\ensuremath{\displaystyle #2}}}}

\newcommand{\SN}[1]{\mathcal{N}\hspace{-2pt}=\hspace{-1pt} #1}
\makeatother

\begin{document}

\begin{titlepage}
% Report number
\vspace*{-3cm} 
\begin{flushright}
{\tt CALT-TH-2022-024}\\
{\tt DESY-22-110}\\
\end{flushright}
\begin{center}
\vspace{1.7cm}
{\LARGE\bfseries Higgs, Coulomb, and Hall--Littlewood\\} 
%{\LARGE\bfseries On the Topic of the Branches of Moduli commonly called by the Names of Higgs and Coulomb, and of Their Connexions to the Indices of Hall and Littlewood, as has been Noticed in a Humble Manner by the Authors, and put in Writing with the Desire that our Lowly Work may prove Beneficial for all Earthly Scholars\\} 
\vspace{1cm}

{\large
Monica Jinwoo Kang$^1$, Craig Lawrie$^2$, Ki-Hong Lee$^3$,\\[0.5em] Matteo Sacchi$^4$, and Jaewon Song$^3$}\\
\vspace{.8cm}
{$^1$ Walter Burke Institute for Theoretical Physics, California Institute of Technology\\
Pasadena, CA 91125, U.S.A.}\par
\vspace{.2cm}
{$^2$ Deutsches Elektronen-Synchrotron DESY,\\ Notkestr.~85, 22607 Hamburg, Germany}\par
\vspace{.2cm}
{$^3$ Department of Physics, Korea Advanced Institute of Science and Technology\\
Daejeon 34141, Republic of Korea}\par
\vspace{.2cm}
{$^4$ Mathematical Institute, University of Oxford\\
Oxford OX2 6GG, United Kingdom}\par
\vspace{.3cm}

\scalebox{.85}{\tt monica@caltech.edu, craig.lawrie1729@gmail.com, khlee11812@gmail.com,}\\ \scalebox{.85}{\tt matteo.sacchi@maths.ox.ac.uk, jaewon.song@kaist.ac.kr}\par
\vspace{1.2cm}
%\vspace{.2cm}
\textbf{Abstract}
\end{center}

The Higgs branch of 4d $\mathcal{N}=2$ SCFTs can be analyzed via the Hilbert series of the Higgs branch or, in special cases, by computing the Hall--Littlewood index. 
For any class $\mathcal{S}$ theory corresponding to a genus-zero Riemann surface, they are conjectured to be identical. We present several families of counterexamples. We find that for any class $\mathcal{S}$ theory with four or more $\mathbb{Z}_2$-twisted punctures, they do not match. We construct 3d mirrors for such theories and analyze their Coulomb branch Hilbert series to compute the Higgs branch Hilbert series of the 4d theory. 
We further construct $a=c$ theories in class $\mathcal{S}$ using the twisted punctures, and these theories, which includes the $\widehat{D}_4(SU(2n+1))$ theories, have  Hall--Littlewood index different from the Hilbert series of the Higgs branch. We conjecture that this is the case for all $a=c$ theories with non-empty Higgs branch, including $\mathcal{N} \ge 3$ SCFTs.

\vspace{1cm}
%{{\bf Key words:} }\\
\vfill 
\end{titlepage}

\tableofcontents
\newpage

\section{Introduction}

Given a quantum field theory, it is important to determine its vacua and spectrum of local operators. This is often challenging due to the complicated nature of the Hilbert space of interacting quantum field theories. Luckily, however, for supersymmetric field theories, we can use an \emph{index} \cite{Witten:1982df} to count subsectors of operators that appear in the theory even when it is strongly-interacting. Utilizing the index, we can understand the refined structure of operators and vacua of supersymmetric theories.

\begin{figure}[H]
    \centering
    \begin{tikzpicture}[scale=.8,every node/.style={scale=.8}]
    \shade[inner color=red!40, outer color=white, opacity = 0.4] (-8,-1.2)--(4,-1.2)--(6,2.4)--(-6,2.4)--(-8,-1.2);
    \draw[white,fill=white] (-8,-1.2)--(-8,-1)--(-6,2)--(-6,2.4)--(-8,-1.2);
    \draw[white,fill=white] (-8,-1.2)--(4,-1.4)--(4,-1)--(-8,-1)--(-8,-1.2);
    \draw[white,fill=white] (4,-1.4)--(4,-1)--(6,2)--(6,2.4)--(4,-1.2);
    \draw[white,fill=white] (6,2.4)--(-6,2.4)--(-6,2)--(6,2)--(6,2.4);
    \draw[red] (-8,-1)--(4,-1)--(6,2)--(-6,2)--(-8,-1);
    \draw[magenta,pattern=coarsedots,pattern color=magenta] (-4.8,1.92) arc[start angle=125, end angle=94,radius=17cm] -- (3.75,2.02) arc[start angle=94,end angle=125,radius=17cm];
    \draw[magenta] (-4.8,1.93)--(-4.8,-1);
    \draw[thick,red] (3.75,2.02) arc[start angle=94,end angle=125,radius=17cm];
    \draw[fill=white](-3,3.9)--(-1,1)--(1,3.9);
    \shade[inner color=blue!60, outer color=white, opacity = 0.4] (-3,3.9)--(-1,1)--(1,3.9);
    \draw[thick,blue!60] (-3,3.9)--(-1,1)--(1,3.9);
    \draw[dashed,blue!60] (-3,3.9)--(1,3.9);
    %\draw[dashed,purple!60] (-1,1)--(-1,3.9);
    \draw[thick,blue,fill=white] (-1,4) ellipse (2cm and .8cm);
    \node[draw=none,opacity=0,thick,scale=0.1,fill=black,label={[label distance=-2mm,color=blue]0:\LARGE{$\mathcal{H}$}}] (H) at (-1.3,2.5) {};
    \node[draw=none,opacity=0,thick,scale=0.1,fill=black,label={[label distance=0mm,color=red]0:\LARGE{$\mathcal{C}$}}] (C) at (3,0) {};
    \node[draw=none,opacity=0,thick,scale=0.1,fill=black,label={[label distance=0mm,color=magenta]0:Mixed}] (M1) at (2,3.5) {};
    \node[draw=none,opacity=0,thick,scale=0.1,fill=black,label={[label distance=-.6mm,color=magenta]0:Branch}] (M2) at (2,3) {};
    %\draw[thick,purple] (-4.8,-1.02) arc[start angle=125, end angle=94,radius=17cm];
    %\draw[thick,dashed,purple] (-4.8,1.92) arc[start angle=125, end angle=94,radius=17cm];
    \end{tikzpicture}
    \caption{Moduli space of 4d $\mathcal{N}\geq 2$ SCFTs. The Coulomb branch $\mathcal{C}$ is depicted as a red plane while the Higgs branch $\mathcal{H}$ is depicted as a blue cone. There might be a mixed branch, depicted as a magenta surface, where the subspace of the Higgs branch is fibered over a subspace of the Coulomb branch depicted as a red curve. }
    \label{fig:4dbranch}
\end{figure}
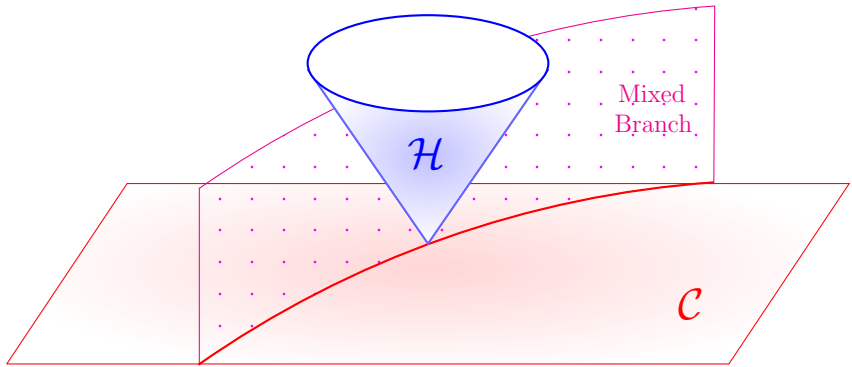

In particular, superconformal field theories (SCFTs) with at least eight supercharges admit both Higgs and Coulomb vacuum moduli. To understand the moduli space of vacua of 4d $\mathcal{N}=2$ SCFTs, we have to understand both the Coulomb branch $\mathcal{C}$, parameterized by vacuum expectation values for the scalar primaries inside of vector multiplets, and the Higgs branch $\mathcal{H}$, parameterized by vacuum expectation values (vevs) of scalar primaries inside of hypermultiplets. A particularly simple moduli space occurs when these two branches meet at a single point, the origin. In this case, moving onto a generic point of the Higgs branch by giving vacuum expectation values to the scalars in the hypermultiplet breaks the gauge group. If the gauge group is entirely broken, then there are no remaining vector multiplets; thus, one cannot move onto the Coulomb branch from such a generic point of the Higgs branch. However, it is possible in some cases, where the gauge group is not entirely broken at a generic point of the Higgs branch, to give vevs to scalars inside the residual vector multiplets instead. The subspace of the moduli space where both vector multiplet and hypermultiplet scalars are given vevs is called the \emph{mixed branch}. We depict in Figure \ref{fig:4dbranch} the Coulomb, Higgs, and mixed branches of a 4d $\mathcal{N}=2$ SCFT. Each component of the moduli space can be parameterized by a collection of local operators of the SCFT, and the spectrum of the local operators of theories are not only interesting in their own right, but also can further refine our understanding of the branch structure.

In fact, the superconformal index \cite{Kinney:2005ej, Romelsberger:2005eg} captures information about the operator spectrum of a given superconformal field theory. It is defined in such a way to count the short multiplets (under the superconformal symmetry) up to recombination, and is invariant under marginal deformations of the theory.\footnote{For a recent pedagogical review of the superconformal index, see \cite{Gadde:2020yah}.} Most importantly, operators in a superconformal theory such as the stress tensor and conserved currents belong to particular short multiplets, so the index serves as an extremely useful tool to uncover the spectrum of a superconformal theory. For a 4d $\SN{2}$ SCFT, the most general superconformal index counts $1/8$-BPS short multiplets, where the contributions only come from the operators that satisfy
\begin{align}
    \delta = \Delta - 2j_2 - 2R + r = 0 \ ,
\end{align}
where $\Delta, j_1, j_2, R, r$ are the Cartans of the bosonic subgroup $SO(4, 2) \times SU(2)_R \times U(1)_r$ of the superconformal group $SU(2, 2|2)$. 
One can consider various special limits of the index to count short multiplets that preserve more supersymmetries \cite{Gadde:2011uv}. In this paper, we focus on
 the Hall--Littlewood (HL) index, which is defined as
\begin{align}\label{eqn:HLdef}
    \operatorname{HL}(\tau)= \text{Tr}_{\text{HL}}\,{(-1)}^F \tau^{2(\Delta-R)}  \,,
\end{align}
where the trace is taken over the states on $S^3$ satisfying 
\begin{align}
    \delta'_\pm = \Delta \pm 2j_1 - 2R - r =0,\quad \delta = 0 .
\end{align}
Therefore, the trace is only over the states with 
\begin{align}
    \Delta = 2R+r\ , \quad j_1 = 0 \ , \quad j_2 = r \,.
\end{align}
Utilizing the state-operator correspondence, this is tantamount to counting all local operators belonging to the $\widehat{\CB}_R$ and $\mathcal{D}_{R(0,j_2)}$ short superconformal multiplets,\footnote{Throughout this paper, we use the notation of Dolan--Osborn \cite{Dolan:2002zh} for $\mathcal{N}=2$ superconformal multiplets.} which form the so-called Hall--Littlewood chiral ring \cite{Beem:2017ooy}.

The Hall--Littlewood chiral ring is closely related to the Higgs branch chiral ring of the theory, which is generated by the $\widehat{\CB}_R$-type multiplets only. The Hilbert series (HS) of the Higgs branch (HB) is defined as 
\begin{equation}\label{HSdef}
    \operatorname{HS}(\tau)= \text{Tr}_{\text{HB}}\,\tau^{2R} \,,
\end{equation}
where the trace is now taken over the Higgs branch chiral ring, which exclusively contains operators corresponding to $\widehat{\CB}_R$ superconformal multiplets; such multiplets satisfy further the shortening condition $\Delta = 2R$. In particular, taking the $r\to 0$ limit of the Hall--Littlewood sector produces the Higgs sector. A comparison between the Hall--Littlewood and the Higgs branch chiral rings is described in Table \ref{tab:HSvsHL}, which demonstrates that the Higgs branch chiral ring is a subring of the Hall--Littlewood chiral ring.

\begin{table}[H]
    \centering
    \renewcommand{\arraystretch}{1.2}
    \begin{threeparttable}
    \begin{tabular}{c|cc}
        \toprule
         & Higgs Branch Sector & Hall--Littlewood Sector \\\midrule
        Condition & $\Delta=2R$, $j_1=j_2=r=0$ & $\Delta=2R+j_2$, $j_1=0$, $j_2=r$ \\
        Multiplet contents & $\widehat{\CB}_R$ & $\widehat{\CB}_R$, $\CD_{R(0, j_2)}$ \\\bottomrule
    \end{tabular}
    \end{threeparttable}
    \caption{The 4d $\mathcal{N}=2$ superconformal multiplets that contribute to the Higgs branch and Hall--Littlewood chiral rings.}
    \label{tab:HSvsHL}
\end{table}

In this paper, we are mainly concerned with the subsector of the 4d $\mathcal{N}=2$ landscape which can be realized via the class $\mathcal{S}$ perspective \cite{Gaiotto:2009hg, Gaiotto:2009we}. The construction of such SCFTs is as follows. Consider the 6d $(2,0)$ SCFT of type $J$, where $J$ is an $ADE$ Lie algebra, and take a twisted-compactification of this theory on a punctured Riemann surface such that eight supercharges are preserved. Class $\mathcal{S}$ theories are typically strongly-coupled non-Lagrangian field theories, and thus there are the usual challenges in determining some of their properties; the advantage of the construction from six dimensions is that many of the physical features and properties in 4d are captured by the choice of Riemann surface. Each $n$-punctured genus $g$ Riemann surface, with $2g - 2 + n > 0$, has a pair-of-pants decomposition in terms of three-punctured spheres, glued together along the punctures. Different pair-of-pants decompositions correspond to different duality frames of the 4d $\mathcal{N}=2$ SCFT, and thus non-perturbative S-duality transformations are simply encoded in the geometry of the compactification space. To understand the space of SCFTs that arise in this way it is necessary to understand the data that specifies each possible puncture, which three-punctured spheres exist, and how they can be gauged together. 

Punctures describe codimension-two defects in the 6d theory; in particular, each nilpotent orbit $\rho: SU(2) \rightarrow J$ constitutes such a defect \cite{Chacaltana:2012zy}. Nilpotent orbits of simple Lie algebras are well-understood \cite{MR417306,MR417307}, and the properties of the associated punctures have been exhaustively worked out in \cite{Chacaltana:2010ks,Chacaltana:2011ze,Chacaltana:2014jba,Chacaltana:2017boe,Chacaltana:2018vhp}. We can also consider twisted defects, which involves a twist by an outer-automorphism of $J$ when encircling the defect. This class of codimension-two defects, which is central to the current work, is known as the twisted punctures, and they are captured by nilpotent orbits $\rho: SU(2) \rightarrow G$, where $G$ is defined such that $G^\vee$ is the subalgebra of $J$ invariant under the action of the outer-automorphism. Twisted punctures are required to come in sets, and inside of each set, they are connected via a twist line. The regular twisted punctures have been enumerated for $G = A_{2n+1}, D_n, E_6$ in \cite{Chacaltana:2012ch,Chacaltana:2013oka,Chacaltana:2015bna,Chacaltana:2016shw, Tachikawa:2010vg, Distler:2021cwz}. Studying the $A_{2n}$ theory with twisted punctures is particularly subtle, and several preliminary steps have appeared in \cite{Chacaltana:2014nya,Tachikawa:2018rgw,Beem:2020pry,Beratto:2020wmn}. 

In \cite{Gadde:2011uv}, it was conjectured that the Higgs branch chiral ring and the Hall--Littlewood chiral ring are identical for all class $\mathcal{S}$ theories obtained from genus-zero Riemann surfaces. Alternatively, this is equivalent to the statement that the Hall--Littlewood chiral ring of such 4d $\mathcal{N}=2$ SCFTs does not contain any $\mathcal{D}_{R(0,j_2)}$-type multiplets. In principle, this conjecture is particularly powerful, as it enables us to calculate the Hilbert series of the Higgs branch via the determination of the Hall--Littlewood index even for non-Lagrangian theories, for which there exists a TQFT approach \cite{Gadde:2011uv, Lemos:2012ph, Mekareeya:2012tn, Beem:2020pry}. However, as we explain in this paper, this conjecture only holds in a more restrictive setting. In particular, we conjecture the following.
\begin{conj}\label{conj:HLneqHS}
Any class $\mathcal{S}$ theory associated to a genus-zero Riemann surface with at least four $\mathbb{Z}_2$-twisted punctures has a Hall--Littlewood index which is different from the Hilbert series of the Higgs branch. 
\end{conj}
Conjecture \ref{conj:HLneqHS}, in the context of class $\mathcal{S}$ construction with a genus zero surface, provides that the expectation of the Hall--Littlewood index being identical to the Hilbert series of the Higgs branch fails. A priori, to verify this conjecture in any given example, it is necessary to determine both the Hall--Littlewood index and the Higgs branch Hilbert series of a 4d $\mathcal{N}=2$ SCFT. In practice, it often suffices to determine the Hall--Littlewood index as we find that, ubiquitously for the A-series and often for the DE-series, the fully flavor-fugacity refined index contains negative coefficients.\footnote{Specifically, we note that in the unrefined index, where the characters of the flavor algebras are equal to the dimension of the relevant representations, we may not observe these negative coefficients; it is thus important that we are considering the fully flavor-refined index.} Since the Hilbert series of the Higgs branch is a Hilbert series, it cannot have any negative coefficients, and thus it suffices to show that the Hilbert series of the Higgs branch and the Hall--Littlewood index are two distinct quantities.

In fact, consistent with what we find in this paper, it has recently been proposed using the Vertex Operator Algebra (VOA) perspective in \cite{Beem:2022mde} that, for class $\mathcal{S}$ theories with two or more twist lines, there is usually a residual unbroken gauge symmetry on the most generic point of the Higgs branch. The gauginos contained in the associated massless vector multiplets can then lead to gauge invariant operators that correspond to $\mathcal{D}$-type multiplets which might make the Hall--Littlewood index differ from the Hilbert series of the Higgs branch, as stated in Conjecture \ref{conj:HLneqHS}. 

\begin{figure}[H]
    \centering
    \begin{tikzpicture}[scale=.6,every node/.style={scale=.8}]
    \draw[thick,red,fill=white] (-1,-3) ellipse (2cm and .8cm);
    \draw[thick,red,fill=white] (-3,-2.9)--(-1,0)--(1,-2.9);
    \shade[inner color=red!60, outer color=white, opacity = 0.4] (-3.75,-3.9)--(-1,0)--(1.75,-3.9);
    \draw[thick,dashed,red] (1,-3) arc (0:180:2cm and .8cm);
    \draw (-3,2.9)--(-1,0)--(1,2.9); 
    \draw[thick,blue,fill=white] (-3,2.9)--(-1,0)--(1,2.9);
    \shade[inner color=blue!60, outer color=white, opacity = 0.4] (-3,2.9)--(-1,0)--(1,2.9);
    \draw[thick,blue,fill=white] (-1,3) ellipse (2cm and .8cm);

    \node[draw=none,opacity=0,thick,scale=0.1,fill=black,label={[label distance=-2mm,color=red]0:\Large{$\mathcal{C}$}}] (C) at (-3.6,-1.7) {};
    \node[draw=none,opacity=0,thick,scale=0.1,fill=black,label={[label distance=-2mm,color=blue]0:\Large{$\mathcal{H}$}}] (H) at (-3.6,1.7) {};
    
    \draw[thick,red,fill=white] (14,-3) ellipse (2cm and .8cm);
    \draw[thick,red,fill=white] (12,-2.9)--(14,0)--(16,-2.9);
    \shade[inner color=red!60, outer color=white, opacity = 0.4] (11.25,-3.9)--(14,0)--(16.75,-3.9);
    \draw[thick,dashed,red] (16,-3) arc (0:180:2cm and .8cm);
    \draw (12,2.9)--(14,0)--(16,2.9); 
    \draw[thick,blue,fill=white] (12,2.9)--(14,0)--(16,2.9);
    \shade[inner color=blue!60, outer color=white, opacity = 0.4] (12,2.9)--(14,0)--(16,2.9);
    \draw[thick,blue,fill=white] (14,3) ellipse (2cm and .8cm);
    
    \node[draw=none,opacity=0,thick,scale=0.1,fill=black,label={[label distance=-2mm,color=red]0:\Large{$\widetilde{\mathcal{H}}$}}] (C2) at (16,-1.7) {};
    \node[draw=none,opacity=0,thick,scale=0.1,fill=black,label={[label distance=-2mm,color=blue]0:\Large{$\widetilde{\mathcal{C}}$}}] (H2) at (16,1.7) {};
    
    \node[draw=none,opacity=0,thick,scale=0.1,fill=black,label={[label distance=0mm]0:$\Large{\xleftrightarrow{\phantom{aaaaa}\text{\Large{mirror}}\phantom{aaaaa}}}$}] (A) at (3.25,0) {};
    \end{tikzpicture}
    \caption{Typical moduli space of 3d $\mathcal{N}=4$ SCFTs have two branches, each of which is a hyperkahler cone: the Higgs branch ($\mathcal{H}$) and the Coulomb branch ($\mathcal{C}$). Three-dimensional mirror refers to a dual gauge theory that flows to the same fixed point in the IR, with the Higgs and Coulomb branch swapped. $\widetilde{\mathcal{H}}$ and $\widetilde{\mathcal{C}}$ denote the Higgs and the Coulomb branch, respectively, of the mirror dual theory, and the colors encode the mirror map $\mathcal{H}=\widetilde{\mathcal{C}}$ and $\mathcal{C}=\widetilde{\mathcal{H}}$.}
    \label{fig:3dbranch}
\end{figure}
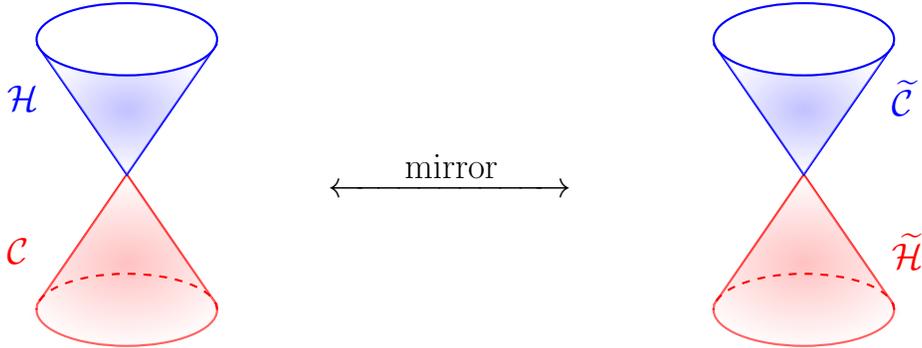

While the observation of the negative coefficients suffices to verify Conjecture \ref{conj:HLneqHS} (and thus that the Hall--Littlewood index cannot be used as a shortcut to the Higgs branch Hilbert series in the presence of multiple twist lines), we want to determine the contents of the Higgs branch. An alternative method for determining the properties of the Higgs branch of a 4d $\mathcal{N}=2$ SCFT is to consider its 3d mirror theory. That is, we consider the compactification first on an $S^1$; this leads to a theory known as the \emph{3d reduction}, where the Higgs branch is the same as the Higgs branch of the 4d theory, and the 3d Coulomb branch is a torus fibration over the 4d Coulomb branch \cite{Seiberg:1996nz}. One can often find another  3d $\mathcal{N}=4$ gauge theory, the \emph{3d mirror}, where the Higgs and Coulomb branches are interchanged \cite{Intriligator:1996ex}. The Hilbert series of the Higgs branch of the 4d SCFT is then equivalent to the Coulomb branch Hilbert series of the 3d mirror. The distinction between the structure of the branches of moduli of the 4d $\mathcal{N}=2$ SCFTs and the $S^1$-reduced 3d $\mathcal{N}=4$ SCFTs can be seen by comparing Figure \ref{fig:4dbranch} with Figure \ref{fig:3dbranch}.

\begin{figure}[H]
    \centering
        \begin{tikzpicture}
            \node[empty] (t3) {$T_{\rho_m}(SU(2n+1))$};
            \node[gauge] (t4) [above right=0.3cm and 1.1cm of t3] {$USp(2n)$};
            \node[empty] (t5) [below right=0.3cm and 1.1cm of t4] {$T_{\sigma_{2k}}(USp'(2n))$};
            \node[empty] (tr) [right=1.3cm of t4] {$\vdots$};
            \node[empty] (tl) [left=1.3cm of t4] {$\vdots$};
            \node[empty] (t3u) [above left=0.3cm and 1.1cm of t4] {$T_{\rho_1}(SU(2n+1))$};
            \node[empty] (t5u) [above right=0.3cm and 1.1cm of t4] {$T_{\sigma_{1}}(USp'(2n))$};
            \node[flavor] (t9) [above=0.5cm of t4] {$2(k-1)$};
            \node[empty] (tw) [right=3.7cm of t4] {$+\quad (k-1)\ \text{free hypermultiplets}$};
            \draw (t3) -- (t4) -- (t5);
            \draw(t3u) -- (t4) -- (t5u);
            \draw (t4) -- (t9);
    	\draw[black,solid] (t4) edge [out=12+45+180,in=78+45+180,loop,looseness=6.5]  node[anchor=south,yshift=-16pt] {$k-1$} (t4);
        \end{tikzpicture}
    \caption{The 3d mirror for the class $\mathcal{S}$ theory of type $A_{2n}$ with $m$ untwisted punctures, labeled by $\rho_1, \cdots, \rho_m$, and $2k$ twisted punctures, labeled by $\sigma_1, \cdots, \sigma_{2k}$, as described in Conjecture \ref{conj:3dmirror}.}
    \label{fig:conjmirror}
\end{figure}
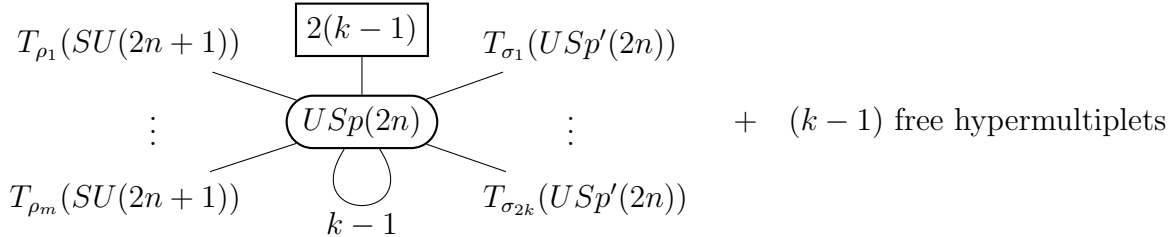

The 3d mirrors of class $\mathcal{S}$ theories generically take the form of star-shaped quivers \cite{Benini:2010uu} where each leg of the star is one of the $T_\rho(G)$ theories of \cite{Gaiotto:2008ak}. Many of our examples lie in the realm of class $\mathcal{S}$ theories of type ${A}_\text{even}$ with twisted punctures, for which the 3d mirrors in the case of spheres with one twist lines have recently been explored in \cite{Beratto:2020wmn}. Once the 3d mirror is known, its Coulomb branch Hilbert series can then be efficiently computed using the techniques of \cite{Cremonesi:2013lqa,Cremonesi:2014kwa,Cremonesi:2014vla,Cremonesi:2014uva,Beratto:2020wmn}. We extend the results of \cite{Beratto:2020wmn} via a conjecture for the Lagrangian description of the 3d mirrors of a class $\mathcal{S}$ theory of type ${A}_\text{even}$ with arbitrary numbers of, twisted and untwisted, punctures. We conjecture the following.
\begin{conj}\label{conj:3dmirror}
  The 3d mirror for a class $\mathcal{S}$ theory of type $A_{2n}$ obtained from a sphere with $m$ untwisted punctures and $2k$ twisted punctures is given by the following Lagrangian quiver. For each untwisted puncture consider the theory $T_{\rho_i}(SU(2n+1))$, where $\rho_i$ is the partition describing the $i$th untwisted puncture; similarly, for each twisted puncture consider $T_{\sigma_j}(USp'(2n))$, where $\sigma_j$ is the C-partition describing the $j$th twisted puncture. Gauge the diagonal $USp(2n)$ (sub-)group of the flavor symmetry of each of these theories; add $2(k-1)$ fundamental hypermultiplets and $k-1$ anti-symmetric hypermultiplets to the introduced $USp(2n)$ gauge node. Finally, include an additional $k-1$ free hypermultiplets.\footnote{The second anti-symmetric power of the fundamental representation of $USp(2n)$ is not an irreducible representation; it is the sum of a $(2n+1)(n-1)$ dimensional representation and a singlet. Throughout this paper, we refer to the $(2n+1)(n-1)$ dimensional irreducible representation as the anti-symmetric representation. With this convention the anti-symmetric representation of $SU(2)$ does not exist.}
\end{conj}
We give a graphical description of the 3d mirror described in Conjecture \ref{conj:3dmirror} in Figure \ref{fig:conjmirror}.  We explain and discuss the content of this conjecture and its derivation in detail in Section \ref{sec:3dmirror}. Given this description for the 3d mirror, we can now compute the Hilbert series of the Higgs branch for the 4d $\mathcal{N}=2$ SCFTs in which we are interested. In fact, we notice that to leading orders the Higgs branch Hilbert series agrees with the Hall--Littlewood index, except for the terms with negative coefficients in the latter. At higher order the two series diverge further. In this way, while we cannot use the tool of the Hall--Littlewood index to directly extract the Higgs branch, we can instead utilize the 3d mirror description from Conjecture \ref{conj:3dmirror}.

The 3d theories arising from dimensional reduction of class $\mathcal{S}$ theories with multiple twist lines have also another interesting feature. In some cases, especially when dealing with theories of $A_{2n}$-type, the 3d reduction is itself Lagrangian.  Nevertheless, in all the examples we consider this turns out to be a \emph{bad} three-dimensional theory in the Gaiotto--Witten sense \cite{Gaiotto:2008ak}. A simple example in twisted $D_n$-type class $\mathcal{S}$ is the $SO(2N)$ superconformal SQCD, which is Lagrangian already in 4d and thus leads to the bad $SO(2N)$ SQCD with $2N-2$ flavors upon 3d reduction. The moduli space of vacua of bad theories has a complicated structure, as investigated for example in \cite{Assel:2017jgo,Assel:2018exy,bad}. In particular, the Coulomb branch of the theory may have several distinct singular loci, from each of which a Higgs branch emanates. Because of this, there is no notion of a globally defined 3d mirror. Instead, we can find different mirror dual descriptions that are valid around each of the singular loci. The 3d mirror, which for the $A_{2n}$-type we propose in Conjecture \ref{conj:3dmirror}, should be understood as the one valid around the singular locus on the Coulomb branch of the 3d reduction of lowest codimension and its Coulomb branch is expected to capture the Higgs branch of the original class $\mathcal{S}$ theory. Similar observations have been made in \cite{Closset:2020afy,Carta:2022spy} for different 4d $\mathcal{N}=2$ SCFTs than those considered in this paper.

We notice that one class of recently explored theories \cite{Kang:2021ccs} with identical central charges, $a=c$, also has a realization in terms of a genus-zero class $\mathcal{S}$ theory with four $\mathbb{Z}_2$-twisted punctures, and thus this class of $a=c$ theories also has HL $\neq$ HS. In fact, it is algorithmic to  explore the space of class $\mathcal{S}$ theories and attempt to find genera and collections of punctures such that $a=c$. It turns out, at least based on the limited scan which we performed for the purposes of this paper, that all such class $\mathcal{S}$ theories either involve a genus $g > 0$ Riemann surface, or else they involve more than four twisted punctures. For class $\mathcal{S}$ compactifications on higher genus Riemann surfaces, it is known that the resulting Hall--Littlewood chiral ring contains $\mathcal{D}$-type multiplets \cite{Gadde:2011uv}. Combining this with Conjecture \ref{conj:HLneqHS} motivates us to tentatively conjecture the following.
\begin{conj} \label{conj:equalac}
Any 4d $\CN \geq 2$ SCFT with $a=c$, such that the Higgs branch is non-trivial, has a  Hall--Littlewood index different from the Hilbert series of the Higgs branch.
\end{conj}
An immediate consequence of this conjecture is that all $\CN=3, 4$ SCFTs have HL $\neq$ HS. This fact is indeed true for $\SN{4}$ super-Yang--Mills theories, and we expect it to be true for $\mathcal{N}=3$ theories due to the additional supercurrents transforming in $\mathcal{D}$-type multiplets. We check this conjecture explicitly for a number of $\SN{3}$ SCFTs where the Macdonald index is known \cite{Agarwal:2021oyl}. It would be interesting to understand how, and whether, the vanishing of $c-a$ in the presence of non-trivial Higgs branch requires the presence of $\mathcal{D}$-type multiplets inside of the Hall--Littlewood chiral ring, however, we leave this question for future work.

The structure of the remainder of this paper is as follows. In Section \ref{sec:Aeven}, we determine the Hall--Littlewood index and the Hilbert series of the Higgs branch for a variety of class $\mathcal{S}$ theories of type ${A}_\text{even}$. We proceed in order of increasing complexity: in Section \ref{sec:A2w4p} we study class $\mathcal{S}$ of type $A_{2}$ on spheres with four-punctures; in Section \ref{sec:A2wmorethan4p} we increase the number of punctures beyond four; and finally we increase the rank to $A_{2n}$ in Section \ref{sec:A2nw4tw}. In all cases, we verify Conjecture \ref{conj:HLneqHS}. In Section \ref{sec:3dmirror}, we provide an a priori motivation for Conjecture \ref{conj:3dmirror}. Next, in Section \ref{sec:OtherTypes}, we show that class $\mathcal{S}$ theories of type $J \neq A_\text{even}$ that permit $\mathbb{Z}_2$-twisted punctures also satisfy Conjecture \ref{conj:HLneqHS}; we look at examples for $G = A_3$, $D_4$, and $E_6$ in Sections \ref{sec:A3}, \ref{sec:D4}, and \ref{sec:E6}, respectively. In Section \ref{sec:ac}, we study Conjecture \ref{conj:equalac}. In Section \ref{sec:acS} we find many new families of 4d $\mathcal{N}=2$ SCFTs with $a=c$ inside class $\mathcal{S}$, and they all involve at least four $\mathbb{Z}_2$-twisted punctures, or have $g > 0$; we also take the Hall--Littlewood index of a collection of $\mathcal{N}=3$ SCFTs in Section \ref{sec:n3scft}, and observe that there are again negative coefficients. In Section \ref{sec:conc}, we conclude and discuss a variety of future directions. To finish, we provide a set of appendices. In Appendix \ref{app:HLeqs}, we summarize the procedure to determine the Hall--Littlewood index of a class $\mathcal{S}$ theory, and similarly in Appendix \ref{app:HSeqs}, we describe how to determine the Hilbert series of the Coulomb or Higgs branch of a 3d $\mathcal{N}=4$ SCFT. We review the algorithm for constructing the 3d mirrors of general class $\mathcal{S}$ theories in Appendix \ref{app:3dmirror}. We include a short summary table of the Hall--Littlewood indices and Higgs branch Hilbert series that we determine for class $\mathcal{S}$ theories of type $A_2$ throughout this paper in Appendix \ref{app:summaryHLHS}. Finally, in Appendix \ref{app:HBHSD4hatSU3}, we explain an explicit, brute-force, calculation of the Higgs branch Hilbert series for the $\widehat{D}_4(SU(3))$ theory.

\section{Class \texorpdfstring{\boldmath{$\mathcal{S}$}}{S} of type \texorpdfstring{\boldmath{$A_\text{even}$}}{Aeven}}\label{sec:Aeven}

In this section, we focus on the study of class $\mathcal{S}$ theories of type $A_{2n}$ on spheres with $\mathbb{Z}_2$-twisted punctures. The aim is to provide evidence for Conjecture \ref{conj:HLneqHS}, which we do by explicitly computing the Hall--Littlewood index for a significant number of examples. We find that the theories with more than two twisted punctures have a Hall--Littlewood index which is distinct from the Hilbert series of the Higgs branch due to the presence of negative coefficients in the flavor-refined Hall--Littlewood index. While this is already an interesting observation in its own right, we would like to determine the Hilbert series of the Higgs branch itself. To this end, we propose and provide evidence for the construction of the 3d mirror duals of the circle reduction of such class $\mathcal{S}$ theories; this proposal provides the content of Conjecture \ref{conj:3dmirror}.

In certain circumstances, which we delineate, the 3d reduction is itself a Lagrangian quiver; though these quivers are often ``bad'', in the sense of \cite{Gaiotto:2008ak}. Bad theories typically have a Coulomb branch with multiple singular loci and from each of these a non-trivial Higgs branch might emanate. This was discussed in \cite{Assel:2017jgo} (see also \cite{bad}) for the case of the bad $U(N)$ SQCD, where it turns out that the Higgs branch at the most singular locus of the Coulomb branch, that is the locus with the highest codimension, contains those at the less singular loci as subvarieties. We see that in our case the 4d Higgs branch is captured by the Higgs branch of the 3d reduction at the most singular locus of its Coulomb branch, similar to what was found in \cite{Closset:2020afy,Carta:2022spy} for different examples. We will compute the Higgs branch Hilbert series either by an explicit (though computationally intensive) computation using software such as {\tt Macaulay2}, or else one can use sequences of known dualities, valid only around the most singular locus of the Coulomb branch, to relate the bad quiver to a good quiver, for which the Hilbert series can be extracted using the standard techniques. When such calculations are possible, the resulting Higgs branch Hilbert series agrees with the Coulomb branch Hilbert series of the proposed 3d mirror.

Another powerful motivation for the 3d mirrors that we propose comes from the comparison with the Hall--Littlewood indices, which we determine. The Coulomb branch Hilbert series of the proposed mirror and the Hall--Littlewood index of the class $\mathcal{S}$ theory agree, modulo the negative terms in the Hall--Littlewood index, up to and including the first order at which the negative coefficients appear.\footnote{As we increase the order, we expect the Hall--Littlewood index to diverge more-and-more from the Hilbert series of the Higgs branch, as the $\mathcal{D}$-type multiplets not only contribute positively at higher orders, but they can form composite operators with the $\mathcal{B}$-type Higgs branch operators.} Furthermore, the dimensions of the Higgs/Coulomb branches of the 3d reduction of the class $\mathcal{S}$ theory and the Coulomb/Higgs branches of the proposed 3d mirror agree.

\subsection{\texorpdfstring{$A_2$}{A2} on four-punctured spheres}\label{sec:A2w4p}

We begin by studying class $\mathcal{S}$ theories of type $A_2$ on spheres with four punctures. There are in total four different possible types of punctures in the $A_2$ theory: two untwisted punctures associated to integer partitions of three,\footnote{There are, of course, three integer partitions of three, but the partition $[3]$ is associated to the trivial puncture, which is equivalent to no puncture.} and two $\mathbb{Z}_2$-twisted punctures associated to integer partitions of two. These four punctures, together with some of their physical properties, are written in Table \ref{tbl:puncts}. Recalling that $\mathbb{Z}_2$-twisted punctures are required to come in pairs connected by twist lines, we see that there are nineteen possible four-punctured spheres.

\begin{table}[H]
    \centering
    \renewcommand{\arraystretch}{1.2}
    \begin{threeparttable}
    \begin{tabular}{ccccc}
        \toprule
        Partition & $\delta a$ & $\delta c$ & $F$ \\\midrule
        $[1^{2n+1}]$ & $\frac{1}{24} n \left(32 n^2+38 n+11\right)$ & $\frac{1}{6} n \left(8 n^2+10 n+3\right)$ & $SU(2n+1)$  \\
        $[2n, 1]$ & $n^2+n+\frac{1}{24}$ & $ n^2+n+ \frac{1}{12}$ & $U(1)$ \\
        $[1^{2n}]_t$ & $\frac{1}{48} n \left(64 n^2+86 n+37\right)$ & $\frac{1}{12} n \left(16 n^2+22 n+9\right)$ & $USp(2n)$ \\
        $[2n]_t$ & $n^3+\frac{3 }{2}n^2+\frac{29 }{48} n$ & $n^3+\frac{3 }{2}n^2+\frac{7}{12}n $ & $\varnothing$ \\\bottomrule
    \end{tabular}
    \end{threeparttable}
    \caption{The maximal and minimal untwisted and twisted punctures in class $\mathcal{S}$ theories of type $A_{2n}$ and their contribution to the central charges and flavor symmetries \cite{Chacaltana:2012zy}. The minimal twisted puncture, $[2n]_t$, is also known as the twisted null puncture. When $n = 1$ these are all the possible punctures. The subscript $t$ emphasizes that the puncture is $\mathbb{Z}_2$-twisted.}
    \label{tbl:puncts}
\end{table}

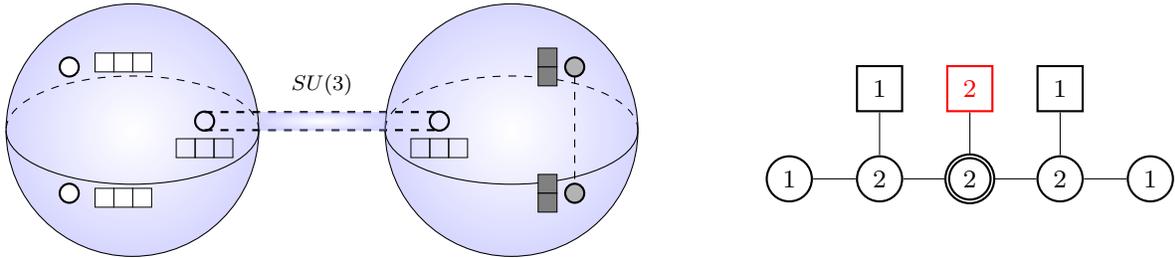
\begin{figure}[H]
    \centering
    \begin{subfigure}[b]{0.54\textwidth}
        \centering
        \begin{tikzpicture}[scale=1.2]
        \shade[inner color=white, outer color=blue!40, opacity = 0.4] (0,0) circle (1.4cm);
        \draw (0,0) circle (1.4cm);
        \draw (-1.4,0) arc (180:360:1.4 and 0.6);
        \draw[dashed] (1.4,0) arc (0:180:1.4 and 0.6);
        \draw[thick,fill=white] (-.7,.7) circle (3pt);
        \node[font=\scriptsize] (P1) at (-.1,.75) {$\ydiagram[*(white)]{3}$};
        \draw[thick] (.8,.1) circle (3pt);
        \node[font=\scriptsize] (P2) at (.8,-.2) {$\ydiagram[]{3}$};
        \draw[thick,fill=white] (-.7,-.7) circle (3pt);
        \node[font=\scriptsize] (P3) at (-.1,-.75) {$\ydiagram[*(white)]{3}$};
        \shade[inner color=white, outer color=blue!40, opacity = 0.4] (4.2,0) circle (1.4cm);
        \draw (4.2,0) circle (1.4cm);
        \draw (2.8,0) arc (180:360:1.4 and 0.6);
        \draw[dashed] (5.6,0) arc (0:180:1.4 and 0.6);
        \draw[thick,fill=gray!60] (4.9,.7) circle (3pt);
        \node[font=\scriptsize] (P4) at (4.6,.7) {$$\ydiagram[*(gray)]{1,1}$$};
        \draw[thick] (3.4,.1) circle (3pt);
        \node[font=\scriptsize] (P5) at (3.4,-.2) {$\ydiagram[]{3}$};
        \draw[thick,fill=gray!60] (4.9,-.7) circle (3pt);
        \node[font=\scriptsize] (P6) at (4.6,-.7) {$\ydiagram[*(gray)]{1,1}$};
        \draw[dashed] (4.9,.6)--(4.9,-.6);
        \draw[thick,dashed] (.8,0)--(3.4,0);
        \draw[thick,dashed] (.8,0.2)--(3.4,0.2);
        \shade[inner color=white, outer color=blue!40, opacity = 0.4] (1.4,0) rectangle ++(1.4,0.2);
        \node[font=\scriptsize] (c) at (2.1,0.5) {$SU(3)$};
        \end{tikzpicture}
        \caption{Each puncture is depicted as a circle; the Young tableau describing the type of puncture is 
        white if the puncture is untwisted, gray if it is twisted, and transparent if it is glued. 
        Twist lines, connecting twisted punctures, are denoted by dashed lines.}\label{fig:ex1a}
    \end{subfigure}\hfill
    \begin{subfigure}[b]{0.43\textwidth}
        \centering
        \raisebox{7mm}{
        \begin{tikzpicture}[scale=1.2,every node/.style={scale=1.2},font=\scriptsize]
        \node[gauge] (p0) at (0,0) {$1$};
        \node[gauge] (p1) at (1,0) {$2$};
        \node[flavor] (p1F) at (1,1) {$1$};
        \node[gauge] (s0) at (4,0) {$1$};
        \node[gauge] (s1) at (3,0) {$2$};
        \node[flavor] (s1F) at (3,1) {$1$};
        \node[gaugeD] (c) at (2,0) {$2$};
        \node[flavor,red] (t0) at (2,1) {$2$};
        \draw (p0)--(p1)--(c)--(s1)--(s0);
        \draw (p1)--(p1F);
        \draw (s1)--(s1F);
        \draw (c)--(t0);
        \end{tikzpicture}}
        \caption{Black single (resp. double) bordered nodes denote unitary (resp. special unitary) groups; red nodes denote special orthogonal groups. Links denote bifundamental (half-)hypermultiplets.
        }\label{fig:ex1b}
    \end{subfigure}
    \caption{In (a), we depict the $A_2$ type class $\CS$ theory on a four punctured sphere with punctures $2\times[1^3]+2\times[2]_t$.   In (b), we show the Lagrangian quiver describing the mirror dual of the 3d reduction. }
    \label{fig:ex1}
\end{figure}

Four-punctured spheres for the $A_2$ theory without twist lines have been enumerated in \cite{Gaiotto:2009we, Chacaltana:2010ks}; in this section, we focus on the contrast between theories with one twist line and theories with two twist lines. We begin with the theory involving two untwisted maximal punctures and two twisted null punctures, denoted by $2\times[1^3]+2\times[2]_t$. This class $\mathcal{S}$ configuration is drawn in Figure \ref{fig:ex1a}, and it has one twist line connecting the two twisted null punctures. The resulting 4d $\mathcal{N}=2$ SCFT has $SU(3)^2$ flavor symmetry,\footnote{In this paper we only focus on the flavor symmetry algebra and we ignore the global structure of the flavor symmetry group. The latter was studied for class $\mathcal{S}$ theories in \cite{Bhardwaj:2021ojs}.} one factor coming from each $[1^3]$ puncture. The Hall--Littlewood index can be computed using the TQFT description \cite{Gadde:2011uv, Lemos:2012ph, Mekareeya:2012tn, Beem:2020pry}, which is given as
\begin{equation}
    \begin{aligned}
        \text{HL}(\tau,a,b)
        =&\,1+\tau^2(\textbf{adj})+\tau^4(\textbf{adj}^2+\textbf{adj}+1)  +\tau^6\Big(\textbf{adj}^3
        +\textbf{adj}^2+2\chi_{\bf{8}}(a)\chi_{\bf{8}}(b)\\
        &+\textbf{adj}+\chi_{\bf{10}}(a)+\chi_{\bf{\overline{10}}}(a)
        +\chi_{\bf{10}}(b)+\chi_{\bf{\overline{10}}}(b)\Big)+\mathcal{O}(\tau^8)\,,
    \end{aligned}\label{eqn:2max2mint}
\end{equation}
where $a,b$ are the fugacities for the $SU(3)$ flavor symmetries. In addition, we have defined the following short-hand notation for future convenience:
\begin{align}\label{eq:adjchar}
    \textbf{adj}^n=\sum_{n_1+n_2+\cdots n_i=n}\chi_{n_1\textbf{adj}}(a_1)\cdots\chi_{n_i\textbf{adj}}(a_i)\,,
\end{align}
where $\chi_{n_i\textbf{adj}}(a_i)$ is the character of the $i$th flavor symmetry in the representation having as its highest weight $n_i$ times that of the adjoint representation. If the theory has no flavor symmetry then $\textbf{adj}^n$ vanishes for all $n$.

The coefficients of the Hall--Littlewood index in equation \eqref{eqn:2max2mint} are characters of the $SU(3)^2$ flavor algebra. In particular, at order $\tau^2$ we can see the moment map operators for the $SU(3)^2$ flavor symmetry appearing in the adjoint representation. All characters appear with positive coefficients, consistent with equation \eqref{eqn:2max2mint} also being the Hilbert series of the Higgs branch.

To verify that the expression in equation \eqref{eqn:2max2mint} indeed coincides with the Hilbert series of the Higgs branch, we explicitly determine the Coulomb branch Hilbert series of the 3d mirror. The latter can be derived following the prescription of \cite{Beratto:2020wmn}, which we review in Appendix \ref{app:3dmirror}. Specifically, to each $[1^3]$ puncture we associate one copy of the $T(SU(3))$ theory, while to each $[2]_t$ puncture we associate one copy of the $T_{[2]}(USp'(2))$ theory.\footnote{We follow the standard nomenclature used in the literature on $T^\sigma_\rho(G)$ theories, where we omit writing maximal partitions. For example, here $T(SU(3))$ stands for $T^{[1^3]}_{[1^3]}(SU(3))$.}  Each of these theories is glued together via a $USp(2)=SU(2)$ gauging, where for the $T(SU(3))$ theories we first have to decompose $SU(3)\supset SU(2)\times U(1)$ resulting in one additional fundamental hypermultiplet attached to each unitary node that is adjacent to the central $SU(2)$ gauge node of the full quiver. The result is drawn in Figure \ref{fig:ex1b}. The Coulomb branch Hilbert series of this quiver can be computed using the techniques of \cite{Cremonesi:2013lqa,Cremonesi:2014kwa,Cremonesi:2014vla,Cremonesi:2014uva} (see Appendix \ref{app:HSeqs} for a review of the relevant formulae) and we find perfect agreement with equation \eqref{eqn:2max2mint}.

\begin{figure}[H]
    \centering
    \begin{subfigure}[b]{0.54\textwidth}
        \centering
        \begin{tikzpicture}[scale=1.2]
        \shade[inner color=white, outer color=blue!40, opacity = 0.4] (0,0) circle (1.4cm);
        \draw (0,0) circle (1.4cm);
        \draw (-1.4,0) arc (180:360:1.4 and 0.6);
        \draw[dashed] (1.4,0) arc (0:180:1.4 and 0.6);
        \draw[thick,fill=white] (-.7,.7) circle (3pt);
        \node[font=\scriptsize] (P1) at (-.1,.75) {$\ydiagram[*(white)]{3}$};
        \draw[thick] (.8,.1) circle (3pt);
        \node[font=\scriptsize] (P2) at (.8,-.2) {$\ydiagram[]{3}$};
        \draw[thick,fill=white] (-.7,-.7) circle (3pt);
        \node[font=\scriptsize] (P3) at (-.1,-.75) {$\ydiagram[*(white)]{3}$};
        \shade[inner color=white, outer color=blue!40, opacity = 0.4] (4.2,0) circle (1.4cm);
        \draw (4.2,0) circle (1.4cm);
        \draw (2.8,0) arc (180:360:1.4 and 0.6);
        \draw[dashed] (5.6,0) arc (0:180:1.4 and 0.6);
        \draw[thick,fill=gray!60] (4.9,.7) circle (3pt);
        \node[font=\scriptsize] (P4) at (4.4,.75) {$$\ydiagram[*(gray)]{2}$$};
        \draw[thick] (3.4,.1) circle (3pt);
        \node[font=\scriptsize] (P5) at (3.4,-.2) {$\ydiagram[]{3}$};
        \draw[thick,fill=gray!60] (4.9,-.7) circle (3pt);
        \node[font=\scriptsize] (P6) at (4.4,-.75) {$\ydiagram[*(gray)]{2}$};
        \draw[dashed] (4.9,.6)--(4.9,-.6);
        \draw[thick,dashed] (.8,0)--(3.4,0);
        \draw[thick,dashed] (.8,0.2)--(3.4,0.2);
        \shade[inner color=white, outer color=blue!40, opacity = 0.4] (1.4,0) rectangle ++(1.4,0.2);
        \node[font=\scriptsize] (c) at (2.1,0.5) {$SU(3)$};
        \end{tikzpicture}
        \caption{$A_2$ with $2 \times [1^3] + 2 \times [1^2]_t$.}\label{fig:ex2a}
    \end{subfigure}\hfill
    \begin{subfigure}[b]{0.43\textwidth}
        \centering
        \begin{tikzpicture}[scale=1.2,every node/.style={scale=1.2},font=\scriptsize]
        \node[gauge] (p0) at (0,0) {$1$};
        \node[gauge] (p1) at (1,0) {$2$};
        \node[flavor] (p1F) at (1,-1) {$1$};
        \node[gauge] (s0) at (4,0) {$1$};
        \node[gauge] (s1) at (3,0) {$2$};
        \node[flavor] (s1F) at (3,-1) {$1$};
        \node[gaugeD] (c) at (2,0) {$2$};
        \node[gauge,red] (t0) at (2,-1) {$3$};
        \node[gauge,red] (t1) at (2,1) {$3$};
        \draw (p0)--(p1)--(c)--(s1)--(s0);
        \draw (p1)--(p1F);
        \draw (s1)--(s1F);
        \draw (c)--(t0);
        \draw (c)--(t1);
        \end{tikzpicture}
        \caption{3d mirror.}\label{fig:ex2b}
    \end{subfigure}
    \caption{The $A_2$ type class $\mathcal{S}$ theory on a four punctured sphere with $2\times[1^3]+2\times[1^2]_t$ is shown in (a), while in (b) we depict the 3d mirror.}
    \label{fig:ex2}
\end{figure}
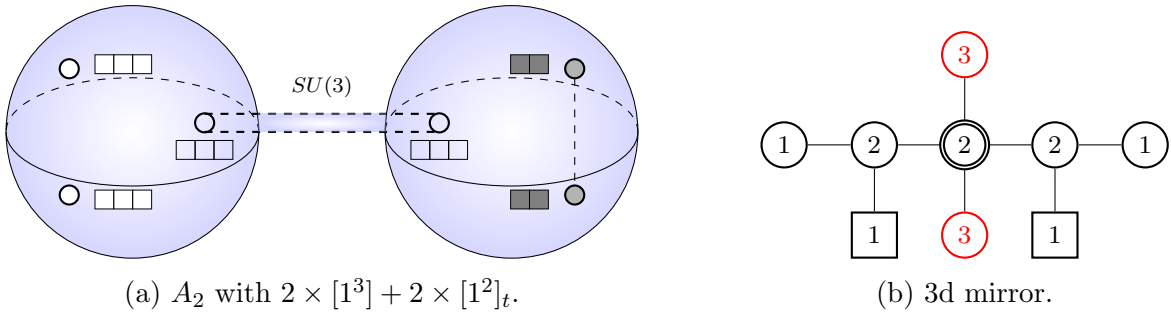

In fact, it is similarly straightforward to show that all nine four-punctured spheres with two untwisted punctures and two twisted punctures have their Hall--Littlewood index equal to the Hilbert series of their Higgs branch. We consider one more example explicitly: the four-punctured sphere with punctures $2\times[1^3]+2\times[1^2]_t$. This configuration is depicted in Figure \ref{fig:ex2a}, and the Hall--Littlewood index of this theory is
\begin{align}
\scalemath{0.97}{
\begin{aligned}
    \text{H} & \text{L}(\tau,a,b,c,d)
    =1+\tau^2(\textbf{adj})+\tau^4(\textbf{adj}^2+\chi_{\bf{8}}(a)+\chi_{\bf{8}}(b)+1)+\tau^6\Big(\textbf{adj}^3+\chi_{\bf{27}}(a)\\
    &\quad +\chi_{\bf{10}}(a)+\chi_{\overline{\bf{10}}}(a)+\chi_{\bf{8}}(a) +\chi_{\overline{\bf{10}}}(b)+\chi_{\bf{8}}(b)+\chi_{\bf{27}}(b)+\chi_{\bf{10}}(b)+\chi_{\bf{3}}(c)+\chi_{\bf{3}}(d) \\
    &\quad
    +2\chi_{\bf{8}}(a)\chi_{\bf{8}}(b)+\chi_{\bf{8}}(a)\chi_{\bf{3}}(c)+\chi_{\bf{8}}(a)\chi_{\bf{3}}(d)+\chi_{\bf{8}}(b)\chi_{\bf{3}}(c)+\chi_{\bf{8}}(b)\chi_{\bf{3}}(d)\Big)+\mathcal{O}\left(\tau^8\right)\,,
\end{aligned}}\label{eq:2max2maxt}
\end{align}
where $a$, $b$ are the fugacities associated to the two $SU(3)$ flavor symmetries, one for each untwisted maximal puncture, and $c$, $d$ are the fugacities for the two $SU(2)$ flavor factors from the twisted punctures.

Again the Hall--Littlewood index has all positive coefficients and thus does not exhibit any evidence that it is not also the Hilbert series of the Higgs branch. Utilizing the 3d mirror we can determine the Higgs branch Hilbert series to confirm that the two indeed coincide. To derive the latter, we consider now two copies of the $T(SU(3))$ theory and two copies of the $T(USp'(2))$ theory, where the latter is an $SO(3)$ gauge theory with two fundamental half-hypermultiplets, and glue them via an $SU(2)$ gauging as in the previous example. The resulting quiver is depicted in Figure \ref{fig:ex2b}; we determine the Hilbert series of the Coulomb branch and observe that it is in perfect agreement with the Hall--Littlewood index in equation  \eqref{eq:2max2maxt}. 

More examples of explicit computations of the Coulomb branch Hilbert series of the 3d mirror to $A_{2n}$ class $\mathcal{S}$ theories on spheres with only one twist line \cite{Beratto:2020wmn} evince that their Hall--Littlewood index and Higgs branch Hilbert series are identical. Instead, we now turn to the study of class $\mathcal{S}$ theories of type $A_2$ with four twisted punctures, to wit, two twist lines. There are five such four-punctured spheres and in each case we verify that the Hall--Littlewood index and the Hilbert series of the Higgs branch are different. 

\begin{figure}[H]
    \centering
    \begin{subfigure}[b]{0.53\textwidth}
        \centering
        \begin{tikzpicture}[scale=1.2]
        \shade[inner color=white, outer color=blue!40, opacity = 0.4] (0,0) circle (1.4cm);
        \draw (0,0) circle (1.4cm);
        \draw (-1.4,0) arc (180:360:1.4 and 0.6);
        \draw[dashed] (1.4,0) arc (0:180:1.4 and 0.6);
        \draw[thick,fill=gray!60] (-.7,.7) circle (3pt);
        \node[font=\scriptsize] (P1) at (-.4,.7) {$\ydiagram[*(gray)]{1,1}$};
        \draw[thick] (.8,.1) circle (3pt);
        \node[font=\scriptsize] (P2) at (.8,-.2) {$\ydiagram[]{3}$};
        \draw[thick,fill=gray!60] (-.7,-.7) circle (3pt);
        \node[font=\scriptsize] (P3) at (-.4,-.7) {$\ydiagram[*(gray)]{1,1}$};
        \draw[dashed] (-.7,.6)--(-.7,-.6);
        \shade[inner color=white, outer color=blue!40, opacity = 0.4] (4.2,0) circle (1.4cm);
        \draw (4.2,0) circle (1.4cm);
        \draw (2.8,0) arc (180:360:1.4 and 0.6);
        \draw[dashed] (5.6,0) arc (0:180:1.4 and 0.6);
        \draw[thick,fill=gray!60] (4.9,.7) circle (3pt);
        \node[font=\scriptsize] (P4) at (4.6,.7) {$$\ydiagram[*(gray)]{1,1}$$};
        \draw[thick] (3.4,.1) circle (3pt);
        \node[font=\scriptsize] (P5) at (3.4,-.2) {$\ydiagram[]{3}$};
        \draw[thick,fill=gray!60] (4.9,-.7) circle (3pt);
        \node[font=\scriptsize] (P6) at (4.6,-.7) {$\ydiagram[*(gray)]{1,1}$};
        \draw[dashed] (4.9,.6)--(4.9,-.6);
        \draw[thick,dashed] (.8,0)--(3.4,0);
        \draw[thick,dashed] (.8,0.2)--(3.4,0.2);
        \shade[inner color=white, outer color=blue!40, opacity = 0.4] (1.4,0) rectangle ++(1.4,0.2);
        \node[font=\scriptsize] (c) at (2.1,0.5) {$SU(3)$};
    \end{tikzpicture}
    \caption{$A_2$ with $4\times [2]_t$.}\label{fig:ex3a}
    \end{subfigure}
    \begin{subfigure}[b]{0.22\textwidth}
    \centering\raisebox{6.5mm}{
        \begin{tikzpicture}[scale=1.3,every node/.style={scale=1.2},font=\scriptsize]
        \node[gauge] (p0) at (0,0) {$1$};
        \node[gauge] (p1) at (1,0) {$1$};
        \node[gauge] (p2) at (0,1) {$1$};
        \node[gauge] (p3) at (1,1) {$1$};
        \node[gaugeD] (c) at (.5,.5) {$3$};
        \draw (p0)--(c)--(p1);
        \draw (p2)--(c)--(p3);
        \end{tikzpicture}}
        \caption{3d reduction.}\label{fig:ex3b}
    \end{subfigure}
    \begin{subfigure}[b]{0.23\textwidth}
    \centering\raisebox{12mm}{
        \begin{tikzpicture}[scale=1.2,every node/.style={scale=1.2},font=\scriptsize]
        \node[gaugeD] (p0) at (0,0) {$2$};
        \node[flavor,red] (p1) at (1,0) {$8$};
        \draw (p0)--(p1);
        \end{tikzpicture}}
        \caption{3d magnetic quiver.}\label{fig:ex3c}
    \end{subfigure}
    \caption{In (a), we show the $A_2$ class $\mathcal{S}$ theory on a sphere with four twisted null punctures. In (b), we depict the Lagrangian quiver describing the 3d reduction. In (c), we portray the magnetic quiver which, together with a free hypermultiplet, is a mirror for the theory in (b) around the most singular locus on its Coulomb branch.}
    \label{fig:ex3}
\end{figure}
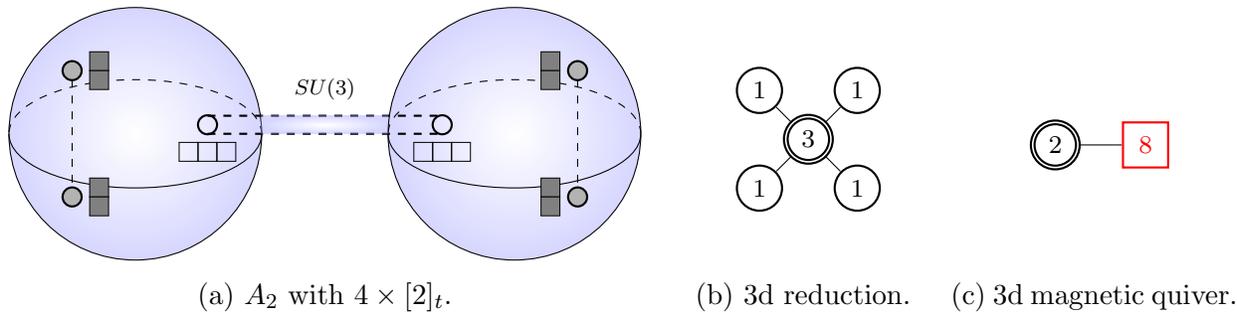

We begin by considering the theory with four twisted null punctures $4\times[2]_t$. This theory can be described as an $SU(3)$ gauging of two copies of a particular three-punctured sphere which has one $[1^3]$ puncture and two $[2]_t$ punctures, as depicted in Figure \ref{fig:ex3a}. The three-punctured sphere in question in fact corresponds to a product theory: it is two copies of the $(A_1, D_4)$ Argyres--Douglas theory \cite{Beem:2020pry}. It is easy to see that the four-punctured sphere thus corresponds to the diagonal $SU(3)$ gauging of four copies of the $(A_1, D_4)$ theory, which is precisely the construction of the $\widehat{D}_4(SU(3))$ SCFT from \cite{Kang:2021lic}. In fact, this theory has identical central charges, a topic to which we return in Section \ref{sec:ac}.

It is straightforward to now use the class $\mathcal{S}$ description to determine the Hall--Littlewood index of the $\widehat{D}_4(SU(3))$ theory. For this theory, we can also use the fact that the $(A_1, D_4)$ theory is the one-instanton theory of $SU(3)$, called $H_2$, where the Higgs branch Hilbert series takes a particularly simple closed form \cite{Benvenuti:2010pq, Keller:2011ek, Keller:2012da}. We can also use the $\CN=1$ Lagrangian description of the $(A_1, D_4)$ theory \cite{Maruyoshi:2016aim, Maruyoshi:2016tqk, Agarwal:2016pjo} to compute the index. Either way, we find
\begin{align}\label{eqn:D4hatSU3HL}
    \begin{split}
        \operatorname{HL}(\tau)=&\,\frac{1+\tau^4-\tau^6}{1-\tau^4}=1+2\tau^4-\tau^6+2\tau^8-\tau^{10}+\mathcal{O}(\tau^{12})\,.
    \end{split}
\end{align}
We immediately notice that there are terms with negative coefficients. Hence, this expression cannot coincide with the Hilbert series of the Higgs branch, as anticipated. We can gain some intuition on the origin of the associated $\mathcal{D}$-type multiplets, i.e., the multiplets which exist inside of the Hall--Littlewood chiral ring but not the Higgs branch chiral ring, by looking at the direct 3d reduction of this theory, which remarkably turns out to be Lagrangian.

The circle compactification of the $(A_1,D_4)$ theory corresponds to the 3d $\mathcal{N}=4$ $U(1)$ gauge theory with three hypermultiplets of unit charge \cite{Xie:2012hs,Buican:2015hsa}. Hence, the 3d reduction of $\widehat{D}_4(SU(3))$ is given by the Lagrangian quiver represented in Figure \ref{fig:ex3b}. It turns out that the matter content is not enough to Higgs the full gauge group of the entire quiver. We can explicitly see this by trying to naively compute the Higgs branch Hilbert series with the standard Molien integral, which assumes that the gauge group is completely Higgsed \cite{Butti:2007jv}. This would give us precisely equation \eqref{eqn:D4hatSU3HL}, which is not a sensible result for a Higgs branch Hilbert series due to the negative terms; thus, our assumption that the gauge group is completely Higgsed at the generic point of the Higgs branch must be violated. Thanks to the simplicity of the quiver, the Hilbert series can still be correctly computed, in a brute-force manner, using \texttt{Macaulay2} \cite{M2} and the result is (see Appendix \ref{app:HBHSD4hatSU3} for more details)
\begin{align}\label{eqn:D4hatSU3HS}
\begin{split}
    \operatorname{HS}(\tau)=&\,\frac{1-\tau^2+\tau^4}{(1-\tau^2)(1-\tau^4)}= 1+2 \tau^4+\tau^6+3 \tau^8+2 \tau^{10}+\mathcal{O}(\tau^{12})\,,
    \end{split}
\end{align} 
which does not have negative terms and thus differs from the Hall--Littlewood index in equation \eqref{eqn:D4hatSU3HL}.

Interestingly, the Higgs branch Hilbert series in equation \eqref{eqn:D4hatSU3HS} coincides with the Higgs branch Hilbert series for a similar 3d quiver gauge theory, where the $SU(3)$ gauge group at the central node is replaced with $SU(2)$, to wit, the affine $D_4$ quiver.\footnote{The 3d rank $n$ affine $J$ quivers are Lagrangian quivers in the shape of the affine Dynkin diagram of $J$, and where each gauge node is $U(n d_i)$ with $d_i$ being the Dynkin label for that particular node of the affine Dynkin diagram. It is clear that there exists a $U(1)$ inside of the gauge group under which no hypermultiplets are charged -- this $U(1)$ thus decouples from the interacting part of the SCFT. In fact, one has the freedom to ``ungauge'' the $U(1)$ on any choice of gauge node; in the case under discussion here, we have decoupled it from the central node to go from $U(2)$ to $SU(2)$.} This suggests that the $U(1)$ which is not completely Higgsed on the Higgs branch corresponds to the $U(1)$ in the decomposition $SU(3)\supset SU(2)\times U(1)$. Furthermore, among the Higgs branch of the $(A_1, D_4)$ theory, given by the closure of the $SU(3)$ minimal nilpotent orbit, only the $SU(2)$ minimal nilpotent orbit that is neutral to this $U(1)$ contributes to the Higgs branch of the $\widehat{D}_4(SU(3))$ theory. Hence, we suspect that the extra $\mathcal{D}$-type multiplets in the Hall--Littlewood index of the 4d theory are due to a $U(1)$ vector multiplet inside the $SU(3)$ gauge group in Figure \ref{fig:ex1a} that remains massless on the Higgs branch of the $\widehat{D}_4(SU(3))$ theory, in accordance with the proposal of \cite{Beem:2022mde}.

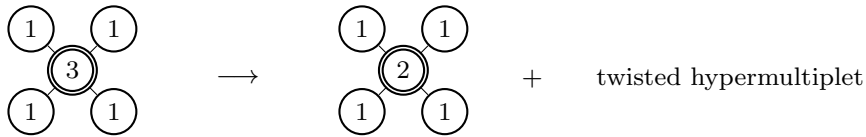
\begin{figure}[H]
    \centering
        \begin{tikzpicture}[scale=1.1,every node/.style={scale=1.2},font=\scriptsize]
        \node[gauge] (p0) at (0,0) {$1$};
        \node[gauge] (p1) at (1,0) {$1$};
        \node[gauge] (p2) at (0,1) {$1$};
        \node[gauge] (p3) at (1,1) {$1$};
        \node[gaugeD] (c) at (.5,.5) {$3$};
        \draw (p0)--(c)--(p1);
        \draw (p2)--(c)--(p3);
        \node[empty] (t) at (2.5,0.4) {$\longrightarrow$};
        \node[empty] (t2) at (8,0.4) {$+\qquad \text{twisted hypermultiplet}$};
        \node[gauge] (p0r) at (4,0) {$1$};
        \node[gauge] (p1r) at (5,0) {$1$};
        \node[gauge] (p2r) at (4,1) {$1$};
        \node[gauge] (p3r) at (5,1) {$1$};
        \node[gaugeD] (cr) at (4.5,.5) {$2$};
        \draw (p0r)--(cr)--(p1r);
        \draw (p2r)--(cr)--(p3r);
        \end{tikzpicture}
    \caption{Around the most singular locus of the Coulomb branch of the bad quiver on the left, one can dualize the theory to the good quiver on the right plus a free twisted hypermultiplet. The Higgs branch Hilbert series, around the most singular locus of the Coulomb branch, can then be computed utilizing the resulting good quiver with the standard Molien integral methods.}
    \label{fig:dual1}
\end{figure}

This observation is not accidental, since we can dualize the theory in Figure \ref{fig:ex3b} so as to obtain the affine $D_4$ quiver theory. Notice that the theory in Figure \ref{fig:ex3b} is ``bad'' in the sense of Gaiotto--Witten \cite{Gaiotto:2008ak}, that is, there are monopole operators whose ultraviolet dimensions violate the unitarity bound. The badness of the quiver is localized at the central $SU(3)$ node, which allows us to manipulate it as follows. First, by a field redefinition we can turn the middle $SU(3)$ gauge node into $U(3)$ while simultaneously making one of the four $U(1)$ gauge nodes a flavor one. The middle $U(3)$ node is bad since it only sees four flavors. As it was discussed in \cite{Assel:2017jgo}, on the most singular locus of the Coulomb branch of $U(3)$ with four flavors, where the Fayet--Iliopoulos (FI) parameter is turned off, the theory is equivalent to $U(2)$ with four flavors, which is now good, and one free twisted hypermultiplet.\footnote{More precisely, by this we mean that the Higgs and Coulomb branch of $U(3)$ with four flavors and $U(2)$ with four flavors and a twisted hypermultiplet are identical locally in this region of the moduli space.} With the same previous operation of selecting the decoupled gauge $U(1)$ we can turn the $U(2)$ node into $SU(2)$ and gauge back the $U(1)$ that we made flavor at the first step. In the end we obtain the affine $D_4$ quiver plus a free twisted hypermultiplet. This operation is depicted in Figure \ref{fig:dual1}.\footnote{Similar manipulations to turn a bad theory into a good one in the vicinity of one of the singular loci of its Coulomb branch have been performed also in \cite{Closset:2020afy,Akhond:2022jts,Carta:2022spy}.}

On the other hand, the Coulomb branch of $U(3)$ with four flavors has another singular locus at a non-vanishing value of the FI parameter, where the theory was argued in \cite{Yaakov:2013fza,Assel:2017jgo} to be dual to $U(1)$ with four flavors and two free twisted hypermultiplets (see also \cite{Dey:2017fqs}). Using this and further dualizing all the $U(1)$ nodes with one flavor into twisted hypermultiplets, we get that our theory in Figure \ref{fig:ex3b} becomes, around this other less singular point of its Coulomb branch, a collection of six free twisted hypermultiplets, whose Higgs branch is trivial. This is expected, since the Higgs branch of the $U(3)$ SQCD with 4 flavors at the less singular locus of its Coulomb branch is contained in the one at the most singular locus and the badness of the quiver theory we are considering resides only in the central gauge node. Hence, the Higgs branch of $\widehat{D}_4(SU(3))$, whose Hilbert series we compute in equation \eqref{eqn:D4hatSU3HS}, is captured by the quiver theory of Figure \ref{fig:ex3b} at the most singular locus of its Coulomb branch, where it is equivalent to the affine $D_4$ quiver plus a free twisted hypermultiplet. The former encodes the Higgs branch of $\widehat{D}_4(SU(3))$, while the latter the massless vector on a generic point of it.
 
Similar considerations were made also in \cite{Closset:2020afy,Carta:2022spy}. In particular in \cite{Closset:2020afy} several examples were discussed where it similarly happens that the 3d reduction of 4d $\mathcal{N}=2$ SCFTs with massless vectors on the Higgs branch typically gives rise to bad quiver theories. Also in those cases it was argued that the relevant Higgs branch geometry is captured by an equivalent good description valid around the most singular point of the Coulomb branch and which also contains free twisted hypermultiplets encoding the massless vector multiplet in 4d. The examples of \cite{Closset:2020afy} arise from 4d $\mathcal{N}=2$ SCFTs that are obtained via conformal gaugings of $\mathcal{D}_p^b(J)$ theories.\footnote{The $\mathcal{D}_p(J)$ theories, and their $\mathcal{D}_p^b(J)$ generalizations, are 4d $\mathcal{N}=2$ SCFTs obtained from the class $\mathcal{S}$ construction with, so-called, irregular punctures \cite{Cecotti:2012jx,Xie:2012hs,Cecotti:2013lda,Wang:2015mra}.}

The same result in equation \eqref{eqn:D4hatSU3HS} for the Hilbert series can be obtained from the 3d mirror of this class $\mathcal{S}$ theory. The naive prescription, summarized in Appendix \ref{app:3dmirror}, would be to glue four copies of the $T_{[2]}(USp'(2))$ theory via a diagonal $SU(2)$ gauging, which results in $SU(2)$ SQCD with four fundamental half-hypermultiplets. This cannot be the mirror since the resulting theory is bad and the Coulomb branch Hilbert series does not reproduce equation \eqref{eqn:D4hatSU3HS} (see \cite{Assel:2018exy} for more details). We find that equation \eqref{eqn:D4hatSU3HS} coincides with the Coulomb branch Hilbert series of $SU(2)$ SQCD with eight fundamental half-hypermultiplets \cite{Cremonesi:2013lqa}, i.e., adding an additional four half-hypermultiplets with respect to the naive prescription. In line with Conjecture \ref{conj:3dmirror}, we propose that to this theory we should also add one free hypermultiplet, whose presence can be deduced by the fact that on the Higgs branch of $\widehat{D}_4(SU(3))$ there is a massless $U(1)$ vector multiplet which in 3d reduces to a twisted hypermultiplet that in the mirror is regarded as an ordinary hypermultiplet.\footnote{More generally, as recently discussed in \cite{Closset:2020afy,Giacomelli:2020ryy,Carta:2021whq,Carta:2021dyx,Carta:2022spy}, every time a 4d SCFT has an unHiggsable sector of rank $r$, that is on the most generic point of its Higgs branch there still lives a rank $r$ SCFT with empty Higgs branch, we obtain $r$ free twisted hypermultiplets upon reduction to 3d.} The result is the rank one $SO(8)$ instanton theory, with the center of mass mode included, which is known to be mirror dual precisely to the rank one affine $D_4$ quiver gauge theory plus a free twisted hypermultiplet. Notice that adding the free hypermultiplet in the mirror dual is crucial, since it allows us to match the dimension of the Coulomb branch of the 3d reduction of $\widehat{D}_4(SU(3))$, which is $1+1+1+1+2=6$,  with the dimension of the Higgs branch of the proposed 3d mirror, which is $\frac{1}{2}\times 2\times 8 - 3 + 1=6$, where the final contribution comes from the free hypermultiplet.

Now that we have a proposal for the 3d mirror, i.e., the Lagrangian quiver given in Figure \ref{fig:ex3c}, we can determine the Coulomb branch Hilbert series. It is equivalent to use either the monopole formula or the Hall--Littlewood formula, both of which are reviewed in Appendix \ref{app:HSeqs}. The monopole formula simply states:
\begin{align}\label{eqn:D4hatmirmon}
    \operatorname{HS}[\text{Figure } \ref{fig:ex3c}](\tau)=\sum_{m=0}^\infty\tau^{4m}P_{SU(2)}(\tau;m)=\frac{1-\tau^2+\tau^4}{(1-\tau^2)(1-\tau^4)}\,.
\end{align}
For the Hall--Littlewood formula, we first determine the Hilbert series for $T_{[2]}(USp'(2))$:
\begin{align}
    \operatorname{HS}[T_{[2]}(USp'(2))](\tau;m)=\tau^m \,.
\end{align}
After the $SU(2)$ gauging and the addition of the four half-hypermultiplets, we find
\begin{align}\label{eqn:D4hatmirHL}
\begin{aligned}
    \operatorname{HS}[\text{Figure } \ref{fig:ex3c}](\tau)&=\sum_{m=0}^\infty\tau^{{\color{blue}4m}-4m}P_{SU(2)}(\tau;m)\operatorname{HS}[T_{[2]}(USp'(2))](\tau;m)^4\\&=\sum_{m=0}^\infty\tau^{4m}P_{SU(2)}(\tau;m)=\frac{1-\tau^2+\tau^4}{(1-\tau^2)(1-\tau^4)}\,.
\end{aligned}
\end{align}
We highlight in blue the contribution of the four extra fundamental half-hypermultiplets and notice that this cancels the contribution of the vector multiplet. As we can see, these two methods for determining the Coulomb branch Hilbert series reproduce the Higgs branch Hilbert series of the 3d reduction, which was given in equation \eqref{eqn:D4hatSU3HS}. This is expected, as the 3d mirror that we have proposed is exactly the known 3d mirror to the affine $D_4$ quiver gauge theory, which, as we have discussed, describes the 3d reduction of the class $\mathcal{S}$ theory around the most singular locus of the Coulomb branch.

As we already mentioned, the structure of the Coulomb branch of bad theories in 3d can be very complicated, see for example \cite{Assel:2017jgo,Assel:2018exy,bad}. In particular there may be more than one singular locus and the theory might not have a well-defined unique mirror dual, rather different magnetic quivers may be needed and which are valid around each singularity (see \cite{Yaakov:2013fza,Assel:2017jgo} for the unitary case\footnote{The reader can find in the ancillary files to \cite{Bourget:2021jwo} a \texttt{Mathematica} code implementing an algorithm to obtain the magnetic quivers of bad linear quiver theories with unitary and special unitary gauge nodes based on the Hanany--Witten brane set-ups \cite{Hanany:1996ie}. An alternative completely field theoretic derivation of these based on the dualization algorithm to derive mirror dualities for linear unitary quivers of \cite{Bottini:2021vms,Hwang:2021ulb} will be given in \cite{bad}.}). The theory in Figure \ref{fig:ex3b} is bad and the one in Figure \ref{fig:ex3c} together with the free hypermultiplet is a mirror dual description that is valid only in a neighbourhood of the most singular locus of the Coulomb branch. In this region, the theory in Figure \ref{fig:ex3b} is also equivalent to the affine $D_4$ quiver plus a free twisted hypermultiplet. There is then another less singular locus where the mirror is just a collection of six free hypermultiplets. For many of the following twisted $A_2$ class $\mathcal{S}$ examples we see 3d mirrors to bad theories which hold only around the most singular locus of the Coulomb branch, where the theory has a good equivalent frame that we can find with suitable manipulations and dualizations. This Coulomb branch captures the full Higgs branch of the 4d SCFT.

We can perform a similar analysis for all the class $\mathcal{S}$ theories of type $A_2$ on a sphere with four twisted punctures.  As the next example, we consider the case of punctures $[1^2]_t+3\times [2]_t$. The Hall--Littlewood index of this theory is written in terms of the characters of its $SU(2)$ flavor algebra as
\begin{align}
    \operatorname{HL}(\tau,a)=&\ 1+\tau^2(\textbf{adj})+\tau^4(\textbf{adj}^2+1)+\tau^5\chi_{2}(a)+\tau^6(\textbf{adj}^3+\textbf{adj}-1)+\mathcal{O}\left(\tau^7\right)\,.
    \label{eq:HL1max3null}
\end{align}
The existence of the $-\tau^6$ term immediately shows that the expression for the Hall--Littlewood index in equation \eqref{eq:HL1max3null} cannot be the Hilbert series of any Higgs branch. 

\begin{figure}[H]
    \centering
    \begin{subfigure}[b]{0.51\textwidth}
        \centering
        \begin{tikzpicture}[scale=1.2]
        \shade[inner color=white, outer color=blue!40, opacity = 0.4] (0,0) circle (1.4cm);
        \draw (0,0) circle (1.4cm);
        \draw (-1.4,0) arc (180:360:1.4 and 0.6);
        \draw[dashed] (1.4,0) arc (0:180:1.4 and 0.6);
        \draw[thick,fill=gray!60] (-.7,.7) circle (3pt);
        \node[font=\scriptsize] (P1) at (-.25,.7) {$\ydiagram[*(gray)]{2}$};
        \draw[thick] (.8,.1) circle (3pt);
        \node[font=\scriptsize] (P2) at (.8,-.2) {$\ydiagram[]{3}$};
        \draw[thick,fill=gray!60] (-.7,-.7) circle (3pt);
        \node[font=\scriptsize] (P3) at (-.4,-.7) {$\ydiagram[*(gray)]{1,1}$};
        \draw[dashed] (-.7,.6)--(-.7,-.6);
        \shade[inner color=white, outer color=blue!40, opacity = 0.4] (4.2,0) circle (1.4cm);
        \draw (4.2,0) circle (1.4cm);
        \draw (2.8,0) arc (180:360:1.4 and 0.6);
        \draw[dashed] (5.6,0) arc (0:180:1.4 and 0.6);
        \draw[thick,fill=gray!60] (4.9,.7) circle (3pt);
        \node[font=\scriptsize] (P4) at (4.6,.7) {$$\ydiagram[*(gray)]{1,1}$$};
        \draw[thick] (3.4,.1) circle (3pt);
        \node[font=\scriptsize] (P5) at (3.4,-.2) {$\ydiagram[]{3}$};
        \draw[thick,fill=gray!60] (4.9,-.7) circle (3pt);
        \node[font=\scriptsize] (P6) at (4.6,-.7) {$\ydiagram[*(gray)]{1,1}$};
        \draw[dashed] (4.9,.6)--(4.9,-.6);
        \draw[thick,dashed] (.8,0)--(3.4,0);
        \draw[thick,dashed] (.8,0.2)--(3.4,0.2);
        \shade[inner color=white, outer color=blue!40, opacity = 0.4] (1.4,0) rectangle ++(1.4,0.2);
        \node[font=\scriptsize] (c) at (2.1,0.5) {$SU(3)$};
        \end{tikzpicture}
        \caption{$A_2$ with $[1^2]_t+3\times[2]_t$.}\label{fig:ex4a}
        \end{subfigure}
    \begin{subfigure}[b]{0.24\textwidth}
    \centering\raisebox{6.5mm}{
        \begin{tikzpicture}[scale=1.1,every node/.style={scale=1.2},font=\scriptsize]
        \node[gauge] (p0) at (1,1) {$2$};
        \node[gauge] (p1) at (0,0) {$1$};
        \node[gauge] (p2) at (2,0) {$1$};
        \node[gaugeD] (c) at (1,0) {$3$};
        \draw (p1)--(c)--(p2);
        \draw (p0)--(c);
        \draw[black,solid,dashed] (p0) edge [out=12+45,in=78+45,loop,looseness=4.5]  (p0);
        \end{tikzpicture}}
        \caption{3d reduction.}\label{fig:ex4b}
    \end{subfigure}
    \begin{subfigure}[b]{0.23\textwidth}
        \centering\raisebox{12mm}{
        \begin{tikzpicture}[scale=1.1,every node/.style={scale=1.2},font=\scriptsize]
        \node[gauge,red] (t0) at (0,0) {$3$};
        \node[gaugeD] (c) at (1,0) {$2$};
        \node[flavor,red] (t1) at (2,0) {$7$};
        \draw (t0)--(c)--(t1);
        \end{tikzpicture}}
        \caption{3d magnetic quiver.}\label{fig:ex4c}
    \end{subfigure}
    \caption{In (a), we show the $A_2$ class $\mathcal{S}$ theory on a sphere with $[1^2]_t+3\times[2]_t$ punctures. In (b), we depict the Lagrangian quiver describing the 3d reduction. In (c), we portray the magnetic quiver which, together with a free hypermultiplet, is a mirror for the theory in (b) around the most singular point on its Coulomb branch. The dashed link in (b) denotes a $U(2)$ adjoint hypermultiplet with the singlet decoupled.}
    \label{fig:ex4}
\end{figure}
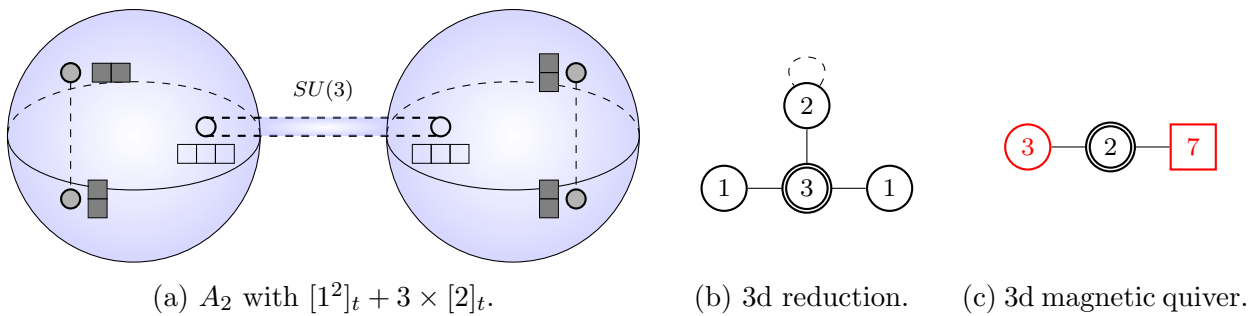

In this case, as in the $\widehat{D}_4(SU(3))$ theory, the negative terms in the Hall--Littlewood index can be explained by a $U(1)$ gauge factor which is not completely Higgsed on the generic point of the Higgs branch of the 3d reduction of the 4d SCFT. We first construct the 3d reduction, which again turns out to be Lagrangian. As depicted in Figure \ref{fig:ex4a}, the four-punctured sphere degenerates into two three-punctured spheres, with punctures $[1^3]+[1^2]_t+[2]_t$ and $[1^3]+2\times[2]_t$.
As we have seen, the latter corresponds to two copies of the $(A_1, D_4)$ theory. The former was shown in \cite{Beem:2020pry} to correspond to the rank two $SU(3)$ instanton SCFT \cite{Buican:2014hfa,Buican:2017fiq,Beem:2019snk}. Thus, the four-punctured sphere that we are considering is nothing other than the common $SU(3)$ gauging of these three theories. The 3d reduction of the rank two $SU(3)$ instanton SCFT is described by the 3d $\mathcal{N}=4$ ADHM $U(2)$ gauge theory with one adjoint and three fundamental hypermultiplets, where the center of mass mode corresponds to the singlet in the $U(2)$ adjoint which decouples as a free hypermultiplet in the infrared. Hence, we see that the 3d reduction of the four-punctured sphere theory in question is given by the quiver in Figure \ref{fig:ex4b}. As in the previous example the matter content is not enough to Higgs the entire gauge group of the quiver; in particular a vector multiplet for a $U(1)$ inside the $SU(3)$ gauge group, that is also the gauge group in 4d, remains massless on the Higgs branch, again in accordance with \cite{Beem:2022mde}. Trying to naively compute the Higgs branch Hilbert series assuming that the gauge group is fully Higgsed would give an expansion with negative coefficients. 

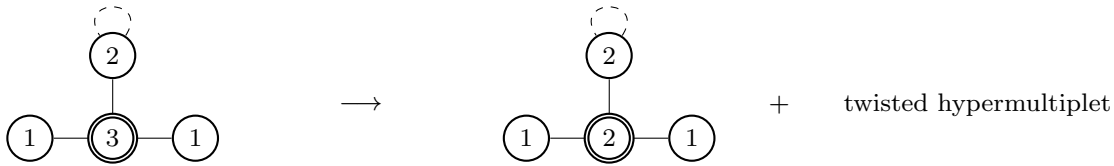
\begin{figure}[H]
    \centering
        \begin{tikzpicture}[scale=1.1,every node/.style={scale=1.2},font=\scriptsize]
        \node[gauge] (p0) at (1,1) {$2$};
        \node[gauge] (p1) at (0,0) {$1$};
        \node[gauge] (p2) at (2,0) {$1$};
        \node[gaugeD] (c) at (1,0) {$3$};
        \draw (p1)--(c)--(p2);
        \draw (p0)--(c);
        \draw[black,solid,dashed] (p0) edge [out=12+45,in=78+45,loop,looseness=4.5]  (p0);
        \node[empty] (t) at (4,0.4) {$\longrightarrow$};
        \node[empty] (t2) at (11,0.4) {$+\qquad \text{twisted hypermultiplet}$};
        \node[gauge] (p0r) at (7,1) {$2$};
        \node[gauge] (p1r) at (6,0) {$1$};
        \node[gauge] (p2r) at (8,0) {$1$};
        \node[gaugeD] (cr) at (7,0) {$2$};
        \draw (p1r)--(cr)--(p2r);
        \draw (p0r)--(cr);
        \draw[black,solid,dashed] (p0r) edge [out=12+45,in=78+45,loop,looseness=4.5]  (p0r);
        \end{tikzpicture}
    \caption{Similarly to Figure \ref{fig:dual1}, the bad quiver on the left can be dualized, around the most singular locus of the Coulomb branch, to the good quiver on the right plus a free twisted hypermultiplet. Around the most singular locus of the Coulomb branch, the Higgs branch Hilbert series is computed from the alternative good description.}
    \label{fig:dual2}
\end{figure}

An accurate way of computing the Higgs branch Hilbert series would be again to use \texttt{Macaulay2}, but the computation seems to be too intensive in this case. Instead, with the exact same operations we did in Figure \ref{fig:dual1} we can show that the theory in Figure \ref{fig:ex4b} is equivalent, around the most singular locus of its Coulomb branch, to a similar quiver theory, but with the middle gauge node being $SU(2)$ rather than $SU(3)$ and an additional free twisted hypermultiplet, as shown in Figure \ref{fig:dual2}. In the latter theory the gauge group is fully Higgsed, so we can compute the Higgs branch Hilbert series with standard methods, finding
\begin{align}
\begin{split}
    \operatorname{HS}(\tau,a)&=\operatorname{PE}\left[\tau^2(\textbf{adj})+\tau^5\chi_{2}(a)-\tau^{12}\right]=\frac{1-\tau^{12}}{(1-\tau^2)(1-a^{\pm2}\tau^2)(1-a^{\pm1}\tau^5)}\\
    &=1+\tau^2(\textbf{adj})+\tau^4(\textbf{adj}^2+1)+\tau^5\chi_{2}(a)+\tau^6(\textbf{adj}^3+\textbf{adj})+\mathcal{O}\left(\tau^7\right)\,,
    \end{split}\label{eq:HS1max3null}
\end{align}
which coincides with equation \eqref{eq:HL1max3null} up to order $\tau^6$, after removing the negative term $-\tau^6$ corresponding to the $\mathcal{D}$-type multiplet. 

On the second singular locus of the Coulomb branch of the quiver in Figure \ref{fig:ex4b}, the theory is not just equivalent to a collection of free fields, unlike in the previous example of $A_2$ with $4 \times [2]_t$. This is because the $U(2)$ gauge theory with one traceless adjoint hypermultiplet and one fundamental hypermultiplet flows in the infrared to the product of a decoupled free twisted hypermultiplet and an interacting SCFT which turns out to have enhanced $\mathcal{N}=8$ supersymmetry \cite{Kapustin:2010xq}.\footnote{This supersymmetry enhancement can be checked using the superconformal index, see e.g.~\cite{Beratto:2020qyk}.}  We still expect that the Higgs branch at this less singular point is a subvariety of the one at the most singular point, since this is the case for the $U(3)$ SQCD with four flavors and the badness of the quiver we are studying again resides only in the central gauge node.

We can also obtain the same result for the Higgs branch Hilbert series in equation \eqref{eq:HS1max3null} using the 3d mirror description. In this case, we propose, following Conjecture \ref{conj:3dmirror}, that the 3d mirror is given by the quiver in Figure \ref{fig:ex4c}. This is obtained by gauging one copy of $T(USp'(2))$ and three copies of $T_{[2]}(USp'(2))$ via the common $SU(2)$ flavor, which would be the usual naive prescription, and in addition including four fundamental half-hypermultiplets on the middle $SU(2)$ node. We also must include one free hypermultiplet, whose presence can be understood as due to the massless $U(1)$ vector multiplet on the Higgs branch of the 4d SCFT and which corresponds to the twisted hypermultiplet in the dual shown in Figure \ref{fig:dual2}. This free hypermultiplet is crucial to match the dimensions of both branches of the mirror theories. 

The dimension of the Higgs branch of the quiver in Figure \ref{fig:ex4b} is $3+3+2\times 3+3-1-1-4-8+1=2$, where the last term takes into account the gauge $U(1)$ which exists on the Higgs branch, and this matches with the dimension of the Coulomb branch of the quiver in Figure \ref{fig:ex4c} which is $1+1=2$. Moreover, the dimension of the Coulomb branch of the quiver in Figure \ref{fig:ex4b} is $1+1+2+2=6$ and it matches with the dimension of the Higgs branch of the quiver in Figure \ref{fig:ex4c} which is $1+\frac{1}{2}\times2\times 7-3+1=6$, where the first term is the dimension of the Higgs branch of the $SO(3)$ gauge theory with two fundamental half-hypermultiplets, shown to be isomorphic to $\mathbb{C}^2/\mathbb{Z}_2$ in \cite{Cremonesi:2014uva}, and the last term is the free hypermultiplet.  We stress once again that the 3d mirror we are proposing is valid only around the most singular locus of the Coulomb branch of the theory in Figure \ref{fig:ex4b}, where it is equivalent to a similar quiver but with an $SU(2)$ in the middle and one additional twisted hypermultiplet, see Figure \ref{fig:dual2}. 

It turns out that the Hall--Littlewood index of any $A_2$ theory on a four-punctured sphere with two twist lines, other than $\widehat{D}_4(SU(3))$, is written in a form similar to equation \eqref{eq:HL1max3null}:
\begin{align}\label{eq:HLA2general}
    \operatorname{HL}(\tau,\bm{a})=&\
    1+\tau^2(\textbf{adj})+\tau^4(\textbf{adj}^2+1)+f(\tau,\bm{a})+\tau^6(\textbf{adj}^3+\textbf{adj}-1)+\mathcal{O}\left(\tau^7\right)\,,
\end{align}
with some extra terms
\begin{align}\label{eq:ffunction}
    \begin{aligned}
        f(\tau,\bm{a})=\begin{cases}
        \tau^5\chi_{2}(a) & 1\times[1^2]_t+3\times[2]_t\\
        \tau^6\chi_{2}(a)\chi_{2}(b) & 2\times[1^2]_t+2\times[2]_t \\
        0 & \text{otherwise}.\end{cases}\,.
    \end{aligned}
\end{align}
They commonly contain a $-\tau^6$ term that indicates that they cannot be the Hilbert series of any Higgs branch. Indeed, for the Hilbert series we get the same result at low orders, but without the negative term $-\tau^6$
\begin{align}\label{eq:HSA2general}
    \operatorname{HS}(\tau,\bm{a})=&\
    1+\tau^2(\textbf{adj})+\tau^4(\textbf{adj}^2+1)+f(\tau,\bm{a})+\tau^6(\textbf{adj}^3+\textbf{adj})+\mathcal{O}\left(\tau^7\right)\,,
\end{align}
where $f(\tau,\bf{a})$ is again defined as in equation \eqref{eq:ffunction}. This is again computed as the Coulomb branch Hilbert series of the 3d mirror which, as in the two explicit examples we just analyzed, is obtained by gluing one copy of the $T_{\rho_t}(USp'(2))$ theory for each twisted puncture of type $\rho_t$ via a common $SU(2)$ gauging and adding four fundamental half-hypermultiplets to this central node and one free hypermultiplet. Moreover, we have checked that the dimensions of the Higgs and Coulomb branches of the 3d reduction of each class $\mathcal{S}$ theory, which in some cases turn out to be Lagrangian, coincide with the dimensions of the Coulomb and Higgs branches, respectively, of the proposed 3d mirror. In 4d, all these theories exhibit an $SU(3)$ gauge theory description where it appears that a $U(1)$ factor is not Higgsed at the generic point of the Higgs branch. Moreover, in all the cases in which the 3d reduction of the class $\mathcal{S}$ theory is Lagrangian, it is bad, and so the mirror should only be regarded as a good description close to the most singular locus of the Coulomb branch, where we can always find an equivalent description which is good, and where the gauge group is fully Higgsed, using manipulations similar to those depicted in Figures \ref{fig:dual1} and \ref{fig:dual2} . In these cases, the expression for the Higgs branch Hilbert series in equation \eqref{eq:HSA2general} can be equivalently obtained from this dual frame.

\subsection{\texorpdfstring{$A_2$}{A2} on spheres with \texorpdfstring{$2k>4$}{2k>4} twisted punctures}\label{sec:A2wmorethan4p}

The discrepancy between the Hall--Littlewood index and the Higgs branch Hilbert series seems to be a general feature of all class $\mathcal{S}$ theories on spheres with two or more twist lines, thus motivating Conjecture \ref{conj:HLneqHS}. We have surveyed the cases for $A_2$ with at most two twist lines in Section \ref{sec:A2w4p}; now we turn our attention to those theories with $k>2$ twist lines.

We first consider the simplest case of type $A_2$ on a sphere with $2k$ twisted null punctures. The Hall--Littlewood index takes the following simple form:
\begin{align}
    \operatorname{HL}(\tau)=&\,\frac{\tau^{6k-8}(1-\tau^2)^{k-2}}{1-\tau^{6k-8}}+\frac{(1-\tau^6)^{k-1}}{1-\tau^4}=1+(1+\delta_{k,2})\tau^4-(k-1)\tau^6+\mathcal{O}(\tau^{8})\,,
    \label{eq:HL2ktwistnull}
\end{align}
which reduces to equation \eqref{eqn:D4hatSU3HL} when $k=2$. Once again, from order $\tau^6$ we start to see negative terms, whose multiplicity depends on the number of twist lines. Again, these negative coefficients signal that this expression cannot coincide with the Hilbert series of the Higgs branch.

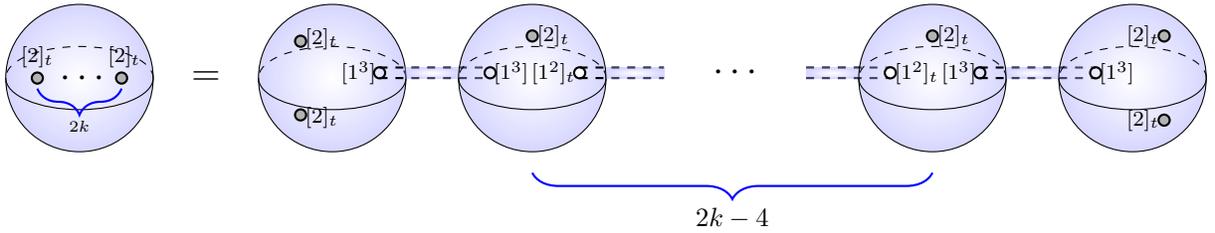
\begin{figure}[H]
    \centering
    \begin{subfigure}{\textwidth}
        \centering
        \begin{tikzpicture}[scale=.7]
        \shade[inner color=white, outer color=blue!40, opacity = 0.4] (-1,0) circle (1.4cm);
        \draw (-1,0) circle (1.4cm);
        \draw (-2.4,0) arc (180:360:1.4 and 0.6);
        \draw[dashed] (.4,0) arc (0:180:1.4 and 0.6);
        \draw[thick,fill=gray!60] (-1.8,0) circle (3pt);
        \node[font=\scriptsize] (P1) at (-1.8,.4) {$[2]_t$};
        \draw[thick,fill=gray!60] (-.2,0) circle (3pt);
        \node[font=\scriptsize] (P2) at (-.15,.4) {$[2]_t$};
        \node[draw=none,opacity=0,thick,scale=0.1,fill=black,label={[label distance=0mm]0:\large{$\cdots$}}] (LC) at (-1.6,0) {};
        \draw[thick,blue,decorate,decoration={brace,amplitude=10pt,mirror}] (-1.8,-.2)--(-.2,-.2) node[black,midway,yshift=-.5cm] {\tiny $2k$};
        \node[draw=none,opacity=0,thick,scale=0.1,fill=black,label={[label distance=0mm]0:\large{$=$}}] (EQ) at (.9,0) {};
        \shade[inner color=white, outer color=blue!40, opacity = 0.4] (3.8,0) circle (1.4cm);
        \draw (3.8,0) circle (1.4cm);
        \draw (2.4,0) arc (180:360:1.4 and 0.6);
        \draw[dashed] (5.2,0) arc (0:180:1.4 and 0.6);
        \draw[thick,fill=gray!60] (3.2,.7) circle (3pt);
        \node[font=\scriptsize] (P4) at (3.6,.75) {$[2]_t$};
        \draw[thick,fill=gray!60] (3.2,.-.7) circle (3pt);
        \node[font=\scriptsize] (P5) at (3.6,-.75) {$[2]_t$};
        \draw[thick] (4.7,.1) circle (3pt);
        \node[font=\scriptsize] (P6) at (4.3,.1) {$[1^3]$};
        \draw[thick,dashed] (4.7,0)--(6.8,0);
        \draw[thick,dashed] (4.7,.2)--(6.8,.2);
        \shade[inner color=white, outer color=blue!40, opacity=.4] (5.2,0) rectangle ++(1,.2);
        \shade[inner color=white, outer color=blue!40, opacity=.4] (7.6,0) circle (1.4cm);
        \draw (7.6,0) circle (1.4cm);
        \draw (6.2,0) arc (180:360:1.4 and 0.6);
        \draw[dashed] (9,0) arc (0:180:1.4 and 0.6);
        \draw[thick,fill=gray!60] (7.6,.8) circle (3pt);
        \node[font=\scriptsize] (P4) at (8,.8) {$[2]_t$};
        \draw[thick] (6.8,.1) circle (3pt);
        \node[font=\scriptsize] (P5) at (7.2,.1) {$[1^3]$};
        \draw[thick,fill=white] (8.5,.1) circle (3pt);
        \node[font=\scriptsize] (P6) at (8,.1) {$[1^2]_t$};
        \draw[thick,dashed] (8.5,0)--(10.1,0);
        \draw[thick,dashed] (8.5,.2)--(10.1,.2);
        \shade[inner color=white, outer color=blue!40, opacity=.4] (9,0) rectangle ++(1.1,.2);
        \node[draw=none,opacity=0,thick,scale=0.1,fill=black,label={[label distance=0mm]0:\large{$\cdots$}}] (C) at (10.8,.1) {};
        \draw[thick,dashed] (12.8,0)--(14.4,0);
        \draw[thick,dashed] (12.8,.2)--(14.4,.2);
        \shade[inner color=white, outer color=blue!40, opacity=.4] (12.8,0) rectangle ++(1,.2);
        \shade[inner color=white, outer color=blue!40, opacity = 0.4] (15.2,0) circle (1.4cm);
        \draw (15.2,0) circle (1.4cm);
        \draw (13.8,0) arc (180:360:1.4 and 0.6);
        \draw[dashed] (16.6,0) arc (0:180:1.4 and 0.6);
        \draw[thick,fill=gray!60] (15.2,.8) circle (3pt);
        \node[font=\scriptsize] (P4) at (15.6,.8) {$[2]_t$};
        \draw[thick,fill=white] (14.4,.1) circle (3pt);
        \node[font=\scriptsize] (P5) at (14.9,.1) {$[1^2]_t$};
        \draw[thick] (16.1,.1) circle (3pt);
        \node[font=\scriptsize] (P6) at (15.7,.1) {$[1^3]$};
        \draw[thick,dashed] (16.1,0)--(18.3,0);
        \draw[thick,dashed] (16.1,.2)--(18.3,.2);
        \shade[inner color=white, outer color=blue!40, opacity=.4] (16.6,0) rectangle ++(1,.2);
        \shade[inner color=white, outer color=blue!40, opacity=.4] (19,0) circle (1.4cm);
        \draw (19,0) circle (1.4cm);
        \draw (19-1.4,0) arc (180:360:1.4 and 0.6);
        \draw[dashed] (19+1.4,0) arc (0:180:1.4 and 0.6);
        \draw[thick,fill=gray!60] (19.6,.8) circle (3pt);
        \node[font=\scriptsize] (P1) at (19.2,.8) {$[2]_t$};
        \draw[thick] (18.3,.1) circle (3pt);
        \node[font=\scriptsize] (P2) at (18.7,.1) {$[1^3]$};
        \draw[thick,fill=gray!60] (19.6,-.8) circle (3pt);
        \node[font=\scriptsize] (P3) at (19.2,-.8) {$[2]_t$};
        \draw[thick,blue,decorate,decoration={brace,amplitude=10pt,mirror}] (7.6,-1.8)--(15.2,-1.8) node[black,midway,yshift=-0.6cm] {\footnotesize $2k-4$};
        \end{tikzpicture}
        \caption{The class $\mathcal{S}$ theory of type $A_2$ on a sphere with $2k$ twisted null punctures degenerates into this linear chain of three punctured spheres. The gauge groups alternate between $SU(3)$ and $SU(2)$ depending on whether the glued punctures are untwisted or twisted maximal punctures.}\label{fig:2ka}
    \end{subfigure}
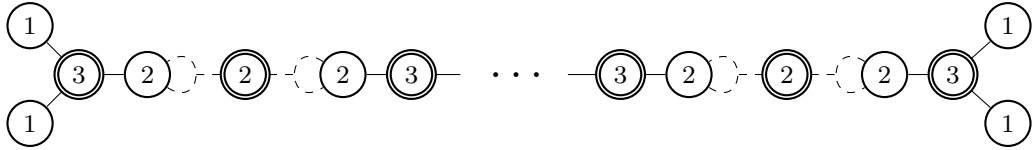
\vspace{3mm}
    \begin{subfigure}{\textwidth}
        \centering
        \begin{tikzpicture}[scale=1.3,every node/.style={scale=1.2},font=\scriptsize]
        \node[gauge] (p0) at (0,1) {$1$};
        \node[gauge] (p1) at (0,0) {$1$};
        \node[gaugeD] (c1) at (.5,.5) {$3$};
        \node[gauge] (p2) at (1.2,.5) {$2$};
        \draw (p0)--(c1)--(p1);
        \draw (c1)--(p2);
        \draw[black,solid,dashed] (p2) edge [out=12+135-180,in=78+135-180,loop,looseness=4.5] (p2);
        \node[gaugeD] (c2) at (2.2,.5) {$2$};
        \draw[dashed] (1.7,.5)--(c2)--(2.7,.5);
        \node[gauge] (p3) at (3.2,.5) {$2$};
        \draw[black,solid,dashed] (p3) edge [out=12+135,in=78+135,loop,looseness=4.5] (p3);
        \node[gaugeD] (c3) at (3.9,.5) {$3$};
        \draw (p3)--(c3)--(4.4,.5);
        \node[draw=none,opacity=0,thick,scale=0.1,fill=black,label={[label distance=0mm]0:\large{$\cdots$}}] (C) at (4.55,.49) {};
        \draw (5.5,.5)--(5.8,.5);
        \node[gaugeD,xshift=6cm] (cR1) at (.5,.5) {$3$};
        \node[gauge,xshift=6cm] (pR2) at (1.2,.5) {$2$};
        \draw (cR1)--(pR2);
        \draw[black,solid,dashed] (pR2) edge [out=12+135-180,in=78+135-180,loop,looseness=4.5] (pR2);
        \node[gaugeD,xshift=6cm] (cR2) at (2.2,.5) {$2$};
        \draw[dashed] (7.25,.5)--(cR2)--(8.3,.5);
        \node[gauge,xshift=6cm] (pR3) at (3.2,.5) {$2$};
        \draw[black,solid,dashed] (pR3) edge [out=12+135,in=78+135,loop,looseness=4.5] (pR3);
        \node[gaugeD,xshift=6cm] (cR3) at (3.9,.5) {$3$};
        \node[gauge] (pR4) at (10,1) {$1$};
        \node[gauge] (pR5) at (10,0) {$1$};
        \draw (pR3)--(cR3)--(pR4);
        \draw (cR3)--(pR5);
        \end{tikzpicture}\vspace{2mm}
        \caption{The 3d reduction of the class $\mathcal{S}$ theory of type $A_2$ on a sphere with $2k$ twisted null punctures. The dashed trivalent link represents a half-hypermultiplet in the $(\bm{3},\bm{2})$ representation of the attached $U(2) \times SU(2)$ gauge nodes. The total number of $SU(3)$ gauge nodes in the quiver is $k-1$.}\label{fig:A22ktnLag}
    \end{subfigure}
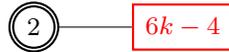
\vspace{3mm}
    \begin{subfigure}{\textwidth}
        \centering
        \begin{tikzpicture}[scale=1.1,every node/.style={scale=1.2},font=\scriptsize]
        \node[gaugeD] (p0) at (0,0) {$2$};
        \node[flavor,red] (p1) at (1.8,0) {$6k-4$};
        \draw (p0)--(p1);
        \end{tikzpicture}
        \caption{The magnetic quiver for the $A_2$ class $\mathcal{S}$ theory on a sphere with $2k$ twisted null punctures, which we propose to be also the mirror around the most singular point of the Coulomb branch of the theory in Figure \ref{fig:A22ktnLag} after the inclusion of $k-1$ free hypermultiplets.}
        \label{fig:A22ktnmirror}
    \end{subfigure}
    \caption{The linear degeneration limit, 3d reduction, and 3d mirror for the class $\mathcal{S}$ theory of type $A_2$ on a sphere with $2k \geq 4$ twisted null punctures.}
    \label{fig:A22k}
\end{figure}

In order to compute the Higgs branch Hilbert series, we go once again to three dimensions. We start by considering the 3d reduction, which can be obtained from the class $\mathcal{S}$ description of the 4d theory as a linear chain of three-punctured spheres. This configuration is depicted in Figure \ref{fig:2ka}. There are two distinct kinds of three-punctured sphere that appear in this degeneration: one with $[1^3] + 2\times [2]_t$, which corresponds to two copies of the $(A_1, D_4)$ theory, and one with $[1^3] + [1^2]_t + [2]_t$, which is nothing other than the rank two $SU(3)$ instanton \cite{Buican:2014hfa,Buican:2017fiq,Beem:2019snk,Beem:2020pry}. The 3d reduction of the former was discussed in Section \ref{sec:A2w4p}, and the 3d reduction of the latter is $U(2)$ with one hypermultiplet in the $\bm{3}$ and three in the $\bm{2}$, as described around Figure \ref{fig:ex4}. As these building blocks are both Lagrangian, the 3d reduction of the $A_2$ theory with $2k \times [2]_t$ is the Lagrangian quiver given in Figure \ref{fig:A22ktnLag}.\\

Figure \ref{fig:A22ktnLag} depicts a somewhat exotic quiver. In particular, the $SU(2)$ gaugings between the twisted maximal punctures in 4d descend to the gaugings of the $SU(2)$ symmetry that rotates the adjoint hypermultiplet of each rank two $SU(3)$ instanton gauge theory. The $SU(3)$ gaugings are standard, however the corresponding gauge groups seem once again not to be fully Higgsed on the Higgs branch, in particular for each $SU(3)$ we have one massless $U(1)$ vector multiplet remaining, which is again compatible with the findings of \cite{Beem:2022mde}. Moreover, each of these $SU(3)$ nodes is bad, however, if we consider the theory around the most singular locus of the Coulomb branch, we can dualize to convert all of them into $SU(2)$ nodes, while also producing $k-1$ free twisted hypermultiplets.

Despite of the complexity of the theory in Figure \ref{fig:A22ktnLag}, the 3d mirror is remarkably simple. Again, the mirror we are going to propose is valid only around the most singular locus of the Coulomb branch. Moreover, this case represents a good starting point to investigate the 3d mirror in the case of multiple twist lines. Recall that, in the case of two twist lines, we modified the standard prescription of gluing one copy of the $T_{[2]}(USp'(2))$ theory for each twisted null puncture by adding four fundamental half-hypermultiplets to the common $SU(2)$ gauge node. We propose that in the case of $k$ twist lines, the number of extra fundamental half-hypermultiplets should be $4(k-1)$. Hence, the 3d mirror of the type $A_2$ theory with $2k$ twisted null punctures is $SU(2)$ SQCD with $2k+4(k-1)=6k-4$ fundamental half-hypermultiplets, which we depict in Figure \ref{fig:A22ktnmirror}. This is the rank one $SO(6k-4)$ instanton. The Coulomb branch Hilbert series was computed in \cite{Cremonesi:2013lqa} and it is given by
\begin{align}\label{eq:HS2ktwistnull}
    \operatorname{HS}(\tau)=&\,\frac{1+\tau^{6k-6}}{(1-\tau^4)(1-\tau^{6k-8})}=\,1+(1+\delta_{k,2})\tau^4+\mathcal{O}(\tau^{8})\,,
\end{align}
which to the lowest orders coincides with the Hall--Littlewood index in equation \eqref{eq:HL2ktwistnull}, except for the negative term at order $\tau^6$ due to the $\mathcal{D}$-type multiplets. At higher orders we start having products of generators as well as relations among them that spoil this naive comparison.

Similarly to the cases with two twist lines, we propose that the full 3d mirror theory is obtained by including an additional $k-1$ free hypermultiplets, one for each massless $U(1)$ vector multiplet on the Higgs branch of the 4d SCFT. These free hypermultiplets correspond to the twisted hypermultiplets in the 3d reduction that we found, after dualizations, for the theory in Figure \ref{fig:A22ktnLag}. In this way, we can match the dimensions of both of the branches between the theory in Figure \ref{fig:A22ktnLag} and the 3d mirror in Figure \ref{fig:A22ktnmirror}. Specifically, the Higgs branch of the theory in Figure \ref{fig:A22ktnLag}, which is also the Higgs branch of the 4d class $\mathcal{S}$ theory, has dimension
\begin{align}
\begin{split}
    \operatorname{dim}_{\mathbb{H}}\mathcal{M}_{\text{HB}}=&\,4\times\underbrace{2}_{(A_1,D_4)}+\,(2k-4)\times\underbrace{5}_{\text{rank 2 }SU(3) \text{ inst.}}-\,(k-1)\times\underbrace{8}_{SU(3)\text{ vect.}}\\
    &\,-(k-2)\times\underbrace{3}_{SU(2)\text{ vect.}}+\underbrace{k-1}_{\text{massless }U(1)^{k-1}\text{ vect.}}=\,1 \,,
\end{split}
\end{align}
which coincides with the dimension of the Coulomb branch of the mirror $SU(2)$ SQCD theory. 
Furthermore, the quaternionic dimension of the Coulomb branch of the 3d reduction in Figure \ref{fig:A22ktnLag} is
\begin{align}
\begin{split}
    \operatorname{dim}_{\mathbb{H}}\mathcal{M}_{\text{CB}}=&\,4\times\underbrace{1}_{(A_1,D_4)}+(2k-4)\times\underbrace{2}_{\text{rank 2 }SU(3) \text{ inst.}}+(k-1)\times\underbrace{2}_{SU(3)\text{ gauge}}\\
    &\,+(k-2)\times\underbrace{1}_{SU(2)\text{ gauge}}=7k-8 \,,
\end{split}
\end{align}
which coincides with the dimension of the Higgs branch of the 3d mirror once we take into account the $k-1$ free hypermultiplets
\begin{align}
    \operatorname{dim}_{\mathbb{H}}\mathcal{M}_{\text{HB}}^{\text{mirror}}=6k-4-3+(k-1)=7k-8\,.
\end{align}

Similar observations hold for $A_2$ theories with $2k$ arbitrary twisted punctures, that is, not just considering twisted null punctures. We have scanned through all 21 examples up to and including $k=4$, and we have found that in all cases the Hall--Littlewood indices display terms with negative coefficients. To low orders, they take the following general form:\footnote{In equations \eqref{eq:HL2ktpunct} and \eqref{eq:HS2ktpunct}, we assume that there are at least three twist lines. The cases where $k=2$ were written in general form in equations \eqref{eq:HLA2general} and \eqref{eq:HSA2general}; it is noteworthy that the occasional extra contributions at order $\tau^5$ for $k=2$, captured by equation \eqref{eq:ffunction}, do not appear below order $\tau^6$ for $k>2$.}
\begin{align}
    \operatorname{HL}(\tau,a)=1+\tau^2(\textbf{adj})+\tau^4(\textbf{adj}^2+1)+\tau^6(\textbf{adj}^3+\textbf{adj}-(k-1))+\mathcal{O}\left(\tau^8\right) \,,
    \label{eq:HL2ktpunct}
\end{align}
where the $\textbf{adj}^n$ is the same shorthand notation that we defined in equation \eqref{eq:adjchar}. Moreover, the Higgs branch Hilbert series computed via the Coulomb branch Hilbert series of the proposed 3d mirror takes a similar form, but without the negative terms at order $\tau^6$
\begin{align}
    \operatorname{HS}(\tau,a)=1+\tau^2(\textbf{adj})+\tau^4(\textbf{adj}^2+1)
    +\tau^6(\textbf{adj}^3+\textbf{adj})+\mathcal{O}\left(\tau^8\right) \,.
    \label{eq:HS2ktpunct}
\end{align}
We stress again that these results are obtained by modifying the ordinary prescription for the 3d mirrors by adding $4(k-1)$ fundamental half-hypermultiplets for the central $SU(2)$ gauge node of the star-shaped quiver. Moreover, adding a further $k-1$ free hypermultiplets allows us to match the dimension of the Higgs branch of the 3d mirror with the dimension of the Coulomb branch of the 3d reduction of the class $\mathcal{S}$ theory. This proposal for the 3d mirror of class $\mathcal{S}$ of type $A_{2n}$ is summarized in Conjecture \ref{conj:3dmirror}.

We summarize our findings for the Hall--Littlewood indices and the Higgs branch Hilbert series of the class $\mathcal{S}$ theories of type $A_2$ on spheres with twisted punctures in Table \ref{tab:HLHSforA2} of Appendix \ref{app:summaryHLHS}.

\subsection{Higher rank \texorpdfstring{$A_{2n}$}{A2n} on spheres with twisted punctures}\label{sec:A2nw4tw}

We next consider some higher rank examples of class $\mathcal{S}$ theories of type $A_{2n}$ with $n>1$ on spheres with multiple twist lines. This serves two purposes. One is to exemplify that the feature of having the Hall--Littlewood index different from the Higgs branch Hilbert series holds for any sphere with multiple twist lines, independently from the rank. The second motivation is to provide evidence for Conjecture \ref{conj:3dmirror}, which gives the prescription for the 3d mirror, in the higher rank case.

\begin{figure}[H]
    \centering
    \begin{subfigure}[b]{0.51\textwidth}
        \centering
        \begin{tikzpicture}[scale=1.2]
        \shade[inner color=white, outer color=blue!40, opacity = 0.4] (0,0) circle (1.4cm);
        \draw (0,0) circle (1.4cm);
        \draw (-1.4,0) arc (180:360:1.4 and 0.6);
        \draw[dashed] (1.4,0) arc (0:180:1.4 and 0.6);
        %\draw[thick,fill=white] (-.7,.7) circle (3pt);
        \node[draw,thick,star,star points=5,star point ratio=2,fill=gray!60,scale=.5] at (-.7, .7) {};
        %\node[font=\scriptsize] (P1) at (-.4,.7) {$\ydiagram[*(gray)]{1,1}$};
        \draw[thick] (.8,.1) circle (3pt);
        %\node[font=\scriptsize] (P2) at (.8,-.2) {$\ydiagram[*(white)]{3}$};
        \node[draw,thick,star,star points=5,star point ratio=2,fill=gray!60,scale=.5] at (-.7,-.7) {};
        %\node[font=\scriptsize] (P3) at (-.4,-.7) {$\ydiagram[*(gray)]{1,1}$};
        \draw[dashed] (-.7,.6)--(-.7,-.6);
        \shade[inner color=white, outer color=blue!40, opacity = 0.4] (4.8,0) circle (1.4cm);
        \draw (4.8,0) circle (1.4cm);
        \draw (3.4,0) arc (180:360:1.4 and 0.6);
        \draw[dashed] (6.2,0) arc (0:180:1.4 and 0.6);
        \node[draw,thick,star,star points=5, star point ratio=2,fill=gray!60,scale=.5] at (5.5, .7) {};
        %\draw[thick,fill=white] (5.5,.7) circle (3pt);
        %\node[font=\scriptsize] (P4) at (4.6,.7) {$$\ydiagram[*(gray)]{1,1}$$};
        \draw[thick] (4,.1) circle (3pt);
        %\node[font=\scriptsize] (P5) at (3.4,-.2) {$\ydiagram[*(white)]{3}$};
        \node[draw,thick,star,star points=5,star point ratio=2,fill=gray!60,scale=.5] at (5.5,-.7) {};
        %\node[font=\scriptsize] (P6) at (4.6,-.7) {$\ydiagram[*(gray)]{1,1}$};
        %\draw[dashed] (5.5,.6)--(5.5,-.6);
        \draw[dashed] (5.5,.6)--(5.5,-.6);
        \draw[thick,dashed] (.8,0)--(4,0);
        \draw[thick,dashed] (.8,0.2)--(4,0.2);
        \shade[inner color=white, outer color=blue!40, opacity = 0.4] (1.4,0) rectangle ++(2,0.2);
        \node[font=\scriptsize] (c) at (2.4,0.5) {$SU(2n+1)$};
    \end{tikzpicture}
    \caption{$A_{2n}$ with $4 \times [2n]_t$.}\label{fig:ex5a}
    \end{subfigure}\hspace{3mm}
    \begin{subfigure}[b]{0.24\textwidth}
        \centering\;\raisebox{6.5mm}{
        \begin{tikzpicture}[scale=1.2,every node/.style={scale=1.2},font=\scriptsize]
        \node[gauge] (p0) at (0,0) {$n$};
        \node[gauge] (p1) at (1.6,0) {$n$};
        \node[gauge] (p2) at (0,1) {$n$};
        \node[gauge] (p3) at (1.6,1) {$n$};
        \node[gaugeD] (c) at (.8,.5) {\fontsize{8pt}{8pt}\selectfont $2n+1$};
        \draw (p0)--(c)--(p1);
        \draw (p2)--(c)--(p3);
        \end{tikzpicture}}
        \caption{3d reduction.}\label{fig:ex5b}
    \end{subfigure}\hspace{-3mm}
    \begin{subfigure}[b]{0.23\textwidth}
        \centering\hspace{-5mm}\raisebox{8mm}{
        \begin{tikzpicture}[scale=1.1,every node/.style={scale=1.2},font=\scriptsize]
        \node[gauge,blue] (p0) at (0,0) {$2n$};
        \node[flavor,red] (p1) at (1,0) {$8$};
        \draw (p0)--(p1);
        \draw[black,solid] (p0) edge [out=12+135,in=78+135,loop,looseness=4.5]  (p0);
        \end{tikzpicture}}
        \caption{3d magnetic quiver.}\label{fig:ex5c}
    \end{subfigure}
    \caption{In (a), we show the $A_{2n}$ class $\mathcal{S}$ theory with four twisted null punctures, the four gray-filled stars denote the twisted null punctures, while the uncolored circles denote the untwisted maximal punctures that are glued together. In (b), we depict the Lagrangian quiver describing the 3d reduction. In (c), we portray the magnetic quiver which, together with a free hypermultiplet corresponding to the singlet part of the antisymmetric, is a mirror for the theory in (b) around the most singular point on its Coulomb branch. In (c), the blue circular node denotes the $USp(2n)$ gauge group, while the arc denotes the antisymmetric hypermultiplet.}
    \label{fig:ex5}
\end{figure}
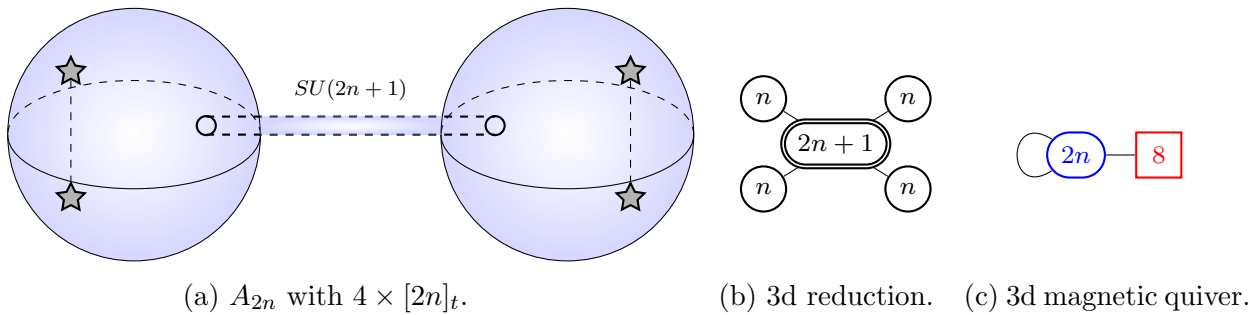

The first family of theories that we consider involves class $\mathcal{S}$ of type $A_{2n}$ on a sphere with four twisted null punctures, $4 \times [2n]_t$, which we depict in Figure \ref{fig:ex5a}. The three-punctured sphere with punctures $[1^{2n+1}]+2\times[2n]_t$ is a product SCFT, given by two copies of the $\mathcal{D}_2(SU(2n+1))$ theory \cite{Beem:2020pry,Beratto:2020wmn}. Thus, the theory with four twisted null punctures can be described as the diagonal $SU(2n+1)$ gauging of four copies of the $\mathcal{D}_2(SU(2n+1))$ theory; the resulting theory is nothing other than the $\widehat{D}_4(SU(2n+1))$ studied in \cite{Kang:2021lic}. This theory is interesting for a variety of reasons. First, as pointed out in \cite{Kang:2021lic}, it has identical central charges, a point which we discuss further in Section \ref{sec:ac}. Moreover, such SCFTs seem to be the only higher rank examples of twisted $A_{2n}$ class $\mathcal{S}$ theory for which the 3d reduction is Lagrangian.\\

While it is computationally infeasible to determine the Hall--Littlewood index for the $\widehat{D}_4(SU(2n+1))$ theory with arbitrarily large $n$, we can study the expansion for low values of $n$ to observe that the index contains terms with negative coefficients and thus satisfies Conjecture \ref{conj:HLneqHS}. For example, for $\widehat{D}_4(SU(5))$, we find that the Hall--Littlewood index is given by
\begin{align}
\scalemath{.97}{
    \operatorname{HL}(\tau)=\frac{1+\tau^4-\tau^6+2\tau^8-2\tau^{10}-\tau^{14}+\tau^{16}}{(1-\tau^4)(1-\tau^8)}=1+2\tau^4-\tau^6+5\tau^8-3\tau^{10}+\mathcal{O}\left(\tau^{12}\right) \,.}
\end{align}
Once again the negative terms that start appearing at order $\tau^6$ prevent this from being also the Hilbert series of the Higgs branch. Similarly, for $\widehat{D}_4(SU(7))$, the first terms in the expansion of the Hall--Littlewood index are
\begin{equation}
    \operatorname{HL}(\tau)=1+2\tau^4-\tau^6+5\tau^8+\mathcal{O}\left(\tau^{10}\right) \,,
\end{equation}
where we again see the ubiquitous negative coefficient at order $\tau^6$.

In order to compute the Higgs branch Hilbert series, we reduce the theory to three dimensions where, as we mentioned above, it becomes Lagrangian. The 3d reduction of $\mathcal{D}_2(SU(2n+1))$ coincides with $U(n)$ SQCD with $2n+1$ fundamental hypermultiplets \cite{Xie:2016uqq};\footnote{See also \cite{Closset:2020afy,Giacomelli:2020ryy,Dey:2021rxw} for the 3d reduction of more general $D_p^b(SU(n))$ theories, \cite{2008arXiv0806.1050B,Nanopoulos:2010bv,Xie:2012hs,Xie:2013jc,DelZotto:2014kka,Buican:2014hfa,Xie:2016uqq,Xie:2017vaf,Buican:2017fiq,Benvenuti:2017kud,Benvenuti:2017bpg,Dey:2020hfe,Beratto:2020wmn,Closset:2020scj,Giacomelli:2020ryy,Carta:2021whq,Xie:2021ewm,Carta:2021dyx,Giacomelli:2022drw} for the 3d mirrors and \cite{Bourget:2020asf,Bourget:2020mez,Bourget:2021csg} for the magnetic quivers \cite{Ferlito:2017xdq,Cabrera:2018jxt,Cabrera:2019izd} of various Argyres--Douglas theories, which can be used to compute their Higgs branch Hilbert series.} thus we see that the reduction of $\widehat{D}_4(SU(2n+1))$ is given by the quiver in Figure \ref{fig:ex5b}. Nevertheless, this description is not particularly useful for computing the Hilbert series since again the gauge group is not fully Higgsed on the generic point of the Higgs branch, as signaled by the negative terms that we find in the Hall--Littlewood index. Instead, it is more convenient to use the 3d mirror, which we propose, in line with Conjecture \ref{conj:3dmirror}, to be $USp(2n)$ SQCD with one antisymmetric hypermultiplet and eight fundamental half-hypermultiplets, together with a free hypermultiplet, as depicted in Figure \ref{fig:ex5c}. Notice that for $n=1$ this reduces to the 3d mirror of $\widehat{D}_4(SU(3))$ that we discuss in Section \ref{sec:A2w4p}, since there is no antisymmetric representation of $SU(2)$. The Coulomb branch Hilbert series of this 3d mirror when $n = 2$ is 
\begin{align}
    \begin{split}
        \operatorname{HS}(\tau)=&\,1+2\tau^4+\tau^6+6\tau^8+4\tau^{10}+\mathcal{O}\left(\tau^{12}\right)\,.
    \end{split}
\end{align}

We also match the dimensions of Higgs and Coulomb branch between the 3d reduction and the 3d mirror in Figures \ref{fig:ex5b} and \ref{fig:ex5c}, respectively. The dimension of the Coulomb branch of the 3d mirror is $n$. As such, the dimension of the Higgs branch of the 3d reduction in Figure \ref{fig:ex5b}, and thus of $\widehat{D}_4(SU(2n+1))$, is $n$. On the other hand, the dimension of the Coulomb branch of the 3d reduction is $6n$, which matches with the dimension of the Higgs branch of the 3d mirror in Figure \ref{fig:ex5c}, where we see that the inclusion of the additional free hypermultiplet is essential. 

The 3d mirror is, in fact, the rank $n$ $SO(8)$ instanton SCFT with the center of mass mode included. Again we stress that this mirror description is valid only around the most singular locus of the Coulomb branch of the theory in Figure \ref{fig:ex5b}. It would be interesting to understand how to relate the quiver in Figure \ref{fig:ex5b} to the known mirror dual to the instanton theory, which is the rank $n$ affine $D_4$ theory with one additional fundamental flavor attached to one of the $U(n)$ nodes \cite{deBoer:1996mp}. Unfortunately, the situation is less clear in the higher rank case than when $n=1$, which we discuss in Section \ref{sec:A2w4p}. Close to the most singular locus of the Coulomb branch we can, similarly to what we did in Figure \ref{fig:dual1}, show that the theory in Figure \ref{fig:ex5b} is equivalent to a similar quiver, but with the $SU(2n+1)$ gauge node replaced by $SU(2n)$ and one additional free twisted hypermultiplet. This is done using the result of \cite{Assel:2017jgo} that on the most singular point of the Coulomb branch of $U(N)$ with $2N-2$ flavors the theory is equivalent to $U(N-1)$ with $2N-2$ flavors and a free twisted hypermultiplet. However, the resulting theory remains bad for $n>1$. This is due to the fact that the badness of the quiver in Figure \ref{fig:ex5b} does not come from the central gauge node only, as happened for $n=1$. This also suggests that the presence of $\mathcal{D}$-type multiplets contributing to the Hall--Littlewood index of the 4d theory is not due only to the fact that the $SU(n)$ gauge group that we also have in 4d is not fully Higgsed on a generic point of the Higgs branch. Whether there exist further dualizations which lead to the known, good, affine quiver remains an open question, which we leave for future work.

\begin{figure}[H]
    \centering
    \begin{subfigure}[b]{.49\textwidth}
        \centering\raisebox{5mm}{
        \begin{tikzpicture}[scale=1.2,every node/.style={scale=1.2},font=\scriptsize]
    	\node[gauge,red] (t0) at (0,0) {$3$};
    	\node[gauge,blue] (t1) at (1,0) {$2$};
    	\node[gauge,red] (t2) at (2,0) {$5$};
    	\node[gauge,blue] (t3) at (3,0) {$4$};
    	\node[flavor,red] (t4) at (4,0) {$7$};
    	\draw (t0)--(t1)--(t2)--(t3)--(t4);
    	\draw[black,solid] (t3) edge [out=12+45,in=78+45,loop,looseness=4.5]  (t3);
	\end{tikzpicture}}
    	\caption{$A_4$ with $[1^4]_t+3\times[4]_t$.}
    	\label{fig:exA41max3nulltmirror}
    \end{subfigure}
    \begin{subfigure}[b]{.49\textwidth}
        \centering
        \begin{tikzpicture}[scale=1.2,every node/.style={scale=1.2},font=\scriptsize]
    	\node[gauge,red] (t0) at (0,0) {$3$};
    	\node[gauge,blue] (t1) at (1,0) {$2$};
    	\node[gauge,red] (t2) at (2,0) {$5$};
    	\node[gauge,blue] (t3) at (3,0) {$4$};
    	\node[flavor,red] (t4) at (4,0) {$13$};
    	\draw (t0)--(t1)--(t2)--(t3)--(t4);
    	\draw[black,solid] (t3) edge [out=12+45,in=78+45,loop,looseness=4.5]  (t3);
    	\draw[black,solid] (t3) edge [out=12+45+180,in=78+45+180,loop,looseness=4.5]  (t3);
	\end{tikzpicture}
    	\caption{$A_4$ with $[1^4]_t+5\times[4]_t$.}
    	\label{fig:exA41max5nulltmirror}
    \end{subfigure}
    \begin{subfigure}[b]{\textwidth}
        \centering
        \begin{tikzpicture}[scale=1.2,every node/.style={scale=1.2},font=\scriptsize]
    	\node[gauge,red] (t0) at (0,0) {$3$};
    	\node[gauge,blue] (t1) at (1,0) {$2$};
    	\node[gauge,red] (t2) at (2,0) {$5$};
    	\node[gauge,blue] (t3) at (3,0) {$4$};
    	\node[flavor,red] (t4) at (3,-1) {$7$};
    	\node[gauge,red] (t5) at (6,0) {$3$};
    	\node[gauge,blue] (t6) at (5,0) {$2$};
    	\node[gauge,red] (t7) at (4,0) {$5$};
    	\draw (t0)--(t1)--(t2)--(t3)--(t7)--(t6)--(t5);
    	\draw (t3)--(t4);
    	\draw[black,solid] (t3) edge [out=12+45,in=78+45,loop,looseness=4.5]  (t3);
	\end{tikzpicture}
    	\caption{$A_4$ with $2\times[1^4]_t+2\times[4]_t$.}
    	\label{fig:exA42max2nulltmirror}
    \end{subfigure}
    \caption{The magnetic quivers for some class $\mathcal{S}$ theories of type $A_4$ with $\geq 4$ twisted punctures. We propose that these describe the mirror duals after the addition of one, two, and one free hypermultiplets, respectively. These mirrors are valid only around the most singular point of the Coulomb branch of the 3d reduction of the class $\mathcal{S}$ theory. Straight lines denote bifundamental half-hypermultiplets, while the arcs represent antisymmetric hypermultiplets. We remind the reader that blue nodes denote $USp$ groups, and red nodes denote $SO$ groups.}
    \label{fig:A4magetic}
\end{figure}
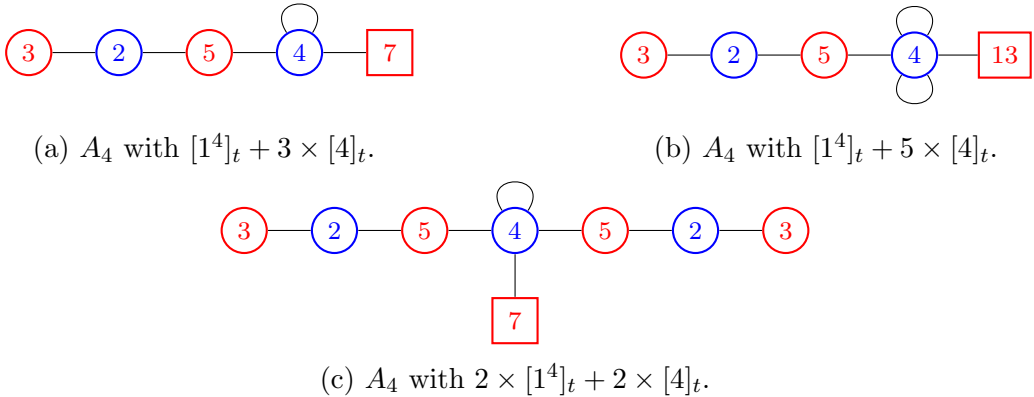

In the remainder of this section we focus on examples where $n=2$. As another example of an $A_4$ theory on a sphere with four twisted punctures, we consider the case of $[1^4]_t+3\times[4]_t$. For the Hall--Littlewood index we find
\begin{align}
    \begin{split}
        \operatorname{HL}(\tau,a)=&\,1+\chi_{\bf{10}}(a)\tau^2+\Big(\chi_{\bf{35'}}(a)+\chi_{\bf{14}}(a)+\chi_{\bf{5}}(a)+1\Big)\tau^4+\Big(\chi_{\bf{84}}(a)+\chi_{\bf{81}}(a)\\
        &+\chi_{\bf{35}}(a)+2\chi_{\bf{10}}(a)-1\Big)\tau^6+\chi_{\bf{4}}(a)\tau^7+O\left(\tau^8\right)\,,
    \end{split}
\end{align}
where $\chi_{\bf{m}}(a)$ denotes the character of the $\bm{m}$-dimensional representation of the $USp(4)$ flavor symmetry. For the Higgs branch Hilbert series we find the same expression without the negative term at order $\tau^6$
\begin{align}
    \begin{split}
        \operatorname{HS}(\tau,a)=&\,1+\chi_{\bf{10}}(a)\tau^2+\Big(\chi_{\bf{35'}}(a)+\chi_{\bf{14}}(a)+\chi_{\bf{5}}(a)+1\Big)\tau^4+\Big(\chi_{\bf{84}}(a)+\chi_{\bf{81}}(a)\\
        &+\chi_{\bf{35}}(a)+2\chi_{\bf{10}}(a)\Big)\tau^6+\chi_{\bf{4}}(a)\tau^7+O\left(\tau^8\right)\,.
    \end{split}
\end{align}
The latter was computed via the Coulomb branch Hilbert series of the 3d mirror, which we derived as usual following the prescription in Conjecture \ref{conj:3dmirror} of adding one antisymmetric, one singlet, and $2(k-1)$ fundamental hypermultiplets to the central $USp(4)$ node of the star-shaped quiver. We depict this Lagrangian quiver in Figure \ref{fig:exA41max3nulltmirror}.

Similarly the Hall--Littlewood index of the $A_4$ theory on a sphere with $2\times[1^4]_t+2\times[4]_t$ punctures is given by
\begin{align}
    \scalemath{.96}{\begin{aligned}
        \operatorname{HL}(\tau,a,b)=&\,1+\Big(\chi_{\bf{10}}(a)+\chi_{\bf{10}}(b)\Big)\tau^2+\Big(\chi_{\bf{35'}}(a)+\chi_{\bf{14}}(a)+\chi_{\bf{5}}(a)+\chi_{\bf{10}}(a)\chi_{\bf{10}}(b)\\
        &+\chi_{\bf{35'}}(b)+\chi_{\bf{14}}(b)+\chi_{\bf{5}}(b)+1\Big)\tau^4+\Big(\chi_{\bf{84}}(a)+\chi_{\bf{81}}(a)+\chi_{\bf{35}}(a)+2\chi_{\bf{10}}(a)\\
        &-1+2\chi_{\bf{10}}(b)+\chi_{\bf{35}}(b)+\chi_{\bf{81}}(b)+\chi_{\bf{84}}(b)+\chi_{\bf{35'}}(a)\chi_{\bf{10}}(b)+\chi_{\bf{10}}(a)\chi_{\bf{35'}}(b)\\
        &+\chi_{\bf{14}}(a)\chi_{\bf{10}}(b)+\chi_{\bf{10}}(a)\chi_{\bf{14}}(b)+\chi_{\bf{10}}(a)\chi_{\bf{5}}(b)+\chi_{\bf{5}}(a)\chi_{\bf{10}}(b)\Big)\tau^6+O\left(\tau^8\right)\,,
    \end{aligned}}
\end{align}
where each term involves the characters of the two flavor $USp(4)$ symmetries, which we call $\chi_{\bf{m}}(a)$ and $\chi_{\bf{n}}(b)$ respectively. Again, the Higgs branch Hilbert series gives rise to the same result but without the $-\tau^6$ term
\begin{align}
    \scalemath{.96}{\begin{aligned}
        \operatorname{HS}(\tau,a,b)=&\,1+\Big(\chi_{\bf{10}}(a)+\chi_{\bf{10}}(b)\Big)\tau^2+\Big(\chi_{\bf{35'}}(a)+\chi_{\bf{14}}(a)+\chi_{\bf{5}}(a)+\chi_{\bf{10}}(a)\chi_{\bf{10}}(b)\\
        &+\chi_{\bf{35'}}(b)+\chi_{\bf{14}}(b)+\chi_{\bf{5}}(b)+1\Big)\tau^4+\Big(\chi_{\bf{84}}(a)+\chi_{\bf{81}}(a)+\chi_{\bf{35}}(a)+2\chi_{\bf{10}}(a)\\
        &+2\chi_{\bf{10}}(b)+\chi_{\bf{35}}(b)+\chi_{\bf{81}}(b)+\chi_{\bf{84}}(b)+\chi_{\bf{35'}}(a)\chi_{\bf{10}}(b)+\chi_{\bf{10}}(a)\chi_{\bf{35'}}(b)\\
        &+\chi_{\bf{14}}(a)\chi_{\bf{10}}(b)+\chi_{\bf{10}}(a)\chi_{\bf{14}}(b)+\chi_{\bf{10}}(a)\chi_{\bf{5}}(b)+\chi_{\bf{5}}(a)\chi_{\bf{10}}(b)\Big)\tau^6+O\left(\tau^8\right)\,.
    \end{aligned}}
\end{align}
This expression is determined via the computation of the Coulomb branch Hilbert series of the 3d mirror which is depicted in Figure \ref{fig:exA42max2nulltmirror}.

Finally, it is interesting to consider an example of an $A_{2n}$ theory with $n>2$ on a sphere with $k>2$ twist lines. Specifically, we consider the $A_4$ theory on a sphere with $[1^4]_t+5\times[4]_t$ punctures. The Hall--Littlewood index is
\begin{align}
    \begin{split}
        \operatorname{HL}(\tau,a)=&\,1+\chi_{\bf{10}}(a)\tau^2+\Big(\chi_{\bf{35'}}(a)+\chi_{\bf{14}}(a)+\chi_{\bf{5}}(a)+1\Big)\tau^4\\
        &+\Big(\chi_{\bf{84}}(a)+\chi_{\bf{81}}(a)+\chi_{\bf{35}}(a)+2\chi_{\bf{10}}(a)-2\Big)\tau^6+O\left(\tau^8\right)\,.
    \end{split}
\end{align}
For the Higgs branch Hilbert series, again computed as the Coulomb branch Hilbert series of the 3d mirror, shown in Figure \ref{fig:exA41max5nulltmirror}, we find the same expression but without the $-2\tau^6$ term:
\begin{align}
    \begin{split}
        \operatorname{HS}(\tau,a)=&\,1+\chi_{\bf{10}}(a)\tau^2+\Big(\chi_{\bf{35'}}(a)+\chi_{\bf{14}}(a)+\chi_{\bf{5}}(a)+1\Big)\tau^4\\
        &+\Big(\chi_{\bf{84}}(a)+\chi_{\bf{81}}(a)+\chi_{\bf{35}}(a)+2\chi_{\bf{10}}(a)\Big)\tau^6+O\left(\tau^8\right)\,.
    \end{split}
\end{align}
In this case, the 3d mirror contains an additional $k-1=2$ copies of antisymmetric and singlet hypermultiplets and $4(k-1)=8$ fundamental half-hypermultiplets for the central $USp(4)$ node of the star-shaped quiver; it is the magnetic quiver, i.e., the interacting sector of the theory without the additional free hypermultiplets, that is shown in Figure \ref{fig:exA41max5nulltmirror}.

In all of the examples which we have studied, which we consider a representative sample of class $\mathcal{S}$ theories of type $A_{2n}$ with $\mathbb{Z}_2$-twisted punctures, we can see that Conjecture \ref{conj:HLneqHS} is satisfied. In particular, each theory with more than one twist line contains a negative coefficient in the Hall--Littlewood index; intriguingly, we note that this negative coefficient always appears at order six, and is a singlet under the flavor symmetry. While these negative terms immediately imply that the Hall--Littlewood index and the Hilbert series of the Higgs branch do not coincide, we also compute the latter explicitly using the proposal for the 3d mirrors of such class $\mathcal{S}$ theories, as given in Conjecture \ref{conj:3dmirror}.

\subsection{The 3d mirrors for twisted \texorpdfstring{\boldmath{$A_\text{even}$}}{Aeven}}\label{sec:3dmirror}

In this section, we spell out a proposal for Lagrangian quivers describing the 3d mirrors of the class $\mathcal{S}$ theories of type $A_{2n}$ on a sphere, with $2k$ twisted punctures and $m$ untwisted punctures. We have summarized this proposal in Conjecture \ref{conj:3dmirror}, which is:
\begin{conj:3dmirror}
  The 3d mirror for a class $\mathcal{S}$ theory of type $A_{2n}$ obtained from a sphere with $m$ untwisted punctures and $2k$ twisted punctures is given by the following Lagrangian quiver. For each untwisted puncture consider the theory $T_{\sigma_i}(SU(2n+1))$, where $\sigma_i$ is the partition describing the $i$th untwisted puncture; similarly, for each twisted puncture consider $T_{\rho_j}(USp'(2n))$, where $\rho_j$ is the C-partition describing the $j$th twisted puncture. Gauge the diagonal $USp(2n)$ (sub-)group of the flavor symmetry of each of these theories; add $2(k-1)$ fundamental hypermultiplets and $k-1$ anti-symmetric hypermultiplets to the introduced $USp(2n)$ gauge node. Finally, include an additional $k-1$ free hypermultiplets.
\end{conj:3dmirror}

A review of the general procedure for the construction of 3d mirrors of class $\mathcal{S}$ theories is the subject of Appendix \ref{app:3dmirror}, while here we only consider the case of type $A_{2n}$ on a sphere with untwisted and twisted punctures. Let us first briefly consider the situation of $k=1$, that is a sphere with only two twisted punctures, connected by a single twist line. The 3d mirrors of such theories have been studied in detail in \cite{Beratto:2020wmn} by adapting the construction of \cite{Benini:2010uu}. To every untwisted puncture we associate one copy of the $T_\rho(SU(2n+1))$ theory \cite{Gaiotto:2008ak}, while to each twisted puncture we associate one copy of the $T_\rho(USp'(2n))$ theory \cite{Cremonesi:2014uva}. These are glued together by gauging a diagonal combination of a common $USp(2n)$ subgroup of their flavor symmetry, so to form a star-shaped quiver with a middle $USp(2n)$ gauge node and one leg for each puncture. For the $T_\rho(SU(2n+1))$ theory, we should gauge the $USp(2n)$ subgroup of its $SU(2n+1)$ flavor symmetry according to the embedding
\begin{equation}\label{eq:SU2n1toUSp2n}
    SU(2n+1) \rightarrow USp(2n) \times U(1) \,,
\end{equation}
under which the fundamental representation decomposes as
\begin{equation}
    {\bf 2n+1} \rightarrow {\bf 2n}^0\oplus {\bf 1}^1 \,.
\end{equation}
This means that the last gauge node of each $T_\rho(SU(2n+1))$ tail is connected both to the middle $USp(2n)$ gauge node and to a flavor node, representing the $U(1)$ symmetry in the decomposition in equation \eqref{eq:SU2n1toUSp2n}, which remains as a flavor symmetry.

For the case of $k>1$ twist lines, we propose that this construction should be modified by adding $2(k-1)$ fundamental, $k-1$ antisymmetric and $k-1$ singlet hypermultiplets, where the latter constitute a free sector of the theory. This is motivated by the many examples that we previously discussed, but we also give the following heuristic argument. Let us consider the case of $k=2$, with the higher $k$ generalization being straightforward. Such a sphere can be constructed, choosing a suitable S-duality frame, by gluing two spheres with one twist line and one maximal untwisted puncture each along the untwisted puncture. In the 3d mirrors, this amounts to gauging a common diagonal combination of the $SU(2n+1)$ Coulomb branch symmetries of the $T(SU(2n+1))$ tails associated to the maximal untwisted punctures that we are gluing. 

Such a gauging is known to lead to a theory with a quantum deformed moduli space of vacua where the $SU(2n+1)\times SU(2n+1)$ symmetry of the two $T(SU(2n+1))$ tails is spontaneously broken to the diagonal subgroup and with the addition of Nambu--Goldstone modes in the adjoint representation of such a diagonal $SU(2n+1)$. This result was derived at the level of the three-sphere partition function in \cite{Benvenuti:2011ga,Nishioka:2011dq,Bottini:2021vms} where it was shown that the one for the gluing of two $T(SU(2n+1))$ theories is proportional, up to a prefactor representing the Nambu--Goldstone modes, to a delta-function identifying the mass parameters of the two $SU(2n+1)$ symmetries, and the physical interpretation was given in \cite{Bottini:2021vms}. In the case of no twisted punctures, since in the two star-shaped quivers that we are gluing the $SU(2n+1)\times SU(2n+1)$ symmetry is gauged, this means that it is Higgsed to its diagonal subgroup and one combination of the two vectors multiplets in the adjoint representation of $SU(2n+1)$ becomes massive by eating the aforementioned Nambu--Goldstone modes. 

\begin{figure}[H]
    \centering
    \begin{tikzpicture}[scale=1.2,every node/.style={scale=1.2},font=\scriptsize]
    \node[gaugeD] (cL) at (0,0) {$2$};
    \node[flavor,red] (r1) at (0,1) {$2$};
    \node[gauge] (n1) at (1,0) {$2$};
    \node[gauge] (n2) at (2,0) {$1$};
    \node[flavor,red] (r2) at (1,-1) {$2$};
    \draw (cL)--(n1)--(n2);
    \draw (cL)--(r1);
    \draw (n1)--(r2);
    \node[gaugeD] (cR) at (6,0) {$2$};
    \node[flavor,red] (rR1) at (6,1) {$2$};
    \node[gauge] (nR1) at (5,0) {$2$};
    \node[gauge] (nR2) at (4,0) {$1$};
    \node[flavor,red] (rR2) at (5,-1) {$2$};
    \draw (cR)--(nR1)--(nR2);
    \draw (cR)--(rR1);
    \draw (nR1)--(rR2);
    \draw[dashed] (3,0) ellipse (2.6cm and 0.5cm);
    \node[draw=none,opacity=0,thick,scale=0.1,fill=black,label={[label distance=0mm]0:\large{$\Longrightarrow$}}] (AR) at (7,0) {};
    \node[gaugeD] (cE) at (9,0) {$2$};
    \node[flavor,red] (pE) at (10,0) {$8$};
    \draw (cE)--(pE);
    \node[draw=none,opacity=0,thick,scale=0.1,fill=black,label={[label distance=0mm]0:\footnotesize{+ 1 free hypermultiplet}}] (AR) at (7.7,-0.8) {};
    \end{tikzpicture}
    \caption{A depiction of the Coulomb gauging of the $SU(3)$ flavor symmetries of two copies of the 3d mirror associated to the class $\mathcal{S}$ theory of type $A_2$ with $[1^3] + 2\times [2]_t$. The result on the right is the 3d mirror for the sphere with $4\times [2]_t$.}
    \label{fig:magneticgauging}
\end{figure}
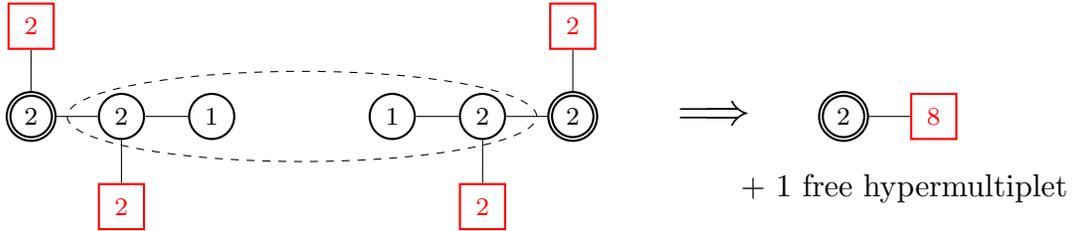

When the glued building blocks have twisted punctures, then the gauge symmetry of the central node of the star shaped-quivers is only the $USp(2n)$ subgroup of $SU(2n+1)$, but since we are still gluing two $T(SU(2n+1))$ tails in the 3d mirror we will have Nambu--Goldstone modes that are in the adjoint representation of $SU(2n+1)$. Hence, these should be decomposed according to the embedding \eqref{eq:SU2n1toUSp2n}:
\begin{equation}\label{eqn:dora}
    {\bf 4n(n+1)} \rightarrow {\bf n(2n+1)}^0\oplus {\bf (n(2n-1)-1)}^0\oplus {\bf 2n}^{\pm 1} \oplus {\bf 1}^0 \,.
\end{equation}
The first term is the adjoint representation of $USp(2n)$, so these modes will again be eaten by the massive vector fields of the Higgsed $USp(2n)$ gauge symmetry. The other terms, on the other hand, remain as massless fields transforming under the surviving $USp(2n)$ gauge symmetry that constitutes the central node of the final star-shaped quiver. These correspond exactly to the additional fields in our Conjecture \ref{conj:3dmirror}. As an example, we depict this Coulomb gauging in a case where $n=1$ in Figure \ref{fig:magneticgauging}; we observe that there are an additional two full fundamental hypermultiplets, as described in equation \eqref{eqn:dora}.

\begin{figure}[H]
    \hspace{9mm}\;\;The brane description\;:\;\qquad
    \raisebox{-10mm}{
    \begin{tikzpicture}[scale=.9,every node/.style={scale=.9},font=\scriptsize]
        \node (D) at (0,.5) {\large{$\otimes$}};
        \draw (2,2)--(2,0);
        \draw (4,2)--(4,0);
        \draw (6,2)--(6,0);
        \draw (8,2)--(8,0);
        \node (D1) at (-.5,.48) {D5};
        \node (D2) at (2.5,1.5) {D5};
        \node (NS2) at (2,2.2) {NS5};
        \node (NS3) at (4,2.2) {NS5};
        \node (NS5) at (6,2.2) {NS5};
        \node (NS6) at (8,2.2) {NS5};
        \draw (.17,.5)--(8,.5);
        \draw (2,1)--(6,1);
        \draw (3.17,1.5)--(4,1.5);
        \node (O) at (3,1.5) {\large{$\otimes$}};
        \node (D3) at (1,.7) {D3};
        \node (D4) at (5,1.2) {D3};
        \node (D5) at (3.5,1.7) {D3};
    \end{tikzpicture}}\\[1mm]
    
    \hspace{100mm}$\Big\updownarrow$\; S-duality\\[-2mm]
    
    \hspace{9mm}$\begin{array}{@{}c}
        \text{The brane configuration}\\
        \text{of the 3d reduction}
    \end{array}$:\qquad\;
    \raisebox{-10mm}{
    \begin{tikzpicture}[scale=.9,every node/.style={scale=.9},font=\scriptsize]
        \draw (0,2)--(0,0);
        \draw (3,2)--(3,0);
        \node (NS1) at (4.5,1.5) {D5};
        \node (NS2) at (1.5,1) {D5};
        \node (NS5) at (6.5,1) {D5};
        \node (NS3) at (8.5,.5) {D5};
        \node (NS2) at (0,2.2) {NS5};
        \node (NS2) at (3,2.2) {NS5};
        \draw (0,.5)--(7.83,.5);
        \draw (2.17,1)--(5.83,1);
        \draw (3,1.5)--(3.83,1.5);
        \node (DT) at (4,1.5) {\large{$\otimes$}};
        \node (DL) at (2,1) {\large{$\otimes$}};
        \node (DR) at (6,1) {\large{$\otimes$}};
        \node (DB) at (8,.5) {\large{$\otimes$}};
        \node (D3) at (.5,.7) {D3};
        \node (D4) at (5,1.2) {D3};
        \node (D5) at (3.5,1.7) {D3};
    \end{tikzpicture}}\\[3mm]
    
    \hspace{20mm}
    \raisebox{-10mm}{
    \begin{tikzpicture}[scale=.9,every node/.style={scale=.9},font=\scriptsize]
        \draw[>=stealth,double,->,thick] (-4.5,1)--(-1.2,1);
        \node (artop) at (-2.9,1.3) {\small{Two copies and}};
        \node (arbot) at (-2.9,.7) {\small{gauging $SU(3)$}};
        \draw (0,2)--(0,0);
        \draw (3,2)--(3,0);
        \draw (7,2)--(7,0);
        \draw (10,2)--(10,0);
        \node (NS1) at (4.5,1.5) {D5};
        \node (NS2) at (1.5,1) {D5};
        \node (NS3) at (4.5,1) {D5};
        \node (NS4) at (4.5,.5) {D5};
        \node (NS5) at (8.5,1) {D5};
        \node (NS9) at (0,2.2) {NS5};
        \node (NS8) at (3,2.2) {NS5};
        \node (NS7) at (7,2.2) {NS5};
        \node (NS6) at (10,2.2) {NS5};
        \draw (0,.5)--(3.83,.5);
        \draw (2.17,1)--(3.83,1);
        \draw (3,1.5)--(3.83,1.5);
        \node (DT) at (4,1.5) {\large{$\otimes$}};
        \node (DL) at (2,1) {\large{$\otimes$}};
        \node (DR) at (4,1) {\large{$\otimes$}};
        \node (DB) at (4,.5) {\large{$\otimes$}};
        \node (DTRl) at (5.5,1.5) {D5};
        \node (DRLl) at (5.5,1) {D5};
        \node (DBRl) at (5.5,.5) {D5};
        \node (DTR) at (6,1.5) {\large{$\otimes$}};
        \node (DRL) at (6,1) {\large{$\otimes$}};
        \node (DRR) at (8,1) {\large{$\otimes$}};
        \node (DBR) at (6,.5) {\large{$\otimes$}};
        \draw (6.17,.5)--(10,.5);
        \draw (6.17,1)--(7.83,1);
        \draw (6.17,1.5)--(7,1.5);
        \draw[dashed, rounded corners=15pt] (3.5,.25) rectangle ++(3,1.5);
        \node (D3) at (.5,.7) {D3};
        \node (D4) at (3.5,1.2) {D3};
        \node (D5) at (3.5,1.7) {D3};
        \node (D3) at (9.5,.7) {D3};
        \node (D4) at (6.5,1.2) {D3};
        \node (D5) at (6.5,1.7) {D3};
    \end{tikzpicture}}\\[-4mm]
    
    \hspace{20mm}
    \raisebox{-10mm}{
    \begin{tikzpicture}[scale=.9,every node/.style={scale=.9},font=\scriptsize]
        \node (artop) at (-2.9,1.3) {\phantom{\small{Two copies and}}};
        \node (arbot) at (-2.9,.7) {\phantom{\small{gauging $SU(3)$} an}};
        \node (ph) at (5,3.1) {\phantom{Two}};
        \draw[thick] (4.9,2.3)--(4.9,2.8);
        \draw[thick] (5.1,2.3)--(5.1,2.8);
        \draw (0,2)--(0,0);
        \draw (3,2)--(3,0);
        \draw (7,2)--(7,0);
        \draw (10,2)--(10,0);
        \node (NS1) at (1.5,1) {D5};
        \node (NS2) at (8.5,1) {D5};
        \node (NS2) at (0,2.2) {NS5};
        \node (NS2) at (3,2.2) {NS5};
        \node (NS2) at (7,2.2) {NS5};
        \node (NS2) at (10,2.2) {NS5};
        \draw (0,.5)--(10,.5);
        \draw (2.17,1)--(7.83,1);
        \draw (3,1.5)--(7,1.5);
        \node (DL) at (2,1) {\large{$\otimes$}};
        \node (DRR) at (8,1) {\large{$\otimes$}};
        \node (D3) at (5,.7) {D3};
        \node (D4) at (5,1.2) {D3};
        \node (D5) at (5,1.7) {D3};
    \end{tikzpicture}}\\[1mm]
    
    \hspace{101mm}$\Big\updownarrow$\; S-duality\\[-2mm]
    
    \hspace{20mm}
    \raisebox{-10mm}{
    \begin{tikzpicture}[scale=.9,every node/.style={scale=.9},font=\scriptsize]
        \draw[>=stealth,double,->,thick] (-4.5,1)--(-1.2,1);
        \node (artop) at (-2.9,1.3) {\small{The 3d mirror}};
        \node (arbot) at (-2.9,.7) {\phantom{\small{gauging $SU(3)$}}};
        \draw (2,2)--(2,0);
        \draw (8,2)--(8,0);
        \node (D1) at (-.5,.5) {D5};
        \node (ND2) at (10.5,.5) {D5};
        \node (D3) at (2.5,1.5) {D5};
        \node (D4) at (7.5,1.5) {D5};
        \node (NS1) at (2,2.2) {NS5};
        \node (NS2) at (8,2.2) {NS5};
        \draw (.17,.5)--(9.83,.5);
        \draw (2,1)--(8,1);
        \draw (3.17,1.5)--(6.83,1.5);
        \node (DL) at (0,.5) {\large{$\otimes$}};
        \node (DR) at (10,.5) {\large{$\otimes$}};
        \node (DBL) at (3,1.5) {\large{$\otimes$}};
        \node (DBR) at (7,1.5) {\large{$\otimes$}};
        \node (D3) at (5,.7) {D3};
        \node (D4) at (5,1.2) {D3};
        \node (D5) at (5,1.7) {D3};
    \end{tikzpicture}}\\[-4mm]
    
    \hspace{20mm}
    \raisebox{-10mm}{
    \begin{tikzpicture}[scale=.9,every node/.style={scale=.9},font=\scriptsize]
        \node (artop) at (-2.9,1.3) {\phantom{\small{The 3d mirror}}};
        \node (arbot) at (-2.9,.7) {\phantom{\small{gauging $SU(3)$} an}};
        \node (ph) at (5,3.1) {\phantom{Two}};
        \draw[thick] (4.9,2.3)--(4.9,2.8);
        \draw[thick] (5.1,2.3)--(5.1,2.8);
        \draw (2,2)--(2,0);
        \draw (8,2)--(8,0);
        \node (D1) at (-.5,.5) {D5};
        \node (ND2) at (10.5,.5) {D5};
        \node (D3) at (-.5,1.5) {D5};
        \node (D4) at (10.5,1.5) {D5};
        \node (NS1) at (2,2.2) {NS5};
        \node (NS2) at (8,2.2) {NS5};
        \draw (.17,.5)--(2,.5);
        \draw[color=blue] (2,.5)--(8,.5);
        \draw (8,.5)--(9.83,.5);
        \draw[color=blue] (2,1)--(8,1);
        \draw (.17,1.45)--(2,1.45);
        \draw[color=red] (.17,1.55)--(9.83,1.55);
        \draw (8,1.45)--(9.83,1.45);
        \node (DL) at (0,.5) {\large{$\otimes$}};
        \node (DR) at (10,.5) {\large{$\otimes$}};
        \node (DBL) at (0,1.5) {\large{$\otimes$}};
        \node (DBR) at (10,1.5) {\large{$\otimes$}};
        \node (D3) at (5,.7) {D3};
        \node (D4) at (5,1.2) {D3};
        \node (D5) at (5,1.7) {D3};
    \end{tikzpicture}}\\[1mm]
    
    \hspace{20mm}
    \begin{tikzpicture}[scale=1.1,every node/.style={scale=1.1},font=\scriptsize]
        \node (artop) at (-2.6,.2) {\phantom{Two copies and now}};
        \node (arbot) at (-2.6,-.2) {\phantom{gauging $SU(3)$}};
        \draw[thick] (.7,.08)--(1.1,.08);
        \draw[thick] (.7,-.07)--(1.1,-.07);
        \node[gaugeD,blue] (DG) at (2,0) {2};
        \node[flavor] (F) at (3,0) {8};
        \draw (DG)--(F);
        \node at (4.5,0) {\small{$+\;$ \color{red}singlet}};
    \end{tikzpicture}
    \caption{We depict the D5-NS5-D3-brane description of the diagonal Coulomb gauging of the $SU(3)$ flavor symmetry of two copies of the 3d mirrors of class $\mathcal{S}$ of type $A_2$ on a sphere with $2\times[2]_t+[1^3]$. In the final step, we use Hanany--Witten moves to translate the D5-branes outside of the region bounded by the NS5-branes, and the red D3-brane then becomes free to move transverse to the page; it thus corresponds to a free singlet. The blue D3-branes are bound to the D5s, and thus contribute an $SU(2)$ gauge algebra. The last node $8$ comes from counting the numbers as half-hypermultiplets.}
    \label{fig:braneCBgauging}
\end{figure}
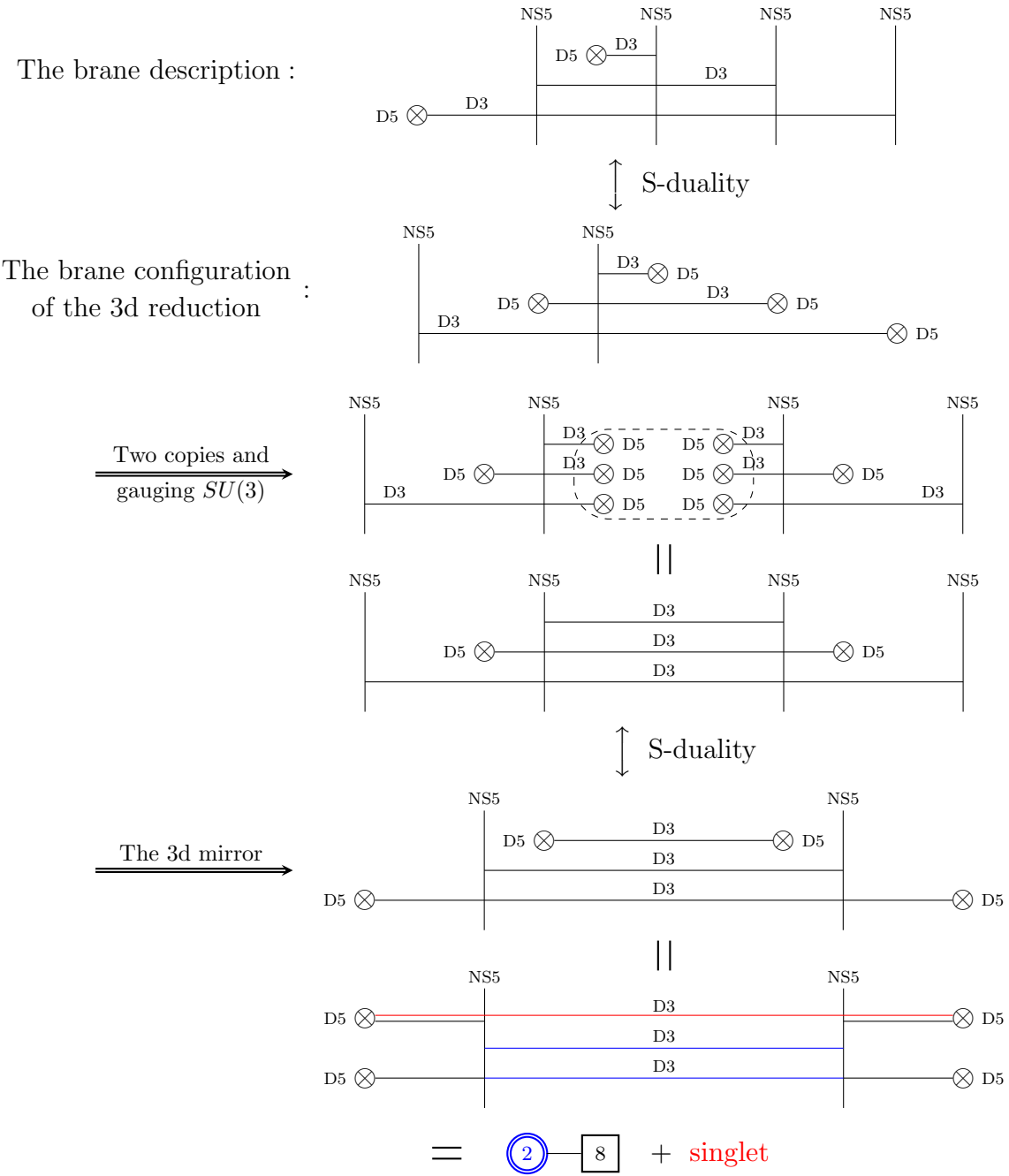

For the simple case of Coulomb gauging shown in Figure \ref{fig:magneticgauging}, the gluing process can be understood via a brane configuration in Type IIB string theory. For example, let us consider the $A_2$ type theory on a sphere with $4\times[2]_t$. It is decomposes into two copies of the $A_2$ theory on a sphere with $2\times[2]_t+[1^3]$. The 3d mirror theory of each building block is a quiver gauge theory as depicted in the left side of Figure \ref{fig:magneticgauging}, where the $USp(2)=SU(2)$ subgroup of the $SU(3)$ flavor symmetry in the $T(SU(3))$ theory is gauged and coupled with two half-hypermultiplets that arise from $2\times[2]_t$. This quiver gauge theory can be engineered via parallel D3-branes suspended between NS5-branes. Each $U(m)$ gauge theory arises from a stack of $m$ D3-branes suspended between two NS5-branes, and a full-hypermultiplet arises from a D3-brane suspended between one NS5-brane and one D5-brane. The specific brane configuration for the 3d mirror of $2\times[2]_t+[1^3]$ is depicted in the first line of Figure \ref{fig:braneCBgauging}. 

To understand the gauging procedure of the diagonal $SU(3)$ Coulomb branch symmetry of two $T(SU(3))$ tails, we first consider the mirror dual of this brane configuration that is obtained by S-duality. It is depicted in the second line of Figure \ref{fig:braneCBgauging}. Here, the $SU(3)$ Coulomb branch symmetry of the $T(SU(3))$ tail is now manifestly described by the flavor symmetry rotating the three D5-branes on the right side. Then, gauging the diagonal part of the flavor $SU(3)$ symmetry corresponds to attaching the D3-branes that were previously ending at the D5-branes. This procedure is depicted in the third and the fourth line of Figure \ref{fig:braneCBgauging}. After taking the S-duality again, we end up with the brane configuration in the fifth line of Figure \ref{fig:braneCBgauging}. It describes the $SU(2)$ gauge theory that arises from two D3-branes suspended between two NS5-branes. There are eight fundamental half-hypermultiplets, each pair comes from a D3-brane ending at each D5-brane. To exhibit the half-hypermultiplets, we move two D5-branes in the interior of the NS5-branes to the exterior, where the Hanany--Witten transition \cite{Hanany:1996ie} creates two additional D3-branes between the NS5-branes and the crossing D5-branes. We end up with the last figure of Figure \ref{fig:braneCBgauging}, where we distinguish the red and blue D3-branes in the sixth figure of Figure \ref{fig:braneCBgauging} by their boundary conditions. The blue branes are stuck at two NS5-branes and carry the vector multiplets. On the other hand, the red D3-brane is suspended between the D5-branes and can freely escape from the NS5-branes and blue D3-branes in the transverse direction to the page. Thus it is associated to a decoupled singlet hypermultiplet. In total, we obtain the 3d mirror of $A_2$ type theory on a sphere with $4\times[2]_t$ that we already discussed in Section \ref{sec:A2nw4tw}.

There is a subtlety in the discussion of the brane system that we now clarify. The brane system engineers a quiver gauge theory where all of the nodes are $U(m)$, and thus the Coulomb gauging studied via the brane system, in fact describes an analogue of Figure \ref{fig:magneticgauging} where all of the gauge nodes are unitary groups. To obtain the result in Figure \ref{fig:magneticgauging} directly, we consider the gauging of the topological $U(1)$ symmetry on both sides of the brane construction, and this converts the appropriate $U(2)$ gauge nodes into $SU(2)$ gauge nodes.

We also mention that a similar modification was already discussed for the case of $D_n$ type class $\mathcal{S}$ theories with $\mathbb{Z}_2$-twisted punctures in \cite{Benini:2010uu} and checked with Hilbert series computations in Appendix A.1.3 of \cite{Cremonesi:2014vla}. In that case, the prescription is to add $k-1$ fundamental hypermultiplets to the middle $SO(2n-1)$ gauge node of the star-shaped quiver. This again can be understood from the decomposition of the adjoint representation of $SO(2n)$ under the embedding of $SO(2n) \rightarrow SO(2n-1)$
\begin{equation}
    {\bf n(2n-1)} \rightarrow{\bf (n-1)(2n-1)} \oplus {\bf 2n-1} \,.
\end{equation}
The first term corresponds to the Nambu--Goldstone modes in the adjoint representation of $SO(2n-1)$ that are eaten in the Higgs mechanism, while the second term represents the additional fundamental hypermultiplet.

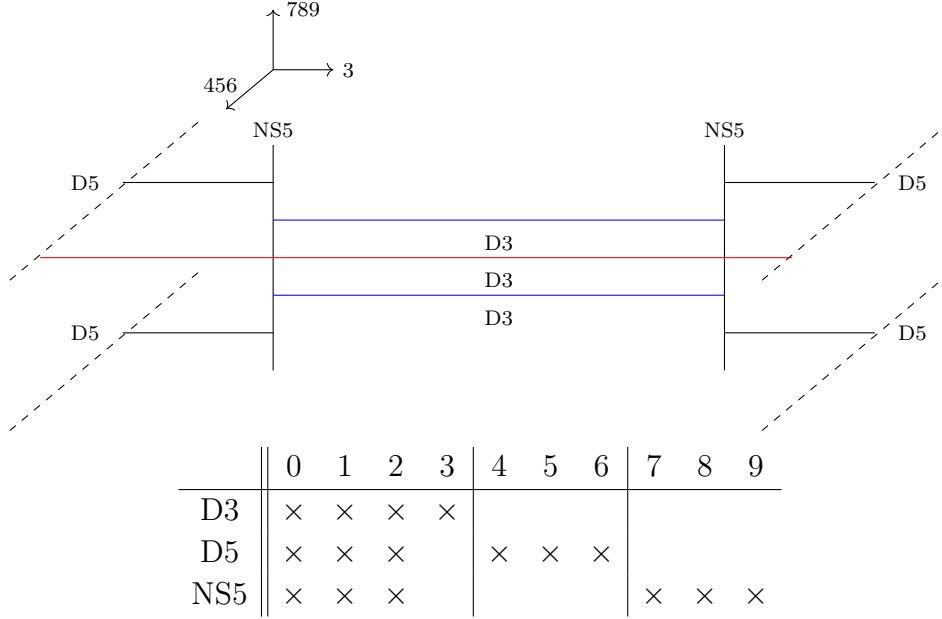
\begin{figure}[H]
    \centering
    \begin{tikzpicture}[font=\scriptsize]]
        \draw[dashed] (1,1.3)--(-1.5,-.8);
        \draw[dashed] (1,3.3)--(-1.5,3-1.8);
        \draw[dashed] (11,1.3)--(10-1.5,-.8);
        \draw[dashed] (11,3.3)--(10-1.5,3-1.8);
        \draw[->] (2,4)--(2,4.8);
        \node (dir1) at (2.4,4.8) {789};
        \draw[->] (2,4)--(2-2.5/4,4-2.1/4);
        \node (dir2) at (1.3,3.8) {456};
        \draw[->] (2,4)--(2.8,4);
        \node (dir3) at (3,4) {3};
        \draw (2,3)--(2,0);
        \draw (8,3)--(8,0);
        \node (D1) at (-.5,.5) {D5};
        \node (ND2) at (10.5,.5) {D5};
        \node (D3) at (-.5,2.5) {D5};
        \node (D4) at (10.5,2.5) {D5};
        \node (NS1) at (2,3.2) {NS5};
        \node (NS2) at (8,3.2) {NS5};
        \draw (0,.5)--(2,.5);
        \draw[color=blue] (2,1)--(8,1);
        \draw (8,.5)--(10,.5);
        \draw[color=blue] (2,2)--(8,2);
        \draw (0,2.5)--(2,2.5);
        \draw[color=red] (-1.1,1.5)--(10-1.1,1.5);
        \draw (8,2.5)--(10,2.5);
        \node (D3) at (5,.7) {D3};
        \node (D4) at (5,1.2) {D3};
        \node (D5) at (5,1.7) {D3};
    \end{tikzpicture}\\[5pt]
    $\begin{array}{c||cccc|ccc|ccc}
         & 0 & 1 & 2 & 3 & 4 & 5 & 6 & 7 & 8 & 9 \\\hline
        \text{D3} & \times & \times & \times & \times &&&&&& \\
        \text{D5} & \times & \times & \times & & \times & \times & \times &&& \\
        \text{NS5} & \times & \times & \times &&&&& \times & \times & \times \\
    \end{array}$
    \caption{A refined description of the last figure in Figure \ref{fig:braneCBgauging}. The blue D3-branes can move freely along $x^{789}$ direction whose position describes the moduli space of Coulomb branch. The red D3-brane decouples from the gauge theory by moving freely in $x^{456}$ direction.}
    \label{fig:my_label}
\end{figure}

Here we are proposing a similar prescription but in the case of twisted $A_{2n}$. Let us stress again that the previous argument is just heuristic, but we confirmed the final result summarized in Conjecture \ref{conj:3dmirror} with a multitude of computations based on the matching of Higgs and Coulomb branch dimensions and on the comparison of the Coulomb branch Hilbert series of the 3d mirror with the Higgs branch Hilbert series of the class $\mathcal{S}$ theory, when this is computable, or with the Hall--Littlewood index otherwise. As we pointed out, when $k>1$ the Hall--Littlewood index and the Hilbert series never match, but they still display the same structure of the positive terms to low orders where the negative terms come from the $\mathcal{D}$-type multiplet generators that are counted by the Hall--Littlewood index but not by the Hilbert series.

We further comment again that generally the 3d reduction of class $\mathcal{S}$ theories with multiple twist lines has a Coulomb branch with a complicated structure since there are multiple singular loci and the mirror that we are constructing is only valid around the most singular locus. In general, we expect this to capture the full Higgs branch of the 4d SCFT, while the Higgs branches emanating at the less singular loci of the Coulomb branch of the 3d reduced theory only capture subsectors of the 4d Higgs branch. We have seen this in explicit examples in the previous sections.

In this paper, we do not consider Riemann surfaces that are not spheres, since for higher genus it is already well known that the Hall--Littlewood index and the Higgs branch Hilbert series are different. In absence of twisted punctures, the 3d mirror for genus $g>0$ is modified by adding $g$ adjoint hypermultiplets to the central gauge node of the star-shaped quiver. It would be interesting to investigate what is the 3d mirror for higher genus $g>0$ and higher number of twisted punctures $2k>0$. Our expectation is that there should be $g$ adjoint hypermultiplets and $k+g-1$ copies of antisymmetric plus singlet plus two fundamental hypermultiplets for the central $USp(2n)$ gauge node of the star-shaped quiver, however we do not test this hypothesis in this paper.

\section{Class \texorpdfstring{\boldmath{$\mathcal{S}$}}{S} of other types} \label{sec:OtherTypes}

We have demonstrated throughout Section \ref{sec:Aeven}, via a large number of examples, that class $\mathcal{S}$ theories of type $A_\text{even}$ on spheres with at least four twisted punctures, have a Hall--Littlewood index that is distinct from the Hilbert series of the Higgs branch of the theory. While class $\mathcal{S}$ theories of type $A_\text{even}$ are special and unique in many ways, it is inessential for the distinction between the Hall--Littlewood index and the Higgs branch Hilbert series.

In this section, we demonstrate that Conjecture \ref{conj:HLneqHS} holds for a variety of examples where $J = A_\text{odd}$, $D_{2n}$, and $E_6$. We find that for $A_\text{odd}$, there appears to be a ubiquitous $-\tau^6$ term in the Hall--Littlewood index, however, for the $D$-series, we find examples where there are no negative coefficients at low orders in the Hall--Littlewood index. In the latter cases, an explicit computation of the Higgs branch Hilbert series, via the known Lagrangian description of the 3d mirror, verifies Conjecture \ref{conj:HLneqHS}.

\subsection{\texorpdfstring{$A_3$}{A3} on a four-punctured sphere}\label{sec:A3}

\begin{figure}[H]
    \centering
    \begin{subfigure}[b]{\textwidth}
        \centering
        \begin{tikzpicture}[scale=1.2,every node/.style={scale=1.2},font=\scriptsize]
            \node[gauge,red] (t0) {$1$};
            \node[gauge,blue] (t1) [right=0.5cm of t0] {$2$};
            \node[gauge,red] (t2) [right=0.5cm of t1] {$3$};
            \node[gauge,blue] (t3) [right=0.5cm of t2] {$4$};
            \node[gauge,red] (t4) [right=0.5cm of t3] {$5$};
            \node[gauge,blue] (t5) [right=0.5cm of t4] {$4$};
            \node[gauge,red] (t6) [right=0.5cm of t5] {$3$};
            \node[gauge,blue] (t7) [right=0.5cm of t6] {$2$};
            \node[gauge,red] (t8) [right=0.5cm of t7] {$1$};
            \node[flavor,blue] (t9) [above=0.5cm of t4] {$2$};
            \draw (t0) -- (t1) -- (t2) -- (t3) -- (t4) -- (t5) -- (t6) -- (t7) -- (t8);
            \draw (t4) -- (t9);
        \end{tikzpicture}
    	\caption{$A_3$ with $2\times[1^4]_t+2\times[4]_t$.}
    	\label{fig:A31}
    \end{subfigure}\vspace{0.5cm}
    \begin{subfigure}[b]{\textwidth}
        \centering
        \begin{tikzpicture}[scale=1.2,every node/.style={scale=1.2},font=\scriptsize]
            \node[gauge,red] (t0) {$1$};
            \node[gauge,blue] (t1) [above right=-0.3cm and 1.1cm of t0] {$2$};
            \node[gauge,red] (t2) [above right=-0.3cm and 1.1cm of t1] {$3$};
            \node[gauge,blue] (t3) [above right=-0.3cm and 1.1cm of t2] {$4$};
            \node[gauge,red] (t4) [above right=-0.3cm and 1.1cm of t3] {$5$};
            \node[gauge,blue] (t5) [below right=-0.3cm and 1.1cm of t4] {$4$};
            \node[gauge,red] (t6) [below right=-0.3cm and 1.1cm of t5] {$3$};
            \node[gauge,blue] (t7) [below right=-0.3cm and 1.1cm of t6] {$2$};
            \node[gauge,red] (t8) [below right=-0.3cm and 1.1cm of t7] {$1$};
            \node[gauge,blue] (t3u) [above left=-0.3cm and 1.1cm of t4] {$4$};
            \node[gauge,red] (t2u) [above left=-0.3cm and 1.1cm of t3u] {$3$};
            \node[gauge,blue] (t1u) [above left=-0.3cm and 1.1cm of t2u] {$2$};
            \node[gauge,red] (t0u) [above left=-0.3cm and 1.1cm of t1u]  {$1$};
            \node[gauge,blue] (t5u) [above right=-0.3cm and 1.1cm of t4] {$4$};
            \node[gauge,red] (t6u) [above right=-0.3cm and 1.1cm of t5u] {$3$};
            \node[gauge,blue] (t7u) [above right=-0.3cm and 1.1cm of t6u] {$2$};
            \node[gauge,red] (t8u) [above right=-0.3cm and 1.1cm of t7u] {$1$};
            \node[flavor,blue] (t9) [above=0.5cm of t4] {$2$};
            \draw (t0) -- (t1) -- (t2) -- (t3) -- (t4) -- (t5) -- (t6) -- (t7) -- (t8);
            \draw (t0u) -- (t1u) -- (t2u) -- (t3u) -- (t4) -- (t5u) -- (t6u) -- (t7u) -- (t8u);
            \draw (t4) -- (t9);
        \end{tikzpicture}
    	\caption{$A_3$ with $4\times[1^4]_t$.}
    	\label{fig:A32}
    \end{subfigure}
    \caption{The magnetic quivers for some class $\mathcal{S}$ theories of type $A_3$ with $\geq 4$ twisted punctures. See Appendix \ref{app:3dmirror} for the procedure for constructing such quivers given the class $\mathcal{S}$ description.}
    \label{fig:A3}
\end{figure}
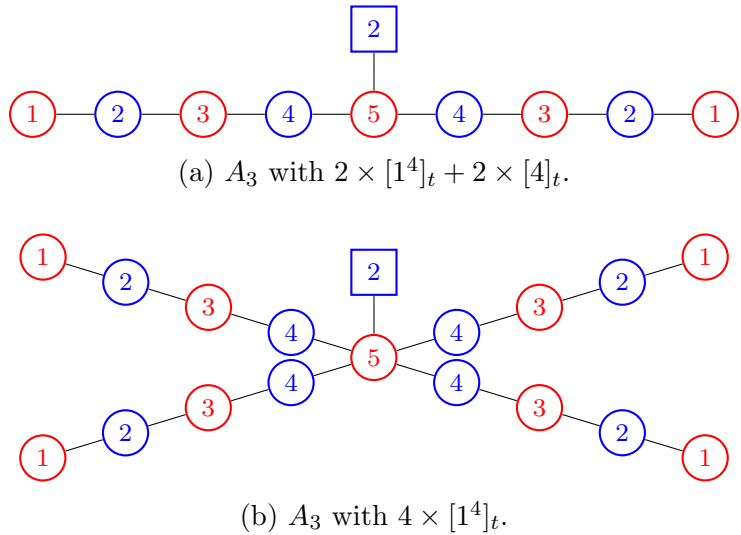

We first consider the $A_3$-type class $\CS$ theories on a four-punctured sphere as representatives of the $A_{\text{odd}}$ theories. At the same time, they can also be considered as $D_3$-type theories. A simple example is $SO(6)$ SQCD with eight fundamental half-hypermultiplets which form an octet under a $USp(8)$ flavor symmetry. This SQCD corresponds to the $A_3$-type class $\CS$ theory on a sphere with two twisted maximal punctures and two twisted minimal punctures. Since there are two twist lines, we expect, in line with Conjecture \ref{conj:HLneqHS}, that the Hall--Littlewood index and the Hilbert series of the Higgs branch are different. This should be the case since we cannot completely Higgs the gauge symmetry and there will be a residual $U(1)$ symmetry unbroken. Indeed, we find that the Hall--Littlewood index is
\begin{align}
\begin{split}
    \text{HL}(\tau;a)=&\,1+\chi_{\mathbf{36}}(a)\tau^2+(\chi_{\mathbf{330}}(a)+\chi_{\mathbf{308}}(a))\tau^4\\
    &+(\chi_{\mathbf{1716}}(a)+\chi_{\mathbf{4914}}(a)+\chi_{\mathbf{825}}(a)-\chi_{\mathbf{42}}(a))\tau^6+O\left(\tau^8\right) \,.
    \end{split}
\end{align}
We can see the $-\tau^6\chi_{\mathbf{42}}(a)$ as clear evidence that the Hall--Littlewood index differs from the Hilbert series of the Higgs branch. Also, one can directly compare the Hall--Littlewood index and the Hilbert series of the Higgs branch by studying the Coulomb branch of the 3d mirror. The 3d mirror can be constructed from the class $\mathcal{S}$ description using the procedure reviewed in Appendix \ref{app:3dmirror}; the result is depicted in Figure \ref{fig:A31}.\footnote{We emphasize that, in this section, we are considering the $D_3$ perspective on these theories, that is, the punctures are associated to $T_\rho(SO(5))$, as opposed to $T_\rho(USp(4))$. The relation between these theories is described in \cite{Cremonesi:2014uva}.} The Hilbert series of the Higgs branch is then found, using the Hall--Littlewood formula for the Coulomb branch Hilbert series of the 3d mirror as reviewed in Appendix \ref{app:HSeqs}, to be
\begin{align}
    \begin{split}
        \text{HS}(\tau;a)=&\,1+\chi_{\mathbf{36}}(a)\tau^2+(\chi_{\mathbf{330}}(a)+\chi_{\mathbf{308}}(a))\tau^4\\
    &+(\chi_{\mathbf{1716}}(a)+\chi_{\mathbf{4914}}(a)+\chi_{\mathbf{825}}(a))\tau^6+O\left(\tau^8\right) \,,
    \end{split}
\end{align}
which at low orders differs from the Hall--Littlewood index exactly by the negative term $-\chi_{\bm{42}}(a)\tau^6$.

Similarly we find a negative term in the Hall--Littlewood index of the $A_3$ theory on a sphere with four twisted maximal punctures, which is a non-Lagrangian theory:
\begin{align}
    \begin{split}
        \text{HL}(\tau;a,b,c,d)=&\,1+\tau^2(\textbf{adj})+\tau^4(\textbf{adj}^2+(\chi_{\mathbf{14}}(a)+\chi_{\mathbf{5}}(a)+\text{permutations}))\\
        &+\tau^6(\textbf{adj}^6+(\chi_{\mathbf{14}}(a)\chi_{\mathbf{10}}(b)+\chi_{\mathbf{10}}(a)\chi_{\mathbf{5}}(b)+\chi_{\mathbf{81}}(a)+\chi_{\mathbf{35}}(a)\\
        &+2\chi_{\mathbf{10}}(a)+\text{permutations})-1)+O\left(\tau^8\right)\,.
    \end{split}
\end{align}
Here the $\textbf{adj}^n$ is defined as in equation \eqref{eq:adjchar} with respect to the $USp(4)^4$ flavor symmetry. The terms given by permutations of $(a,b,c,d)$, which correspond to the fugacities of each flavor $USp(4)$ group respectively, are abbreviated. On the other hand, the Hilbert series of the Higgs branch obtained from the 3d mirror theory depicted in Figure \ref{fig:A32} is exactly
\begin{align}
    \begin{split}
        \text{HS}(\tau;a,b,c,d)=\text{HL}(\tau;a,b,c,d)+\tau^6+O\left(\tau^8\right)\,.
    \end{split}
\end{align}
At low orders, the Hall--Littlewood index mismatches with the Hilbert series of the Higgs branch by precisely the $-\tau^6$ term. It is algorithmic, though somewhat tedious, to iterate over all other four-punctured spheres of type $A_3$ with four $\mathbb{Z}_2$-twisted punctures and verify that the Hall--Littlewood index always contains terms with negative coefficients at sixth order in $\tau$. Similarly, we see that the Hilbert series of the Higgs branch begins to diverge from the Hall--Littlewood index at $\tau^6$, where the distinction is that the negative term is absent.

\subsection{\texorpdfstring{$D_4$}{D4} on a four-punctured sphere}\label{sec:D4}

In this section, we consider the $D$-type class $\CS$ theories on four-punctured spheres. We start with an example of $D_4$-type theory that admits a Lagrangian description.\footnote{When considering the $D_4$ theory we not only have $\mathbb{Z}_2$-twisted punctures but also $S_3$-twisted punctures \cite{Chacaltana:2016shw,Distler:2021cwz}. While we do not comment further on this exceptional case in this paper, it is natural to expect that there will be distinctions between the Hall--Littlewood index and the Higgs branch Hilbert series once one includes sufficiently many $S_3$-twisted punctures.} This is $SO(8)$ SQCD with twelve fundamental half-hypermultiplets that are in the fundamental representation of a $USp(12)$ flavor symmetry. This theory is realized by the $D_4$-type class $\mathcal{S}$ theory on a sphere with two twisted maximal punctures and two twisted minimal punctures. In the class $\mathcal{S}$ description only the $USp(6)\times USp(6)$ subgroup of the $USp(12)$ flavor symmetry is manifest. The Hall--Littlewood index of this theory is given by
\begin{align}
\begin{split}
    \text{HL}(\tau;a)=&\,1+\chi_{\mathbf{78}}(a)\tau^2+(\chi_{\mathbf{1365}}(a)+\chi_{\mathbf{1650}}(a))\tau^4+(\chi_{\mathbf{12376}}(a)+\chi_{\mathbf{51051}}(a)\\
    &+\chi_{\mathbf{13650}}(a))\tau^6+(\chi_{\mathbf{75582}}(a)+\chi_{\mathbf{543400}}(a)+\chi_{\mathbf{551616'}}(a)+\chi_{\mathbf{247247}}(a)\\
    &+\chi_{\mathbf{49686}}(a)-\chi_{\mathbf{429'}}(a))\tau^8+O\left(\tau^{10}\right) \,.
\end{split}
\end{align}
It differs from the Higgs branch Hilbert series as it has a negative term: $-\chi_{\mathbf{429'}}(a)\tau^8$. This is expected since we cannot completely Higgs the gauge symmetry and there is residual $U(1)$ gauge symmetry on the Higgs branch. We note that, in contrast to the $A$-type theories, the negative term first appears at order eight, rather than order six. The unrefined Hilbert series of the Higgs branch of this theory, which can be determined directly from the Lagrangian description in 4d, is
\begin{align}
    \begin{split}
        \operatorname{HS}(\tau)=1+78\tau^2+3015\tau^4+77077\tau^6+1467533\tau^8+O\left(\tau^{10}\right)\,.
        \end{split}
\end{align}
In terms of the contributing representations in the refined index, the coefficient of the $\tau^8$ term can be decomposed as
\begin{align}
    \bm{75582}+\bm{543400}+\bm{551616'}+\bm{247247}+\bm{49686}+\bm{1}+\bm{1}\,,
\end{align}
which differs from that of the Hall--Littlewood index by $\mathbf{1}+\mathbf{1}+\mathbf{429}'$.

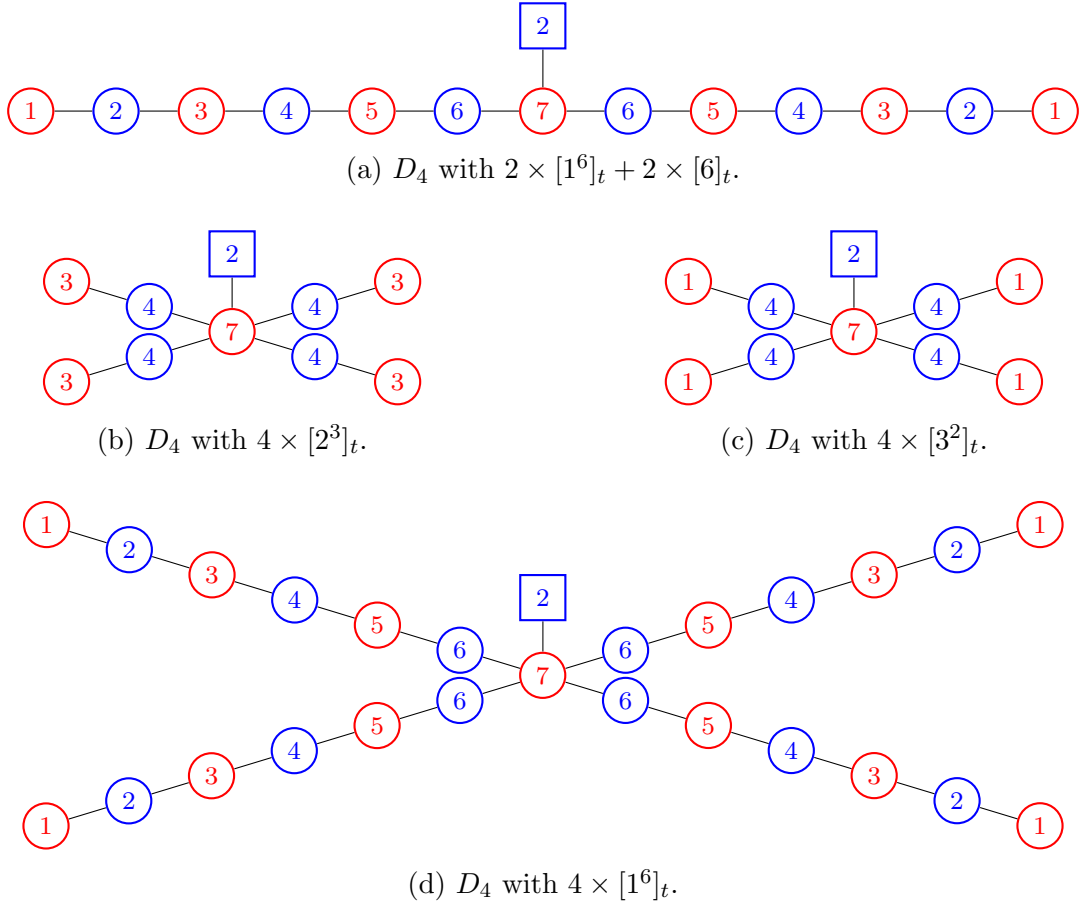
\begin{figure}[H]
    \centering
    \begin{subfigure}[b]{\textwidth}
        \centering
        \begin{tikzpicture}[scale=1.2,every node/.style={scale=1.2},font=\scriptsize]
            \node[gauge,red] (tm2) {$1$};
            \node[gauge,blue] (tm1) [right=0.5cm of tm2] {$2$};
            \node[gauge,red] (t0) [right=0.5cm of tm1] {$3$};
            \node[gauge,blue] (t1) [right=0.5cm of t0] {$4$};
            \node[gauge,red] (t2) [right=0.5cm of t1] {$5$};
            \node[gauge,blue] (t3) [right=0.5cm of t2] {$6$};
            \node[gauge,red] (t4) [right=0.5cm of t3] {$7$};
            \node[gauge,blue] (t5) [right=0.5cm of t4] {$6$};
            \node[gauge,red] (t6) [right=0.5cm of t5] {$5$};
            \node[gauge,blue] (t7) [right=0.5cm of t6] {$4$};
            \node[gauge,red] (t8) [right=0.5cm of t7] {$3$};
            \node[gauge,blue] (t9) [right=0.5cm of t8] {$2$};
            \node[gauge,red] (t10) [right=0.5cm of t9] {$1$};
            \node[flavor,blue] (tf) [above=0.5cm of t4] {$2$};
            \draw (tm2) -- (tm1) -- (t0) -- (t1) -- (t2) -- (t3) -- (t4) -- (t5) -- (t6) -- (t7) -- (t8) -- (t9) -- (t10);
            \draw (t4) -- (tf);
        \end{tikzpicture}
    	\caption{$D_4$ with $2\times[1^6]_t+2\times[6]_t$.}
    	\label{fig:D40}
    \end{subfigure}\vspace{0.5cm}
    \begin{subfigure}[b]{.49\textwidth}
        \centering
            \begin{tikzpicture}[scale=1.2,every node/.style={scale=1.2},font=\scriptsize]
            \node[gauge,red] (t2) {$3$};
            \node[gauge,blue] (t3) [above right=-0.3cm and 1.1cm of t2] {$4$};
            \node[gauge,red] (t4) [above right=-0.3cm and 1.1cm of t3] {$7$};
            \node[gauge,blue] (t5) [below right=-0.3cm and 1.1cm of t4] {$4$};
            \node[gauge,red] (t6) [below right=-0.3cm and 1.1cm of t5] {$3$};
            \node[gauge,blue] (t3u) [above left=-0.3cm and 1.1cm of t4] {$4$};
            \node[gauge,red] (t2u) [above left=-0.3cm and 1.1cm of t3u] {$3$};
            \node[gauge,blue] (t5u) [above right=-0.3cm and 1.1cm of t4] {$4$};
            \node[gauge,red] (t6u) [above right=-0.3cm and 1.1cm of t5u] {$3$};
            \node[flavor,blue] (t9) [above=0.4cm of t4] {$2$};
            \draw (t2) -- (t3) -- (t4) -- (t5) -- (t6);
            \draw (t2u) -- (t3u) -- (t4) -- (t5u) -- (t6u);
            \draw (t4) -- (t9);
        \end{tikzpicture}
    	\caption{$D_4$ with $4\times[2^3]_t$.}
    	\label{fig:D41}
    \end{subfigure}
    \begin{subfigure}[b]{.49\textwidth}
        \centering
        \begin{tikzpicture}[scale=1.2,every node/.style={scale=1.2},font=\scriptsize]
            \node[gauge,red] (t2) {$1$};
            \node[gauge,blue] (t3) [above right=-0.3cm and 1.1cm of t2] {$4$};
            \node[gauge,red] (t4) [above right=-0.3cm and 1.1cm of t3] {$7$};
            \node[gauge,blue] (t5) [below right=-0.3cm and 1.1cm of t4] {$4$};
            \node[gauge,red] (t6) [below right=-0.3cm and 1.1cm of t5] {$1$};
            \node[gauge,blue] (t3u) [above left=-0.3cm and 1.1cm of t4] {$4$};
            \node[gauge,red] (t2u) [above left=-0.3cm and 1.1cm of t3u] {$1$};
            \node[gauge,blue] (t5u) [above right=-0.3cm and 1.1cm of t4] {$4$};
            \node[gauge,red] (t6u) [above right=-0.3cm and 1.1cm of t5u] {$1$};
            \node[flavor,blue] (t9) [above=0.4cm of t4] {$2$};
            \draw (t2) -- (t3) -- (t4) -- (t5) -- (t6);
            \draw (t2u) -- (t3u) -- (t4) -- (t5u) -- (t6u);
            \draw (t4) -- (t9);
        \end{tikzpicture}
    	\caption{$D_4$ with $4\times[3^2]_t$.}
    	\label{fig:D42}
    \end{subfigure}\vspace{0.5cm}
    \begin{subfigure}[b]{\textwidth}
        \centering
        \begin{tikzpicture}[scale=1.2,every node/.style={scale=1.2},font=\scriptsize]
            \node[gauge,red] (tm2) {$1$};
            \node[gauge,blue] (tm1) [above right=-0.3cm and 1.1cm of tm2] {$2$};
            \node[gauge,red] (t0) [above right=-0.3cm and 1.1cm of tm1] {$3$};
            \node[gauge,blue] (t1) [above right=-0.3cm and 1.1cm of t0] {$4$};
            \node[gauge,red] (t2) [above right=-0.3cm and 1.1cm of t1] {$5$};
            \node[gauge,blue] (t3) [above right=-0.3cm and 1.1cm of t2] {$6$};
            \node[gauge,red] (t4) [above right=-0.3cm and 1.1cm of t3] {$7$};
            \node[gauge,blue] (t5) [below right=-0.3cm and 1.1cm of t4] {$6$};
            \node[gauge,red] (t6) [below right=-0.3cm and 1.1cm of t5] {$5$};
            \node[gauge,blue] (t7) [below right=-0.3cm and 1.1cm of t6] {$4$};
            \node[gauge,red] (t8) [below right=-0.3cm and 1.1cm of t7] {$3$};
            \node[gauge,blue] (t9) [below right=-0.3cm and 1.1cm of t8] {$2$};
            \node[gauge,red] (t10) [below right=-0.3cm and 1.1cm of t9] {$1$};
            \node[gauge,blue] (t3u) [above left=-0.3cm and 1.1cm of t4] {$6$};
            \node[gauge,red] (t2u) [above left=-0.3cm and 1.1cm of t3u] {$5$};
            \node[gauge,blue] (t1u) [above left=-0.3cm and 1.1cm of t2u] {$4$};
            \node[gauge,red] (t0u) [above left=-0.3cm and 1.1cm of t1u]  {$3$};
            \node[gauge,blue] (tm1u) [above left=-0.3cm and 1.1cm of t0u] {$2$};
            \node[gauge,red] (tm2u) [above left=-0.3cm and 1.1cm of tm1u]  {$1$};
            \node[gauge,blue] (t5u) [above right=-0.3cm and 1.1cm of t4] {$6$};
            \node[gauge,red] (t6u) [above right=-0.3cm and 1.1cm of t5u] {$5$};
            \node[gauge,blue] (t7u) [above right=-0.3cm and 1.1cm of t6u] {$4$};
            \node[gauge,red] (t8u) [above right=-0.3cm and 1.1cm of t7u] {$3$};
            \node[gauge,blue] (t9u) [above right=-0.3cm and 1.1cm of t8u] {$2$};
            \node[gauge,red] (t10u) [above right=-0.3cm and 1.1cm of t9u] {$1$};
            \node[flavor,blue] (tf) [above=0.4cm of t4] {$2$};
            \draw (tm2) -- (tm1) -- (t0) -- (t1) -- (t2) -- (t3) -- (t4) -- (t5) -- (t6) -- (t7) -- (t8) -- (t9) -- (t10);
            \draw (tm2u) -- (tm1u) -- (t0u) -- (t1u) -- (t2u) -- (t3u) -- (t4) -- (t5u) -- (t6u) -- (t7u) -- (t8u) -- (t9u) -- (t10u);
            \draw (t4) -- (tf);
        \end{tikzpicture}
    	\caption{$D_4$ with $4\times[1^6]_t$.}
    	\label{fig:D43}
    \end{subfigure}
    \caption{The magnetic quivers for some class $\mathcal{S}$ theories of type $D_4$ with four $\mathbb{Z}_2$-twisted punctures.}
    \label{fig:D4}
\end{figure}

Another example we consider is the $D_4$ theory on a sphere with four $[2^3]_t$ punctures. The Hall--Littlewood index of this theory is expressed in the characters of $SU(2)^4$ as
\begin{align}
    \begin{split}
        \text{HL}(\tau;a,b,c,d)=&\,1+\tau^2(\textbf{adj})+\tau^4\left(\textbf{adj}^2+(\chi_{\mathbf{5}}(a)+\text{permutations})+5\right)\\
        &+\tau^6\left(\textbf{adj}^3+(\chi_{\mathbf{7}}(a)+\chi_{\mathbf{5}}(a)\chi_{\mathbf{3}}(b)+\chi_{\mathbf{5}}(a)+6\chi_{\mathbf{3}}(a)\right.\\
        &+\left.\text{permutations})\right)+\tau^8\left(2\,\textbf{adj}^4+(\chi_{\mathbf{9}}(a)+2\chi_{\mathbf{5}}(a)\chi_{\mathbf{5}}(b)+\chi_{\mathbf{7}}(a)\right.\\
        &+\chi_{\mathbf{5}}(a)\chi_{\mathbf{3}}(b)+12\chi_{\mathbf{5}}(a)+7\chi_{\mathbf{3}}(a)\chi_{\mathbf{3}}(b)+\chi_{\mathbf{3}}(a)+\text{permutations})\\
        &\left.+19)\right)+O\left(\tau^{10}\right)\,,
    \end{split}
\end{align}
where the $(a,b,c,d)$ are the fugacities of $SU(2)^4$ flavor symmetry and we omit the terms that are obtained by permuting $(a,b,c,d)$. Note that the Hall--Littlewood index does not exhibit any negative term and thus it is not immediately apparent that this expression is distinct from that of the Hilbert series of the Higgs branch. In order to compare the two, we directly compute the Hilbert series of the Higgs branch from the 3d mirror theory that is depicted in Figure \ref{fig:D41}. Thus, we find that the Hilbert series of the Higgs branch is
\begin{align}
    \begin{split}
        \operatorname{HS}(\tau;a,b,c,d)=\operatorname{HL}(\tau;a,b,c,d)+\tau^8+O\left(\tau^{10}\right)\,.
    \end{split}
\end{align}
A similar comparison between the Hall--Littlewood index and the Hilbert series of the Higgs branch can be done for the $D_4$ theory on a sphere with four $[3^2]_t$ punctures. It also has a $SU(2)^4$ flavor symmetry and its Hall--Littlewood index is expressed in terms of the characters of this flavor $SU(2)^4$ as
\begin{align}
    \begin{split}
        \text{HL}(\tau;a,b,c,d)=&\,1+\tau^2(\textbf{adj})+\tau^4(\textbf{adj}^2+6)+\tau^6(\textbf{adj}^3+(\chi_{\mathbf{4}}(a)\chi_{\mathbf{2}}(b)\chi_{\mathbf{2}}(c)\chi_{\mathbf{2}}(d)\\
        &+5\chi_{\mathbf{2}}(a)\chi_{\mathbf{2}}(b)\chi_{\mathbf{2}}(c)\chi_{\mathbf{2}}(d)+8\chi_{\mathbf{3}}(a)+\text{permutations})+1)\\
        &+\tau^8(\textbf{adj}^4+8\textbf{adj}^2+4\textbf{adj}+22+(\chi_{\mathbf{6}}(a)\chi_{\mathbf{2}}(b)\chi_{\mathbf{2}}(c)\chi_{\mathbf{2}}(d)\\
        &+\chi_{\mathbf{4}}(a)\chi_{\mathbf{4}}(b)\chi_{\mathbf{2}}(c)\chi_{\mathbf{2}}(d)+\chi_{\mathbf{3}}(a)\chi_{\mathbf{3}}(b)\chi_{\mathbf{3}}(c)\chi_{\mathbf{3}}(d)\\
        &+5\chi_{\mathbf{4}}(a)\chi_{\mathbf{2}}(b)\chi_{\mathbf{2}}(c)\chi_{\mathbf{2}}(d))+3\chi_{\mathbf{3}}(a)\chi_{\mathbf{3}}(b)\\
        &+10\chi_{\mathbf{2}}(a)\chi_{\mathbf{2}}(b)\chi_{\mathbf{2}}(c)\chi_{\mathbf{2}}(d)+\text{permutations}))+O\left(\tau^{10}\right)\,.
    \end{split}
\end{align}
Again each of $(a,b,c,d)$ is the fugacity of one of the $SU(2)$ flavor symmetries and the permutation terms are omitted. We also find the Hilbert series of the Higgs branch, again from the Coulomb branch of the 3d mirror which is shown in Figure \ref{fig:D42}, for this case is 
\begin{align}
    \begin{split}
        \text{HS}(\tau;a,b,c,d)=\text{HL}(\tau;a,b,c,d)+2\tau^8+O\left(\tau^{10}\right)\,.
    \end{split}
\end{align}
While we observe no negative coefficient in the Hall--Littlewood index, we see that the Higgs branch Hilbert series differs from the Hall--Littlewood index at order $\tau^8$, consistent with Conjecture \ref{conj:HLneqHS}.

Lastly, we consider the $D_4$ theory on a sphere with four twisted maximal punctures. Due to the computational complexity, here we only consider the unrefined versions of the Hall--Littlewood index and the Hilbert series of the Higgs branch, however this is enough to see the difference between them. The Hall--Littlewood index is
\begin{align}
    \begin{split}
        \text{HL}(\tau)=1+84\tau^2+3567\tau^4+102088\tau^6+2215184\tau^8+O\left(\tau^{10}\right)\,.
    \end{split}
\end{align}
On the other hand, the Hilbert series of the Higgs branch, obtained from the 3d mirror theory described in Figure \ref{fig:D43}, is exactly
\begin{align}
    \text{HS}(\tau)=\text{HL}(\tau)+\tau^8+O\left(\tau^{10}\right)\,.
\end{align}

To summarize this section, we find that every twisted $D_4$ theory with two twist lines has Hall--Littlewood index and Higgs branch Hilbert series distinguished by the coefficient at $O\left(\tau^{8}\right)$. We have focused on four examples herein, however, it is straightforward to extend this to all four-punctured $\mathbb{Z}_2$-twisted spheres of type $D_4$. It is also possible to check that the Hall--Littlewood index and the Higgs branch Hilbert series are different even if we add more punctures or for higher rank $D_n$ theories.

\subsection{\texorpdfstring{$E_6$}{E6} on a four-punctured sphere}\label{sec:E6}

We now consider a short example for a class $\mathcal{S}$ theory of exceptional type. This provides some evidence that Conjecture \ref{conj:HLneqHS} holds beyond just the classical algebras. We consider the class $\mathcal{S}$ theory of type $E_6$ on a sphere with punctures $F_4 + F_4(a_1) + F_4(a_3) + B_3$. Recall that $\mathbb{Z}_2$-twisted punctures of type $E_6$ are labeled via the Bala--Carter notation \cite{MR417306,MR417307} for the nilpotent orbits of $F_4$ \cite{Chacaltana:2015bna}. The theory corresponds to four copies of the rank one $E_6$ Minahan--Nemeschansky theory \cite{Minahan:1996fg}, gauged together along the diagonal $SU(6)$ subgroup of the four $E_6$ flavor groups. After gauging, there exists a residual $SU(2)_{6}^4$ flavor symmetry, arising from the commutant of the gauged $SU(6)$ inside of each of the $E_6$ factors, however, the class $\mathcal{S}$ description makes manifest only the diagonal $SU(2)_{24}$ subgroup.

Instead of computing the Hall--Littlewood index from the class $\mathcal{S}$ description, where the full superconformal flavor symmetry is obscured, we can use the fact that the Higgs branch of each Minahan--Nemeschansky theory is given by the minimal nilpotent orbit of $E_6$. Considering the diagonal $SU(6)$ gauging of four such Higgs branches we find that the fully flavor-fugacity refined Hall--Littlewood index is
\begin{align}
    \begin{split}
        \operatorname{HL}(\tau,a,b,c,d)
        &=1+\tau^2(\chi_3(a))+\tau^4(6+\chi_5(a)+\chi_2(a)\chi_2(b) +\chi_3(a)\chi_3(b))\\
        &~+\tau^6(-1+6\chi_3(a)+\chi_7(a) +3\chi_2(a)\chi_2(b)+\chi_2(a)\chi_4(b)+\chi_3(a)\chi_5(b) \\  &~+\chi_2(a)\chi_2(b)\chi_3(c)+\chi_3(a)\chi_3(b)\chi_3(c)) 
        +(\text{permutations})+O(\tau^8)\,,
    \end{split}
\end{align}
where $a$, $b$, $c$, and $d$ are the fugacities of the four $SU(2)$ factors. We observe that the Hall--Littlewood index contains a term $-\tau^6$, which demonstrates that this expression cannot coincide with the Hilbert series of any Higgs branch, let alone the Higgs branch of this theory in particular. Interestingly, the $-\tau^6$ term is hidden when considering the Hall--Littlewood index as computed directly from the class $\mathcal{S}$ description, as there are additional flavor singlets under the diagonal $SU(2)$ flavor that is manifest from that perspective.

\section{\texorpdfstring{\boldmath{$a=c$}}{a=c} theories}\label{sec:ac}

Four-dimensional conformal field theories have two independent central charges (or conformal anomalies) $a$ and $c$. For $\CN \ge 3$ SCFTs, extended superconformal symmetry forces them to be equal \cite{Aharony:2015oyb}. It was not clear whether it is possible to have 4d genuinely $\CN=1, 2$ SCFTs with equal central charges $a=c$. Such theories would appear to be particularly special from the perspective of a putative holographic dual description in terms of supergravity in AdS$_5$. There, the difference between central charges $c-a$ appears in the four-derivative correction to the effective action \cite{Anselmi:1998zb, Henningson:1998gx}. There is no obvious reason for such a term to vanish in view of the effective theory for finite $c$ or the rank of gauge group in the case of gauge theory. Surprisingly, it has been recently found that there exist families of genuinely $\CN=1, 2$ SCFTs with $a=c$ \cite{Kang:2021lic, Kang:2021ccs} for every value of $N$, where $N$ scales with the rank of the gauge group.\footnote{In the context of isolated hypersurface singularities, for example to describe the non-Higgsable $(G, G^\prime)$ Argyres--Douglas SCFTs with eight supercharges, $a=c$ theories have also recently been discussed in \cite{Carta:2021whq,Closset:2021lwy}.} 

In Section \ref{sec:acS}, we find new examples of genuinely $\CN=2$ theories that have equal central charges $a=c$ in class $\CS$. Class $\CS$ realizations of $a=c$ theories means that we also have holographic duals for such theories in AdS$_5$ \cite{Gaiotto:2009gz, Nishinaka:2012vi}, which we leave for future work. Such theories include the $\widehat{D}_4 (SU(2N+1))$ theory, which we discuss in Section \ref{sec:Aeven}. We find that all the $a=c$ theories realized in class $\CS$ theories have either $g=0$ with more than one twist line or have higher genus. 
This motivates us to conjecture that any $a=c$ theory with non-trivial Higgs branch must have Hall--Littlewood index different from the Higgs branch Hilbert series (HL $\neq$ HS). We also test this conjecture in this section. 
A corollary of this conjecture is that any $\CN=3$ SCFT has HL $\neq$ HS, which we discuss in Section \ref{sec:n3scft}. The $\mathcal{N}=3$ superconformal algebra has enhanced supersymmetry from $\mathcal{N}=2$, and, from the $\mathcal{N}=2$ perspective, the additional supersymmetry currents transform in $\mathcal{D}_{1/2(0,0)}$-type multiplets. As such multiplets contribute to the Hall--Littlewood chiral ring, but not to the Higgs branch chiral ring, this provides a convincing motivation for the belief that HL $\neq$ HS in the $\mathcal{N}=3$ setting.\footnote{This is enough to show that the Hall--Littlewood chiral ring is different from the Higgs branch chiral ring, but we cannot rule out (albeit extremely unlikely) accidental agreement of the HL index and the Higgs branch Hilbert series.}

\subsection{New classes of \texorpdfstring{$a=c$}{a=c} theories in class \texorpdfstring{$\mathcal{S}$}{S}} \label{sec:acS}

Let us describe new $\CN=2$ SCFTs in class $\CS$ with $a=c$. We do not claim we have the full list of all such theories. The central charges for any theory in class $\CS$ is determined by 1) a choice of $J \in ADE$ that labels the 6d $\CN=(2, 0)$ theory, 2) a choice of Riemann surface (equivalently, the genus $g$) that the 6d theory is being compactified upon, and 3) a set of punctures put on the Riemann surface. Each untwisted puncture is labeled by an embedding of $SU(2)$ into $J \in ADE$. A twisted puncture is labeled by an embedding of $SU(2)$ into $G \in BCFG$, where $G$ labels the Langlands dual of the outer-automorphism invariant subgroup of $J$; see Table \ref{tab:GammaG}. 

\begin{table}[H]
    \centering
    \begin{tabular}{c|ccccc}
        $J$ & $A_{2n-1}$ & $A_{2n}$ & $D_{n+1}$ & $D_4$ & $E_6$ \\
        \hline
        $o$ & $\mathbb{Z}_2$ & $\mathbb{Z}_2$ & $\mathbb{Z}_2$ & $\mathbb{Z}_3$ & $\mathbb{Z}_2$ \\
        $G$ & $B_n$ & $C_n$ & $C_n$ & $G_2$ & $F_4$ \\
        $G^\vee$ & $C_n$ & $B_n$ & $B_n$ & $G_2$ & $F_4$ 
    \end{tabular}
    \caption{Upon outer-automorphism twist $o$, simply-laced $J$ changes to $G$. It is given by the Langlands-dual of the group $G^\vee$, which is the invariant subgroup of $J$ under the twist.}
    \label{tab:GammaG}
\end{table}

The central charges for the class $\CS$ theory can be simply computed by combining the contribution from the `background geometry' and the local contributions from the punctures. Equivalently, we can rewrite the central charges in terms of the effective number of vector multiplets and hypermultiplets using
\begin{align}
    a = \frac{1}{24}(5 n_v + 2 n_h) \, \qquad c = \frac{1}{12} (2 n_v + n_h) \ , 
\end{align}
or equivalently
\begin{align}
    n_v = 4(2a - c) \ , \qquad n_h = 4(5c-4a) \ . 
\end{align}
If there exists a duality-frame such that the class $\mathcal{S}$ theory obtained in this way has a weakly-coupled Lagrangian description, then these are the numbers of each kind of multiplet in that description. The numbers of effective multiplets are given in terms of $J$ and the punctured Riemann surface as \cite{Chacaltana:2012zy}
\begin{align}
    \begin{split}
    n_v &= \sum_i n_v (\rho_i) + (g-1) \left(\frac{4}{3} h^\vee (J) \textrm{dim}J + \textrm{rank}J \right) \ , \\
    n_h &= \sum_i n_h (\rho_i) + (g-1) \left(\frac{4}{3} h^\vee (J) \textrm{dim}J \right) \ , 
    \end{split}
\end{align}
where $h^\vee(J)$ denotes the dual Coxeter number of $J$. Furthermore, $n_v(\rho_i)$ and $n_h(\rho_i)$ denote the effective numbers of vector multiplets and hypermultiplets for punctures labeled by $\rho_i$, respectively. They are given by
\begin{align}
    n_v(\rho) &= 8 (\rho_J \cdot \rho_J - \rho_G \cdot \frac{h}{2} ) + \frac{1}{2} \textrm{dim} G_{1/2} \ , \\
    n_h(\rho) &= 8 (\rho_J \cdot \rho_J - \rho_G \cdot \frac{h}{2} ) + \frac{1}{2} (\textrm{rank}J - \textrm{dim} G_0) \ , 
\end{align}
where $\rho_J$ and $\rho_G$ denote the Weyl vectors of $J$ and $G$ (for the twisted punctures), respectively, and $h=\rho(\sigma_3)$. We also decomposed the Lie algebra of $G$ into 
\begin{align}
    G =  \bigoplus_{j \in \frac{1}{2}\mathbb{Z}} G_j \ , 
\end{align}
where $j$ is the eigenvalue of the action $h/2$. 
From these formulae, one can search for $a=c$, or equivalently $n_h = n_v$, theories. 

We categorize the $a=c$ theories that we find according to the behavior of the large $N$ limits, where we consider the 6d $(2,0)$ SCFT generating the class $\mathcal{S}$ theories to be of type $J=SU(N)$ or $J=SO(2N)$. Namely, 1) we can take the large $N$ limit without changing geometry or the numbers punctures, 2) we can take the large $N$ limit while increasing the genus, 3) we can take the large $N$ limit while increasing the number of punctures, or 4) we can have sporadic cases that work only for finite $N$. We do not claim our list is exhaustive, but it is interesting to see that there exist many examples of theories with $a=c$ without any known symmetry constraints. 

\paragraph{Large $N$ limit without changing geometry}
We find one class of class $\CS$ theories for which we can take $N$ to be large while keeping fixed the geometry:
\begin{align}
\begin{array}{c|c|c|c}
        J & g & \textrm{punctures} & a=c \\
        \hline
        A_{2N} & 0 & 4 \times [2N]_t &  2N(2N+1) \\
        D_N & 1 & k \times ([2N-2]_t + [2N-4, 2]_t) & 2k(N^2-N-1)
\end{array} \,.
\end{align}
Here $k$ in the second entry can be an arbitrary positive integer, since the contributions from the pair of twisted punctures in the parenthesis add up such that $a=c$. The first entry is identical to the $\widehat{D}_4 (SU(2N+1))$ theory \cite{Kang:2021lic}. The second theory has a $U(1)^k$ (manifest) flavor symmetry. Notice that the central charge for both theories scales as $O(N^2)$. We have already verified that the $\widehat{D}_4 (SU(2N+1))$ theory has HL $\neq$ HS and the second example has $g=1$. 
The class $\CS$ realizations of these theories in the large $N$ limit also implies the existence of a holographic dual description in M-theory \cite{Gaiotto:2009gz, Nishinaka:2012vi}, where our field theory analysis forbids certain 4-derivative corrections. It would be interesting to study this further. 

\paragraph{Large $N$ limit upon modifying geometry:}
There are $a=c$ theories for which one can take the large $N$ limit while simultaneously scaling the genus with order $g=O(N)$:
\begin{align} \label{eq:aeqcLargeNg}
    \begin{array}{c|c|c|c}
        J & g & \textrm{punctures} & a=c \\
        \hline
        A_{2N} & 2N+2 & 2 \times [1^{2N+1}] & \frac{8}{3} N (N+1)^2 (2N+1) \\
        A_{2N} & 2N+1 & 2 \times [1^{2N+1}] + 4 \times [2N]_t & \frac{2}{3}N(N+1) (8N^2 + 12N+7) \\
        A_{2N} & 2N & 2 \times [1^{2N+1}] + 8 \times [2N]_t & \frac{4}{3} N (N+1) (4N^2+6N+5 ) \\
        A_{N-1} & N+1 & 2 \times [1^N] & \frac{1}{3}N(N+1)^2 (N-1) 
    \end{array} \,.
\end{align}
For these theories, we can take the large $N$ limit (in the sense of increasing the rank of $J$), but we have to scale genus as well. This results in the scaling of central charge as $a=c=O(N^4)$. Notice that these examples have maximal punctures carrying $SU(2N+1)$ or $SU(N)$ so that in the large $N$ limit, the flavor symmetry also grows. Each of these four families of class $\mathcal{S}$ theories involves a compactification on a Riemann surface of genus $g > 0$, and as such it is clear that they all have HL $\neq$ HS, regardless of the presence of twisted punctures.

\paragraph{Large $N$ limit with fixed genus while scaling the number of punctures:}
We can also have another class of $a=c$ theories with a large $N$ limit where the genus $g$ is fixed, but where the number of punctures scales with $N$:
\begin{align}
    \scalemath{0.87}{\begin{array}{c|c|c|c}
        J & g & \textrm{punctures} & a=c \\
        \hline
%        A_{2N} & 0 & 2 \times [1^{2N+1}] + 4(2N+2)\times [2N])t & \frac{16}{3} N(N+1)^2 (4N+5)  \\
        A_{2N} & k & 2 \times [1^{2N+1}] + 4(2N+2-k) \times [2N]_t & \frac{2}{3}N(N+1) (8N^2 + 18N - 3k + 10) \\
        D_N & 0 & 2N [2N-2]_t + k ([2N-2]_t + [2N-4, 2]_t) & \frac{1}{3} N^2 (6 k-N)+ N (N-2 k)-\frac{2}{3} (3 k+N) \\
        D_N & 2 & 2N \times [2N-4, 2]_t & \frac{1}{2}N(13N^2-15N-N) 
    \end{array}} \,.
\end{align}
The first entry reduces to the first three cases of equation \eqref{eq:aeqcLargeNg} when we choose $g=2N$, $2N+1$, and $2N+2$, respectively. The (manifest) flavor symmetry of the first theory is $SU(2N+1)$. For the second theory, we have to scale $k$ as $O(N)$ for it to be well-defined otherwise the central charges become negative for large enough $N$. Since this class of theories involves scaling the number of punctures (simultaneously modifying the flavor symmetry), there is no guarantee that there exists a good holographic description in terms of supergravity. 

\paragraph{Sporadic examples with fixed $J$:}
Lastly, we list some sporadic theories where $a=c$ with a fixed, low-rank, choice of $J$:
\begin{align}
    \begin{array}{c|c|c|c}
        J & g & \textrm{punctures} & a=c \\
        \hline \hline
        A_2 & 0 & 4 \times [2]_t + k \times ([1^2]_t + [2]_t) & 7k+4 \\
        A_4 & 0 & 4 \times [4]_t + k \times ([2, 1^2]_t + [t]_t) & \frac{125}{4}k+12 \\
        A_2 & 1 & k \times ([1^2] + [2]_t) & 7k \\
        A_4 & 1 & k \times ([2, 1^2] + [4]_t) & \frac{125}{4} k \\
        \hline
        A_2 & k+1 & 4k \times [1^2]_t & 24k \\
        A_4 & k+1 & 4k \times [2, 1^2]_t & 113k
    \end{array} \,.
\end{align}
It is noteworthy that they all belong to the $A_{2N}$ type with twisted punctures. 

In this section, we have considered new $a=c$ theories that can be realized in class $\CS$. They either have at least two twist lines (four twisted punctures) on a $g=0$ surface or higher genus. This means that each of these theories has non-trivial Higgs branch but that the Hall--Littlewood index is distinct from the Higgs branch Hilbert series. We have directly verified these statements for $g=0$ with at least two twist lines in Sections \ref{sec:Aeven} and \ref{sec:OtherTypes}. To summarize, we have found many examples satisfying, and no counterexamples to, Conjecture \ref{conj:equalac}.

\subsection{\texorpdfstring{$\mathcal{N}=3$}{N=3} SCFTs} \label{sec:n3scft}

Genuinely $\CN=3$ SCFTs have been first constructed in \cite{Garcia-Etxebarria:2015wns}. These theories cannot have a manifestly $\CN=3$ (or even $\CN=2$) supersymmetric Lagrangian description.\footnote{Some of the $\CN=3$ theories can be realized as a renormalization group fixed point of certain $\CN=1$ Lagrangian gauge theories \cite{Zafrir:2020epd}.} 
The `Higgs branch operator' of an $\CN=3$ SCFT actually combines with the Coulomb branch operator (in the $\CN=2$ sense) to form a single $\CN=3$ multiplet, parametrizing the `Coulomb branch' of the $\CN=3$ SCFT. Therefore, we always have an Abelian gauge multiplet in the low-energy effective theory on the `Higgs branch'. Even though $\CN=3$ theories are not conventional gauge theories, it is natural to expect that the situation is similar to the case where we cannot completely Higgs the gauge group so that the Hall--Littlewood chiral ring differs from the Higgs branch chiral ring. To wit, there are $\mathcal{D}_{R(0,j_2)}$-type multiplets in any $\mathcal{N}=3$ theory. We then test Conjecture \ref{conj:equalac} via an explicit computation of the Hall--Littlewood index for a number of example theories with $\mathcal{N}=3$ supersymmetry. 

Since $\CN=3$ SCFTs are non-Lagrangian (except for a particular example studied in \cite{Zafrir:2020epd}), we do not have a direct way of computing the superconformal index. Instead, we use the SCFT/VOA correspondence \cite{Beem:2013sza}, where a particular sector (referred to as the Schur sector) of any 4d $\CN=2$ SCFT is encoded in a vertex operator algebra (VOA). In this correspondence, the vacuum character of the VOA associated to an $\CN=2$ SCFT is identical to the Schur index of the 4d theory. Moreover, one can reconstruct the Higgs branch out of the associated variety \cite{arakawa2015associated} of the VOA \cite{Beem:2017ooy, Xie:2017vaf}. Since the Schur sector contains all the operators that are counted by the Hall--Littlewood index, it indeed captures the Hall--Littlewood index, and even more, the Macdonald index \cite{Song:2017oew, Beem:2019tfp}. The associated VOAs for the $\CN=3$ SCFTs have been identified in \cite{Nishinaka:2016hbw, Lemos:2016xke, Bonetti:2018fqz}. Using these results, the Macdonald indices for $\CN=3$ SCFTs have been computed in \cite{Agarwal:2021oyl}. 

Using those results, one can simply take the Hall--Littlewood limit of the Macdonald index to obtain the following: 
\begin{align}
\begin{split}
    \operatorname{HL}_{\mathbb{Z}_3} &= 
     1+\tau ^2+\tau ^3 \left(z^3+\frac{1}{z^3}-\frac{1}{z}\right) +\tau ^4 \left(1-z^2\right)  \\
     & \qquad +\tau ^5 \left(z^3+\frac{1}{z^3}-\frac{1}{z}\right) 
      +\tau ^6 \left(z^6+\frac{1}{z^6}-\frac{1}{z^4}-z^2+1\right)+\mathcal{O}\left(\tau ^7\right) \,,
\end{split} \label{eqn:HLofMDZ3}\\
\begin{split}
    \operatorname{HL}_{\mathbb{Z}_4} 
    & = 1+\tau ^2-\frac{\tau ^3}{z}+\tau ^4 \left(z^4+\frac{1}{z^4} +1\right)\\
     & \qquad  +\tau ^5 \left(-z^3-\frac{1}{z}\right) +\tau ^6 \left(z^4+\frac{1}{z^4}+1\right)  +\mathcal{O}\left(\tau ^7\right) \,,
\end{split} \label{eqn:HLofMDZ4}\\
\begin{split}
    \operatorname{HL}_{\mathbb{G}_{(3, 1, 2)}}
    & = 1+\tau ^2+\tau ^3 \left(z^3+\frac{1}{z^3}-\frac{1}{z}\right) +\tau ^4 \left(2-z^2\right)\\
     & \qquad +\tau ^5 \left(2 z^3+\frac{2}{z^3}-\frac{2}{z}\right) +\tau ^6 \left(2 z^6+\frac{2}{z^6}-\frac{2}{z^4}-3 z^2+3\right)  +\mathcal{O}\left(\tau ^7\right) \,.
\end{split} \label{eqn:HLofMDG312}
\end{align}
The first two examples in equations \eqref{eqn:HLofMDZ3} and \eqref{eqn:HLofMDZ4}, labeled $\mathbb{Z}_3$ and $\mathbb{Z}_4$, are the rank-one theories obtained via $\mathbb{Z}_{3}$ and $\mathbb{Z}_4$ S-folding, respectively. The third example in equation \eqref{eqn:HLofMDG312}, labeled $\mathbb{G}(3, 1, 2)$, is a rank-two S-fold theory. Here $z$ is the fugacity for the $U(1)$ symmetry that commutes with $\CN=2$ supersymmetry. We see that the Hall--Littlewood indices for the $\CN=3$ SCFTs cannot be identical to the Higgs branch Hilbert series since they have negative coefficients. This verifies Conjecture \ref{conj:equalac} for a small sample of $\mathcal{N}=3$ SCFTs; this is in agreement with expectations as such theories possess operators transforming in $\mathcal{D}_{R(0,j_2)}$-type multiplets. In particular, the term $-z^{-1}\tau^3$ comes from the $\mathcal{D}_{1/2(0,0)}$ multiplets which contains the extra supersymmetry current from the $\mathcal{N}=2$ perspective.

\section{Discussion}\label{sec:conc}

Throughout this paper, we have provided ample evidence that any class $\mathcal{S}$ theory obtained from a genus-zero Riemann surface compactification with at least four $\mathbb{Z}_2$-twisted punctures has a Hall--Littlewood index which is distinct from the Hilbert series of the Higgs branch chiral ring. This may seem to be concerning as it means that the Hall--Littlewood chiral ring cannot be straightforwardly used to understand the Higgs branch of the SCFT. To alleviate this difficulty, we resort to the 3d mirror description which allows us to study the 4d Higgs branch by using a Lagrangian quiver theory. To this end, we provide a proposal for a Lagrangian description of the 3d mirrors of class $\mathcal{S}$ theories of type $A_\text{even}$ with four or more twisted punctures that has not been discussed previously.  

Our three-dimensional analysis suggests that the 3d reduction of class $\mathcal{S}$ theories with multiple twist lines is usually ``bad'' in the sense of \cite{Gaiotto:2008ak}. For bad theories, the moduli space is typically characterized by a Coulomb branch with several distinct singular loci, from each of which a different Higgs branch cone may emanate. Such a structure was investigated in \cite{Assel:2017jgo,bad} for theories with unitary gauge groups, similar to those encountered in this paper. In particular, in \cite{bad}, it is shown that each singular locus is associated to a different value of the FI parameter which in turn corresponds to a different monopole operator taking a vacuum expectation value, so that the ultraviolet gauge theory is led to flow to the SCFT located at the specific singular locus. In many cases, the Higgs branch at the most singular locus, that is the one of highest codimension, contains those at the less singular loci as subvarieties. Because of this complicated structure to their moduli spaces, bad theories usually do not admit a globally defined 3d mirror, that is, a theory that captures the full moduli space upon swapping Higgs and Coulomb branches. Instead, we find several distinct mirrors that are only valid locally around each of the singular loci. 

In this paper, we did not attempt to give a complete characterization of the moduli space of the bad theories arising from the 3d reduction. Rather, the 3d mirrors of class $\mathcal{S}$ theories with multiple twist lines, which we proposed here for type $A_\text{even}$, but which were already discussed for type $D$, have to be intended as valid only around the most singular locus. Nevertheless, their Coulomb branch still seems to capture the 4d Higgs branch completely, which was our main focus in this paper. It would be interesting to determine the 3d mirrors also at the less singular loci, for example following the technique of \cite{bad} to study the effect of the Coulomb gauging of two $T(G)$ theories in the star-shaped quivers associated to the three-punctured spheres that we glue to form the Riemann surface with multiple twist lines.

\begin{figure}[H]
\centering
    \begin{tikzpicture}[scale=1.4]
        \shade[inner color=white, outer color=blue!40, opacity = 0.4] (0,0) ellipse (1.2cm and 0.75cm);
        \draw (0,0) ellipse (1.2cm and 0.75cm);
        \draw[thick,fill=gray!60] (-.6,.35) circle (2.5pt);
        \draw[thick,fill=gray!60] (.6,.35) circle (2.5pt);
        \node[draw,thick,star,star points=5, star point ratio=2,fill=gray!60,scale=.4] at (-.6,-.35) {};
        \node[draw,thick,star,star points=5, star point ratio=2,fill=gray!60,scale=.4] at (.6,-.35) {};
        \draw[dashed] (-.6,.27)--(-.6,-.27);
        \draw[dashed] (.6,.27)--(.6,-.27);
    \end{tikzpicture}\quad\raisebox{26pt}{$\longleftrightarrow$}\quad
    \raisebox{21pt}{\begin{tikzpicture}[scale=1.2,every node/.style={scale=1.2},font=\scriptsize]
        \node[gauge] (g) at (1.5,0) {$SO(4)$};
        \node[flavor] (f1) at (0,0) {$Sp(1)$};
        \node[flavor] (f2) at (3,0) {$Sp(1)$};
        \draw (f1)--(.85,0)--(g);
        \draw (f2)--(2.15,0)--(g);
    \end{tikzpicture}}\\[5pt]
    \hspace{-16pt}$\big\downarrow$\qquad\qquad\qquad\qquad\qquad\quad\qquad $\big\downarrow$\\[5pt]\quad
    \begin{tikzpicture}[scale=1.4]
        \shade[inner color=white, outer color=blue!40, opacity = 0.4] (0,0) ellipse (1.2cm and 0.75cm);
        \draw (0,0) ellipse (1.2cm and 0.75cm);
        \draw (-.45,.05) arc (-130:-50:0.7);
        \draw (.3,.-.05) arc (50:130:0.45);
        \draw[thick,fill=white] (-.8,0) circle (2.5pt);
        \draw[thick,fill=white] (.8,0) circle (2.5pt);
    \end{tikzpicture}\quad\raisebox{26pt}{$\longleftrightarrow$}\quad
    \begin{tikzpicture}[scale=1.3,every node/.style={scale=1.2},font=\scriptsize]
        \node[gauge] (g1) at (1.5,.5) {$SU(2)$};
        \node[gauge] (g2) at (1.5,-.5) {$SU(2)$};
        \node[flavor] (f1) at (0,0) {$SU(2)$};
        \node[flavor] (f2) at (3,0) {$SU(2)$};
        \draw (f1)--(.85,0)--(g1);
        \draw (f1)--(.85,0)--(g2);
        \draw (f2)--(2.15,0)--(g1);
        \draw (f2)--(2.15,0)--(g2);
    \end{tikzpicture}
    \caption{Class $\CS$ realization of the $SO(4)$ SQCD with 4 half-hypermultiplets (top left) and its quiver diagram (top right). Alternatively, they can be realized as the type $A_1$ class $\CS$ theory (bottom left and right).}
    \label{fig:so4}
\end{figure}
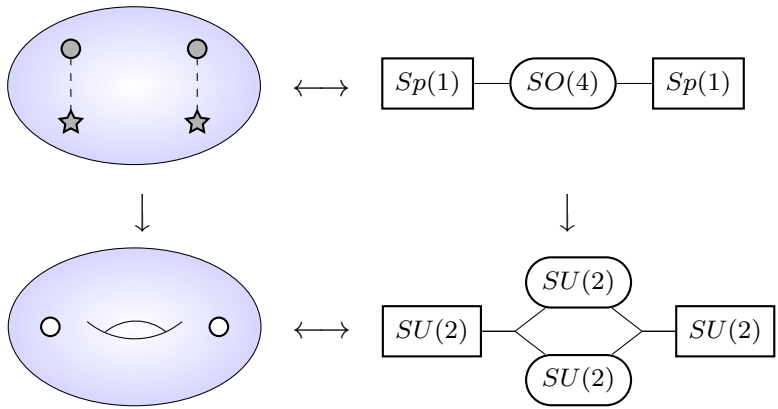

By combining the techniques of the Hall--Littlewood index and 3d mirrors, we were able to verify that having four (or more) $\mathbb{Z}_2$-twisted punctures or two twist (or more) lines on a genus zero Riemann surface give rise to $\mathcal{D}$-type multiplets so that HL $\neq$ HS. One may wonder why this is the case. One observation is that for the case of a $D_2$ class $\CS$ theory with twisted punctures, it is always possible to map such theory to an $A_1$ class $\CS$ theory \cite{Hollands:2010xa, Hollands:2011zc, Lemos:2012ph}. This can be done via taking the Riemann surface to its covering space as in the Figure \ref{fig:so4}. Namely, when we have $g-1$ twist lines for the $D_2$ type theory, it gets mapped to the genus-$g$ $A_1$ type theory. Even though this is not a generic picture, this may be a hint towards the origin of the $\mathcal{D}$-type multiplets, as also suggested in \cite{Beem:2022mde}. 
Alternatively, consider codimension two defects in the 6d $\CN=(2, 0)$ theory corresponding to the twisted punctures. By considering an `OPE' of four codimension two defects, we may obtain an extra operator that descends to the $\mathcal{D}$-type multiplet in 4d. This may also explain why having a single twist line does not give rise to HL $\neq$ HS. 

In this paper, we observe that some of the $a=c$ theories studied in \cite{Kang:2021ccs}, specifically the $\widehat{D}_4(SU(2n+1))$ SCFTs, have realizations inside of class $\mathcal{S}$ of type $A_{2n}$ with four twisted null punctures. From this perspective it is straightforward to use the class $\mathcal{S}$ machinery to compute the Hall--Littlewood index, however this is not necessary as the Hall--Littlewood index for the $\widehat{\Gamma}(G)$ theories can be determined from the constituent $\mathcal{D}_p(G)$ building blocks. A priori, the same methods can be used to work out the full superconformal index for both the $\mathcal{N}=2$ $\widehat{\Gamma}(G)$ SCFTs with $a=c$ and for the broad class of $\mathcal{N}=1$ SCFTs with $a=c$ that were constructed and studied in a similar manner in \cite{Kang:2021lic}. In \cite{OPSPEC}, the superconformal indices of these $a=c$ theories, together with what we can learn about the spectrum of local operators from the index, are studied. This provides a broad generalization of the observation from the Hall--Littlewood index that there exists $\mathcal{D}$-type multiplets in the spectrum, as we observed in this paper.

We have also constructed several new examples of $a=c$ theories in class $\CS$. We have verified that all of these theories have non-trivial Higgs branch and distinct Hall--Littlewood index and Higgs branch Hilbert series.
In addition, since holographic duals for class $\CS$ theories have been established \cite{Gaiotto:2009gz}, one can try to construct the gravity duals of such theories including the $\widehat{D}_4(SU(2N+1))$ theory. It is especially interesting to search for the mechanism behind the $a=c$ property (or the absence of certain four-derivative terms in 5d supergravity) in the holographic setup. 

In another vein, it has been well-established that there exists a small but significant subset of 4d $\mathcal{N}=2$ SCFTs that can be obtained both from the class $\mathcal{S}$ construction, and from the perspective of a $T^2$ compactification of a 6d $(1,0)$ SCFT \cite{DelZotto:2015rca,Ohmori:2015pua,Ohmori:2015pia,Baume:2021qho}. This raises the natural question of whether any of the theories with Hall--Littlewood index different from the Higgs branch Hilbert series studied in this paper lie within this subset of theories. This is particularly interesting, as the available tools to directly study the Higgs branch of the 6d $(1,0)$ SCFTs are limited. The ability to use the Hall--Littlewood index of a dual class $\mathcal{S}$ description, which has the same Higgs branch as the 6d SCFT because it arises from a torus-compactification, is thus useful. For example, see \cite{Distler:2022yse} where the Hall--Littlewood index of the dual class $\mathcal{S}$ description provides a verification of the spectrum of high dimension Higgs branch operators read off from the tensor branch description of rank $N$ $(SO(4k), SO(4k))$ conformal matter theories.

When considering a torus-compactification of a 6d $(1,0)$ SCFT, any dual class $\mathcal{S}$ description is typically of the following form.\footnote{There can, of course, also exist ``accidental'' overlap between class $\mathcal{S}$ and 6d $(1,0)$ on a torus. A class $\mathcal{S}$ theory can be related to other class $\mathcal{S}$ theories via partial closure of the punctures; similarly a 6d $(1,0)$ SCFT can be related to other 6d $(1,0)$ SCFTs by Higgs branch renormalization group flows. An overlap between class $\mathcal{S}$ and 6d $(1,0)$ on $T^2$ is considered accidental if their Higgs flows do not overlap: i.e., the class $\mathcal{S}$ theories related by partial puncture closure and the 6d $(1,0)$ SCFTs related by Higgs branch flows on $T^2$ do not overlap. For example, the 6d $(1,0)$ rank one E-string compactified on a torus gives rise to the 4d $\mathcal{N}=2$ Minahan--Nemeschansky theory of type $E_8$ \cite{Ganor:1996xd}; this particular theory has a wide variety of constructions inside of class $\mathcal{S}$, not all of these are related to the rank two $E_8$ Minahan--Nemeschansky theory by partial puncture closure.} Consider the family of 6d $(1,0)$ SCFTs known as rank $N$ $(G, G)$ conformal matter, which are engineered in M-theory as the worldvolume theory on a stack of $N$ M5-branes probing a $\mathbb{C}^2/\Gamma_G$ orbifold \cite{DelZotto:2014hpa}.\footnote{We note that these 6d $(1,0)$ SCFTs here are constructed 
via atomic construction of elliptic curves \cite{Heckman:2015bfa,Heckman:2013pva,DelZotto:2014hpa}, with an extension of anomaly matching using the Dynkin index \cite{Esole:2020tby}, where this extension arose from the wide zoo of data from geometric perspectives in \cite{Esole:2019asj,Esole:2018mqb,Esole:2018csl,Esole:2017hlw,Esole:2017qeh,Esole:2017rgz,Esole:2017kyr,Esole:2014dea,Esole:2014hya,Esole:2014bka,Esole:2011sm}.} 
The dual description of the $T^2$ compactification of these conformal matter theories is in terms of class $\mathcal{S}$ of type $G$ on a sphere with two maximal and $N$ simple punctures. When the $G \times G$ flavor group of the 6d SCFT is not simply-connected, we can consider the compactification on a torus with a non-trivial Stiefel--Whitney twist \cite{Ohmori:2018ona,Giacomelli:2020jel,Giacomelli:2020gee,Heckman:2022suy}; in this case, the resulting 4d $\mathcal{N}=2$ SCFT has a dual class $\mathcal{S}$ description of type $\widetilde{G}$ on a sphere with two twisted maximal punctures and $N$ simple punctures. Here $\widetilde{G}$ is defined such that $(\mathbb{C}^2/\Gamma_{\widetilde{G}})/\mathbb{Z}_t = \mathbb{C}^2/\Gamma_{G}$, where $t$ is the order of the Stiefel--Whitney twist, and the order of the outer-automorphism of the twisted punctures is similarly fixed, see \cite{Ohmori:2018ona} for more details. Partial closure of the untwisted maximal punctures in the class $\mathcal{S}$ description yields new 4d $\mathcal{N}=2$ SCFTs, and these can equivalently be obtained by first performing a Higgs branch renormalization group flow in 6d, and then taking the torus-compactification; see \cite{Baume:2021qho} for a wealth of examples.

As we can see, the class $\mathcal{S}$ description of these torus compactifications of 6d $(1,0)$ SCFTs always involves at most two $\mathbb{Z}_2$-twisted punctures. Therefore, by Conjecture \ref{conj:HLneqHS}, their Hall--Littlewood index is equal to the Hilbert series of their Higgs branch. Furthermore, the Higgs branch is unchanged under the dimensional reduction on the torus, and thus one can directly access the Higgs branch chiral ring of the 6d theory from a study of the Hall--Littlewood index of the dual class $\mathcal{S}$ description. This is particularly relevant as the study of the Higgs branches of the 6d $(1,0)$ SCFTs is an ongoing and active area of research \cite{Anderson:2013rka,Anderson:2017rpr,Hanany:2018uhm,Hanany:2018vph,Cabrera:2019izd,Baume:2021chx,Distler:2022yse,DKL}.

\subsection*{Acknowledgements}
The authors thank Chris Beem, Jacques Distler, Simone Giacomelli, Noppadol Mekareeya, and Sara Pasquetti for discussions.
M.J.K.~and C.L.~thank the Korea Advanced Institute of Science and Technology for hospitality during the middle stages of this work; K.H.L.~thanks the California Institute of Technology for hospitality during the later stages of this work.
M.J.K.~is supported by a Sherman Fairchild Postdoctoral Fellowship and the U.S.~Department of Energy, Office of Science, Office of High Energy Physics, under Award Number DE-SC0011632.
C.L.~acknowledges support from DESY (Hamburg, Germany), a member of the Helmholtz Association HGF. K.H.L.~and J.S.~are supported by the National Research Foundation of Korea (NRF) grant NRF-2020R1C1C1007591. 
M.S.~is supported by the ERC Consolidator Grant \#864828 “Algebraic Foundations of Supersymmetric Quantum Field Theory (SCFTAlg)” and the grant \#494786 from the Simons Foundation.
The work of J.S.~is also supported by POSCO Science Fellowship of POSCO TJ Park Foundation and a Start-up Research Grant for new faculty provided by KAIST.  
	
\appendix 

\section{The formulae for the Hall--Littlewood index}\label{app:HLeqs}

In this appendix, we review the general formulae used to compute the Hall--Littlewood index of any 4d $\mathcal{N}=2$ SCFT arising in class $\CS$. This method was introduced in \cite{Gadde:2011uv,Lemos:2012ph,Chacaltana:2014jba,Chacaltana:2015bna}, and we largely follow the exposition in those references throughout this appendix.

Consider a class $\mathcal{S}$ theory of type $J$ associated to an $n$-punctured genus $g$ Riemann surface. Each puncture is labeled via a nilpotent orbit of $G$, where $G$ is the Langlands dual of the invariant subalgebra of $J$ under the outer-automorphism associated to the twisting of the puncture. If the puncture is untwisted then, since $J$ is simply-laced, $G = J$. A nilpotent orbit $\rho_i$ corresponds to an $SU(2)$ embedding $SU(2)\times F_i \subset G$. The punctures, and thus the nilpotent orbits, of simple classical Lie groups are captured by integer partitions \cite{Gaiotto:2009we,Lemos:2012ph}. More generally, nilpotent orbits are understood through the Bala--Carter classification \cite{MR417306,MR417307}, which are explicitly listed for the exceptional algebras in the class $\mathcal{S}$ context in the appendix of \cite{Chacaltana:2012zy}. 

The Hall--Littlewood index of a class $\mathcal{S}$ theory is computed by the general formula
\begin{align}
    \text{HL}=\sum_{\mathbf{m}}\frac{\prod_{i=1}^{n}\mathcal{K}^G_{\rho_i}(\mathbf{x}_i)\psi_G(\tau;\mathbf{a}_i(\tau,\mathbf{x}_i);\mathbf{m})}{(\mathcal{K}^J_{\text{null}}(\tau)\psi_J(\tau;\mathbf{a}_{\text{triv}}(\tau);\mathbf{m}))^{2g-2+n}} \,,
    \label{eq:HLindex}
\end{align}
where the sum is taken over every representation, with highest weight vector $\mathbf{m}$, of $J$ that is invariant under the outer-automorphism twist. The variables $\mathbf{a}_i$ are the fugacities of $G$ that are mapped to fugacities $(\tau,\mathbf{x}_i)$ of $SU(2)\times F_i$ with the explicit mapping $\mathbf{a}\rightarrow \mathbf{a}_i(\tau,\mathbf{x}_i)$ on the $i$th puncture. The $\mathbf{a}_{\text{triv}}$ corresponds to the fugacity mapping of the untwisted null puncture. $\mathcal{K}_{\rho_i}(\mathbf{x}_i)$ is given as
\begin{align}\label{eqn:appK}
\begin{split}
    \mathcal{K}^G_{\rho_i}(\mathbf{x}_i)=\text{PE}\left[\sum_{\alpha=1}^{k}\tau^{2(n_\alpha+1)}\chi_{\mathbf{R}_{\alpha}}(\mathbf{x}_i)\right]\,,
\end{split}
\end{align}
when the adjoint representation of $G$ is decomposed under the $SU(2)$ embedding by
\begin{align}
    \textbf{adj}\longrightarrow\oplus_{\alpha=1}^{k}(n_\alpha,\mathbf{R}_\alpha)\,.
\end{align}
Here $(n,\mathbf{R})$ is the tensor product of the representation with dimension $n$ in $SU(2)$ and the representation $\mathbf{R}$ in $F_i$ on the $i$th puncture. The $\text{PE}$ refers to the plethystic exponential
\begin{align}
    \text{PE}\left[f(x)\right]=\text{exp}\left[\sum_{k=1}^{\infty}\frac{f(x^k)}{k}\right]\,.
\end{align}
$\psi_G$ denotes the Hall--Littlewood polynomial of $G$, which is written as
\begin{align}\label{eqn:appHL}
\begin{split}
    &\psi_G(\tau;\mathbf{a};\mathbf{m})=C_G(\tau;\mathbf{m})\Psi_G(\tau;\mathbf{a};\mathbf{m})\,,\\
    &C_G(\tau;\mathbf{m})=\left(\sum_{\omega\in W,\omega(\mathbf{m})=\mathbf{m}}\frac{\tau^{2l(\omega)}}{(1-\tau^2)^{\text{rk}}}\right)^{-1/2}\,,\\
    &\Psi_G(\tau;\mathbf{a};\mathbf{m})=\sum_{\omega\in W}\mathbf{a}^{\omega(\mathbf{m})}\prod_{\mathbf{\alpha}\in R^+}\frac{1-\tau^2\mathbf{a}^{-\omega(\mathbf{\alpha})}}{1-\mathbf{a}^{-\omega(\mathbf{\alpha})}}\,,
    \end{split}
\end{align}
where the $R^+$, $W$, $l(\omega)$, and $\text{rk}$ refer to the positive roots, the Weyl group, the length of the Weyl group element $\omega$, and the rank of $G$, respectively. The denominator of equation \eqref{eq:HLindex} is a normalization factor that corresponds to the structure constant of the two-dimensional TQFT that is determined in \cite{Lemos:2012ph,Song:2015wta}. Note that we put the additional $(1-\tau^2)^{\text{rk}/2}$ in the normalization factor $C(\tau)$ compared with \cite{Chacaltana:2013oka,Chacaltana:2014jba,Chacaltana:2015bna,Chacaltana:2017boe,Chacaltana:2018vhp} in order to absorb the overall factor $\mathcal{A}(\tau)$ there. 

\section{The formulae for the Hilbert series}\label{app:HSeqs}

In this appendix, we review the relevant formulae used in the main text to compute the Coulomb branch Hilbert series of the 3d mirrors of class $\mathcal{S}$ theories. All the equations are taken from the original references \cite{Cremonesi:2013lqa,Cremonesi:2014kwa,Cremonesi:2014vla,Cremonesi:2014uva} (see also Appendix A of \cite{Beratto:2020wmn}).

Given a 3d $\mathcal{N}=4$ quiver gauge theory, we can compute its Coulomb branch Hilbert series using the monopole formula \cite{Cremonesi:2013lqa}. For simplicity, we only discuss the case of a simple gauge group $G$, with the generalization to a quiver theory being straightforward. The formula is
\begin{align}\label{eq:monformula}
    \operatorname{HS}(\tau)=\sum_{\mathbf{m}}\tau^{2\Delta(\mathbf{m})}z^{\sum_im_i}P_G(\tau;\mathbf{m}) \,.
\end{align}
The ingredients in this formula are as follows. The sum is over magnetic fluxes $\mathbf{m}$ living in the coweight lattice of $G$, and which label the different monopole sectors. $\Delta(\mathbf{m})$ is the dimension of the monopole operator with magnetic flux $\mathbf{m}$ and it is given by
\begin{align}\label{eq:mondim}
    \Delta(\mathbf{m})=\frac{1}{2}\sum_i\sum_{w_i\in \mathcal{R}_i}|w_i(\mathbf{m})|-\sum_{\mathbf{\alpha}\in\Delta_+(G)}|\mathbf{\alpha}(\mathbf{m})| \,,
\end{align}
where $i$ runs over all the hypermultiplets in the theory which transform in representations $\mathcal{R}_i$ with weights $w_i$ of the gauge group $G$, and where $\mathbf{\alpha}$ are the positive roots of $G$. Finally, $P_G(\tau;\mathbf{m})$ is a dressing factor that counts monopole operators, dressed with the scalar in the $\mathcal{N}=2$ adjoint chiral contained in the $\mathcal{N}=4$ vector multiplet, that are invariant under the residual gauge group in the monopole background
\begin{align}\label{eq:Pfactor}
    P_G(\tau;\mathbf{m})=\prod_a\frac{1}{1-\tau^{2d_a}} \,.
\end{align}
Here $d_a$ are the degrees of the independent Casimir invariants of the residual gauge group. The formula in equation \eqref{eq:monformula} can be further refined with a fugacity $z$ for the topological symmetry when the gauge group admits one. This is done by inserting a factor of the form $z^{\sum_im_i}$ in the summation. The topological symmetry for a gauge group $G$ is the center $Z(G^\vee)$ of the dual group $G^\vee$. For example, $U(N)$ gauge groups have a $U(1)$ topological symmetry, $SO(N)$ gauge groups have a $\mathbb{Z}_2$ topological symmetry so $z^2=1$, while $USp(2N)$ gauge groups do not have any topological symmetry, thus $z=1$.

The monopole formula can be effectively used to compute the Coulomb branch Hilbert series only when the theory is not bad, that is when all monopole operators have non-negative scaling dimensions: $\Delta(\mathbf{m})>0$. For many of the 3d mirror theories considered in the main text this condition is not satisfied. Instead, we can extract the Coulomb branch Hilbert series of the $T_\rho(G)$ theories, refined by background magnetic fluxes for their global symmetries, even when they are bad using the Hall--Littlewood formula of \cite{Cremonesi:2014kwa,Cremonesi:2014vla,Cremonesi:2014uva}.\footnote{The Hall--Littlewood formula actually applies to both the Higgs and the Coulomb branch Hilbert series of the more general $T_\rho^\sigma(G)$.}$^,$\footnote{We note that, despite the similar nomenclature, the Hall--Littlewood formula used to compute the Hilbert series reviewed in this appendix is distinct from the formula used to compute the Hall--Littlewood index reviewed in Appendix \ref{app:HLeqs}.} In favorable cases in which the central gauge node of the star-shaped quiver is not bad, one can then glue these building blocks together using the monopole formula for this node. Moreover, the Hall--Littlewood formula allows us to refine the Hilbert series of the $T_\rho(G)$ theories with fugacities for the enhanced topological symmetry, when such exists.

The general formula for the Coulomb branch Hilbert series of a $T_\rho(G^\vee)$ theory is
\begin{align}\label{eq:HSHLformula}
    \operatorname{HS}[T_\rho(G^\vee)](\tau;\mathbf{x};\mathbf{m})=\tau^{\delta_{G^\vee}(\mathbf{m})}(1-\tau^2)^{\operatorname{rk}(G)}K_\rho^G(\tau;\mathbf{x})\Psi_G(\tau;\mathbf{a}(\tau,\mathbf{x});\mathbf{m}) \,.
\end{align}
Many of the ingredients are the same we saw in Appendix \ref{app:HLeqs} for the definition of the Hall--Littlewood index of a class $\mathcal{S}$ theory, so here we will only focus on the new features. $\mathbf{m}$ are the background magnetic fluxes for the flavor symmetry $G^\vee$ of the $T_\rho(G^\vee)$ theory. This will constitute, after gauging, the central node of the star-shaped quiver, so we will sum over them as in the monopole formula in equation \eqref{eq:monformula} when doing such a gauging. The exponent $\delta_{G^\vee}(\mathbf{m})$ is defined as
\begin{align}
\delta_{G^\vee}(\mathbf{m})=\sum_{\mathbf{\alpha}\in\Delta_+(G^\vee)}\mathbf{\alpha}(\mathbf{n}) \,.
\end{align}
For the groups that are relevant to us, it takes the following values
\begin{subequations}
\begin{align}
&\delta_{U(N)}(\mathbf{m})=\sum_{j=1}^N(N+1-2j)m_j \,,  \\ &\delta_{SO(2N)}(\mathbf{m})=\sum_{j=1}^N(2N-2j)m_j\,, \\ &\delta_{SO(2N+1)}(\mathbf{m})=\sum_{j=1}^N(2N+1-2j)m_j\,, \\ &\delta_{USp(2N)}(\mathbf{m})=\sum_{j=1}^N(2N+2-2j)m_j\,.
\end{align}
\end{subequations}
Finally, $K_\rho^G(\tau;\mathbf{x})$ is the same factor that we defined in equation \eqref{eqn:appK} and $\Psi_G(\tau;\mathbf{a}(\tau,\mathbf{x});\mathbf{m})$ is the un-normalized Hall--Littlewood polynomial associated with the group $G$ that we define in equation \eqref{eqn:appHL}.

To illustrate the application of this Hall--Littlewood formula, we consider an explicit example from Section \ref{sec:Aeven}. For the case of the class $\mathcal{S}$ theory of type $A_2$ on a sphere with punctures $2\times[1^3]+2\times[1^2]_t$ and whose 3d mirror is depicted in Figure \ref{fig:ex2b}, we first consider the Coulomb branch Hilbert series of the $T(SU(3))$ and the $T(USp'(2))$ theories
\begin{align}
\begin{split}
    &\operatorname{HS}[T(SU(3))](\tau;\mathbf{x};\mathbf{m})=\tau^{2m_1-2m_3}(1-\tau^2)^2\operatorname{PE}\left[\chi_{\bf 8}^{SU(3)}(\mathbf{x})\tau^2\right]\Psi_{SU(3)}(\tau;\mathbf{x};\mathbf{m}) \,,\\
    &\operatorname{HS}[T_{[1^2]}(USp'(2))](\tau;v;n)=\tau^{2n}(1-\tau^2)\operatorname{PE}\left[\chi_{\bf 3}^{SU(2)}(v)\tau^2\right]\Psi_{SU(2)}(\tau;v;n)\,.
\end{split}\label{eq:HSTSU3TUSp2}
\end{align}
The former depends on the fugacities $\mathbf{x}$ for its $SU(3)$ topological symmetry and on the background magnetic fluxes $\mathbf{m}$ for its flavor symmetry. The latter depends on the fugacity $y$ for its $SU(2)$ topological symmetry and on the background magnetic flux $n$ for its flavor symmetry, and for $n=0$ it coincides with the Hilbert series for $\mathbb{C}^2/\mathbb{Z}_2$:
\begin{align}\label{eq:TuspC2Z2}
    \operatorname{HS}[T_{[1^2]}(USp'(2))](\tau;v;n=0)=\frac{1-\tau^4}{(1-\tau^2)(1-v^{\pm1}\tau^2)} \,.
\end{align}

In order to get the Coulomb branch Hilbert series of the quiver in Figure \ref{fig:ex2b} we then have to multiply two copies of the Hilbert series of $T(SU(3))$ and two copies of the Hilbert series of $T_{[1^2]}(USp'(2))$ and sum over the magnetic fluxes of the $SU(2)$ central gauge node as in the monopole formula in equation \eqref{eq:monformula}. We find
\begin{align}
\begin{aligned}
    \operatorname{HS}&[\text{Figure}\ref{fig:ex2b}](\tau;\mathbf{x};\mathbf{y};v;w) \\
    =\,&\sum_{m=0}^\infty\tau^{-4m}P_{SU(2)}(\tau;m)\operatorname{HS}[T(SU(3))](\tau;\mathbf{x};m,0,-m)\operatorname{HS}[T(SU(3))](\tau;\mathbf{y};m,0,-m) \\
    &\qquad\qquad \times \operatorname{HS}[T_{[1^2]}(USp'(2))](\tau;v;m)\operatorname{HS}[T_{[1^2]}(USp'(2))](\tau;w;m) \,,
\end{aligned}
\end{align}
where we recall that we are gauging an $SU(2)$ subgroup of the $SU(3)$ flavor symmetries and the dressing factor for $SU(2)$ is given by
\begin{align}
    P_{SU(2)}(\tau;m)=\begin{cases}(1-\tau^2)^{-1} & m\neq 0 \\ (1-\tau^4)^{-1} & m=0 \,. \end{cases}
\end{align}
Putting everything together, the Coulomb branch Hilbert series is
\begin{align}
\scalemath{0.97}{
\begin{aligned}
    \text{H} & \text{S}(\tau,a,b,c,d)
    =1+\tau^2(\textbf{adj})+\tau^4(\textbf{adj}^2+\chi_{\bf{8}}(a)+\chi_{\bf{8}}(b)+1)+\tau^6\Big(\textbf{adj}^3+\chi_{\bf{27}}(a)\\
    &\quad +\chi_{\bf{10}}(a)+\chi_{\overline{\bf{10}}}(a)+\chi_{\bf{8}}(a) +\chi_{\overline{\bf{10}}}(b)+\chi_{\bf{8}}(b)+\chi_{\bf{27}}(b)+\chi_{\bf{10}}(b)+\chi_{\bf{3}}(c)+\chi_{\bf{3}}(d) \\
    &\quad
    +2\chi_{\bf{8}}(a)\chi_{\bf{8}}(b)+\chi_{\bf{8}}(a)\chi_{\bf{3}}(c)+\chi_{\bf{8}}(a)\chi_{\bf{3}}(d)+\chi_{\bf{8}}(b)\chi_{\bf{3}}(c)+\chi_{\bf{8}}(b)\chi_{\bf{3}}(d)\Big)+\mathcal{O}\left(\tau^8\right)\,.
\end{aligned}}\label{eq:2max2maxtHS}
\end{align}
As explained in the main text, this result coincides with the Hall--Littlewood index of the same class $\mathcal{S}$ theory given in equation \eqref{eq:2max2maxt}, as expected since this theory has only one twist line. We recall that $\textbf{adj}^n$ was defined in equation \eqref{eq:adjchar}, so from the term of order $\tau^2$ we can see that the global symmetry is $SU(3)^2 \times SU(2)^2$, as expected from the class $\mathcal{S}$ picture.

It is instructive to see how this $SU(3)^2 \times SU(2)^2$ symmetry arises from the 3d mirror in Figure \ref{fig:ex2}. 
Each of the legs of the star-shaped quiver in Figure \ref{fig:ex2b}, each of which is associated with one of the punctures of the class $\mathcal{S}$ theory, carries a global symmetry that is determined by the corresponding partition with the usual rule. For example, the puncture $[1^3]$ is associated with the $T(SU(3))$ theory, which possesses an $SU(3)$ global symmetry on its Coulomb branch. This arises at low energies from an enhancement of the manifest $U(1)^2$ topological symmetry due to the fact that each unitary gauge node is balanced, that is, it sees a number of flavors which is twice the number of colors \cite{Gaiotto:2008ak}. Similarly, each $T_{[1^2]}(USp'(2))$ theory carries a $SU(2)$ global symmetry on its Coulomb branch. Again this is enhanced at low energies, since the manifest topological symmetry for an $SO(3)$ gauge group is only $\mathbb{Z}_2$. This enhancement is more difficult to see than in the previous case since the theory is bad, and as such it is harder to identify the monopole operators that provide the required extra moment maps. Nevertheless, as discussed in \cite{Cremonesi:2014uva}, the $T_{[1^2]}(USp'(2))$ is self-mirror and both its Higgs and Coulomb branches are isomorphic to $\mathbb{C}^2/\mathbb{Z}_2$, and thus they both carry an $SU(2)$ symmetry. This can be seen from the Hilbert series in equation \eqref{eq:2max2maxtHS}, which remarkably can be computed incorporating the fugacity for the $SU(2)$ using the techniques of \cite{Cremonesi:2014kwa,Cremonesi:2014vla,Cremonesi:2014uva}.

\section{The 3d mirrors of class \texorpdfstring{\boldmath{$\mathcal{S}$}}{S}: a review}\label{app:3dmirror}

To understand the Higgs branch of a class $\mathcal{S}$ theory, it is often useful to understand the 3d $\mathcal{N}=4$ mirror of the 4d $\mathcal{N}=2$ SCFT. This is obtained first by performing the circle-reduction of the 4d theory, and then taking the mirror that swaps the Coulomb and Higgs branches between the 3d reduction and the 3d mirror. When the 3d mirror is a Lagrangian theory, it is straightforward to determine the Coulomb branch Hilbert series, which is identical to the Higgs branch Hilbert series of the original class $\mathcal{S}$ theory. The process by which the 3d mirror is determined from the data of the class $\mathcal{S}$ theory, to wit, the punctured Riemann surface, was put forth in \cite{Benini:2010uu}, and has been further extended to the case of twisted $A_\text{even}$ theories in \cite{Beratto:2020wmn}. The exposition in this appendix largely follows these references.

As we have discussed, any $n$-punctured genus-$g$ Riemann surface admits a pair-of-pants decomposition into three-punctured spheres, glued together along the punctures. In a similar manner, the construction of the 3d mirror is a two-step process. First, it is necessary to determine how to construct the 3d mirror for each three-punctured sphere, and then it is necessary to understand how each individual 3d mirror is connected together via the pair-of-pants decomposition to form the 3d mirror of any arbitrary class $\mathcal{S}$ theory.

Consider a class $\mathcal{S}$ theory of type $J$. In Appendix \ref{app:HLeqs}, we reviewed that, for an outer-automorphism $o$ of $J$, each $o$-twisted puncture is associated to a nilpotent orbit of $G$, where $G$ is the Langlands dual of the invariant subalgebra of $J$ under the action of $o$.\footnote{In this appendix, we consider only untwisted and $\mathbb{Z}_2$-twisted punctures.}$^{,}$\footnote{We abuse notation slightly and use $o$ to refer to both the outer-automorphism group, and the action of the outer-automorphism group on $J$.} When $J$, and thus $G$, is a classical Lie algebra then each nilpotent orbit can be captured by a particular integer partition, subject to certain constraints.\footnote{We note that there is not quite a one-to-one correspondence between nilpotent orbits and partitions; some partitions correspond to more than one nilpotent orbit, and some further labelling of such partitions is thus necessary, see \cite{MR1251060} for details. For the ease of this appendix, we suppress this point.} We summarise the tuples of $J$, $o$, $G$, and the data describing the puncture for the class $\mathcal{S}$ theories relevant to this paper in Table \ref{tbl:part}.

\begin{table}[H]
    \centering
    \renewcommand{\arraystretch}{1.2}
    \begin{threeparttable}
    \begin{tabular}{ccccc}
        \toprule
        $J$ & $o$ & $G$ & Puncture data & Mirror of maximal puncture \\\midrule
        \multirow{2}{*}{$A_{2n}$} & --- & $A_{2n}$ & Partition of $2n+1$ & $T(SU(2n+1))$ \\
         & $\mathbb{Z}_2$ & $C^\prime_n$ & C-partition of $2n$ & $T(USp'(2n))$ \\
        \multirow{2}{*}{$A_{2n-1}$} & --- & $A_{2n-1}$ & Partition of $2n$ & $T(SU(2n))$ \\
         & $\mathbb{Z}_2$ & $B_n$ & B-partition of $2n+1$ & $T(USp(2n))$  \\
         \multirow{2}{*}{$D_{n}$} & --- & $D_{n}$ & D-partition of $2n$ & $T(SO(2n))$ \\
         & $\mathbb{Z}_2$ & $C_{n-1}$ & C-partition of $2n-2$ & $T(SO(2n-1))$ \\
         \bottomrule
    \end{tabular}
    \end{threeparttable}
    \caption{In this table, we write $J$, $o$, $G$, the data describing the punctures, and the $T(G^\vee)$ 3d $\mathcal{N}=4$ theory associated to the maximal $o$-twisted puncture. A B/D-partition is a partition such that every even element of the partition appears with even multiplicity; a C-partition is a partition such that every odd element of the partition appears with even multiplicity.}
    \label{tbl:part}
\end{table}

There exists a partial ordering on the set of nilpotent orbits of $G$, and there is a unique maximal element under this partial ordering; this corresponds to the trivial embedding $SU(2) \rightarrow G$. The puncture associated to this maximal nilpotent orbit is known as the \emph{maximal puncture}, and it carries flavor symmetry $G$. To the maximal puncture is associated the 3d $\mathcal{N}=4$ SCFT known as $T(G^\vee)$ \cite{Benini:2010uu,Beratto:2020wmn}. The $T(G^\vee)$ theories, explored in \cite{Gaiotto:2008ak} (see also \cite{Feng:2000eq}) and many subsequent works, have mirrors $T(G)$, and thus when $G$ is simply-laced they are self-mirror.\footnote{We emphasize that the special case of $G = USp'(2n)$ is self-dual: $USp'(2n)^\vee = USp'(2n)$. All other choices of $G$ obey the normal rules of group theory.} In particular, they have global symmetry $G^\vee \times G$, where the first factor is the global symmetry on the Higgs branch and the latter is the global symmetry on the Coulomb branch.

Partial closure of the maximal puncture corresponds to giving a nilpotent vacuum expectation value, inside the nilpotent orbit associated to the partially-closed puncture, to the moment map operator associated to the $G$ flavor symmetry. Similarly, the 3d $\mathcal{N}=4$ SCFT that is relevant for the Higgsing of the flavor symmetry of the maximal puncture via a nilpotent orbit $\rho$ involves Higgsing the Coulomb branch flavor symmetry by the same choice of nilpotent orbit. The resulting theories are known as $T_\rho(G^\vee)$; these are the principle building blocks of the 3d mirrors of the class $\mathcal{S}$ theories obtained from three-punctured spheres.

As $\mathbb{Z}_2$-twisted punctures come in pairs, there are two distinct kinds of three-punctured spheres that we must consider: those with three untwisted punctures, and those with two twisted punctures and one untwisted puncture. In the former case, the 3d mirror is constructed by gauging together the Higgs branch flavor symmetry $J$ of the three $T_{\rho_i}(J)$ theories associated to each puncture. The situation is slightly more complicated when there are twisted punctures. In that case, we gauge the diagonal $G^\vee$ subgroup of the $G^\vee \times G^\vee \times J$ Higgs branch flavor symmetry of the product theory 
\begin{align}
    T_{\rho_1}(G^\vee) \times T_{\rho_2}(G^\vee) \times T_{\rho_3}(J) \,.
\end{align} 
Then, we must consider the branching
\begin{equation}
    J \rightarrow G^\vee \times \widetilde{G} \,,
\end{equation}
where the commutant $\widetilde{G}$ is a Higgs branch flavor symmetry that persists in the 3d mirror.

\begin{figure}[H]
    \centering
    \begin{tikzpicture}[scale=1.2,every node/.style={scale=1.2},font=\scriptsize]
        \draw (0,2)--(0,0);
        \draw (2,2)--(2,0);
        \draw (4,2)--(4,0);
        \node (C) at (6.6,1.09) {$\cdot$};
        \node (C) at (6.4,1.06) {$\cdot$};
        \node (C) at (6.2,1.04) {$\cdot$};
        \draw (8,2)--(8,0);
        \draw (10,2)--(10,0);
        \node (NS1) at (0,2.2) {NS5};
        \node (NS2) at (2,2.2) {NS5};
        \node (NS3) at (4,2.2) {NS5};
        \node (NS4) at (6,2.2) {\large{$\cdots$}};
        \node (NS5) at (8,2.2) {NS5};
        \node (NS6) at (10,2.2) {NS5};
        \draw (0,.4)--(12,.4);
        \node (D1) at (12.28,0.4) {\tiny $\otimes$ D5};
        \node (D2) at (12.28,0.6) {\tiny $\otimes$ D5};
        \node (D3) at (12.28,0.8) {\tiny $\otimes$ D5};
        \node (D3) at (12.28,1.3) {\tiny $\otimes$ D5};
        \node (D3) at (12.28,1.5) {\tiny $\otimes$ D5};
        \node (N1) at (1,.6) {$N_1$ D3};
        \draw (2,.6)--(12,.6);
        \node (N2) at (3,.8) {$N_2$ D3};
        \draw (4,.8)--(12,.8);
        \node (N3) at (4.8,1) {$N_3$ D3};
        \draw (8,1.3)--(12,1.3);
        \draw (10,1.5)--(12,1.5);
        % \draw (10,1.5)--(12,1.5);
        \node (N3) at (9,1.15) {$\vdots$};
        \node (N4) at (11,1.15) {$\vdots$};
        \node (N5) at (11,1.8) {$n$ D3};
        \node (N1) at (9,1.6) {$N_{\ell -1}$ D3};
        % \node (N5) at (11,1.8) {$2n+1$ D3};
    \end{tikzpicture}
    \caption{The D3-D5-NS5 brane system in Type IIB description the 3d $\mathcal{N}=4$ SCFT known as $T_\rho(SU(n))$. The number of D3-branes in each interval is fixed by $\rho$ following equation \eqref{eqn:NjU}.}
    \label{fig:ellbranes}
\end{figure}
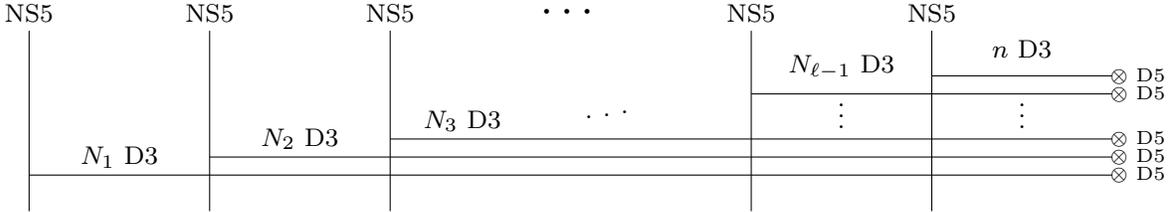

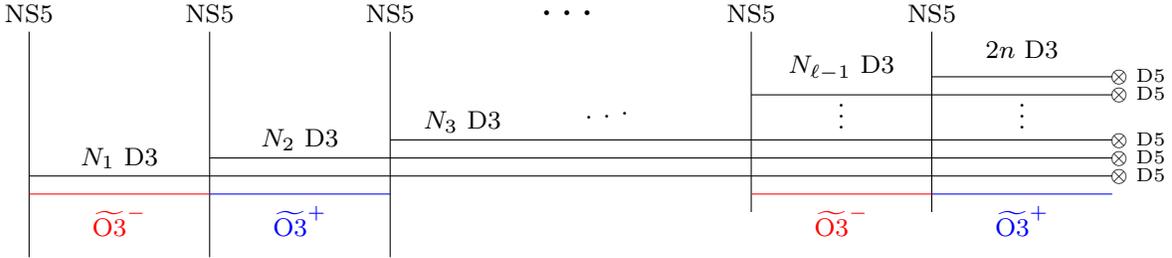
\begin{figure}[H]
    \centering
    \begin{tikzpicture}[scale=1.2,every node/.style={scale=1.2},font=\scriptsize]
        \draw (0,2)--(0,-0.5);
        \draw (2,2)--(2,-0.5);
        \draw (4,2)--(4,-0.5);
        \node (C) at (6.6,1.09) {$\cdot$};
        \node (C) at (6.4,1.06) {$\cdot$};
        \node (C) at (6.2,1.04) {$\cdot$};
        \draw (8,2)--(8,0);
        \draw (10,2)--(10,0);
        \node (NS1) at (0,2.2) {NS5};
        \node (NS2) at (2,2.2) {NS5};
        \node (NS3) at (4,2.2) {NS5};
        \node (NS4) at (6,2.2) {\large{$\cdots$}};
        \node (NS5) at (8,2.2) {NS5};
        \node (NS6) at (10,2.2) {NS5};
        \draw (0,.4)--(12,.4);
        \node (D1) at (12.28,0.4) {\tiny $\otimes$ D5};
        \node (D2) at (12.28,0.6) {\tiny $\otimes$ D5};
        \node (D3) at (12.28,0.8) {\tiny $\otimes$ D5};
        \node (D3) at (12.28,1.3) {\tiny $\otimes$ D5};
        \node (D3) at (12.28,1.5) {\tiny $\otimes$ D5};
        \node (N1) at (1,.6) {$N_1$ D3};
        \draw (2,.6)--(12,.6);
        \node (N2) at (3,.8) {$N_2$ D3};
        \draw (4,.8)--(12,.8);
        \node (N3) at (4.8,1) {$N_3$ D3};
        \draw (8,1.3)--(12,1.3);
        \draw (10,1.5)--(12,1.5);
        \node (N3) at (9,1.15) {$\vdots$};
        \node (N4) at (11,1.15) {$\vdots$};
        \node (N1) at (9,1.6) {$N_{\ell -1}$ D3};
        \node (N5) at (11,1.8) {$2n$ D3};
        \draw[red] (0,0.2)--(2,0.2);
        \node[red] (N1) at (1,-0.1) {$\widetilde{\text{O3}}^-$};
        \draw[blue] (2,0.2)--(4,0.2);
        \node[blue] (N1) at (3,-0.1) {$\widetilde{\text{O3}}^+$};
        \draw[red] (8,0.2)--(10,0.2);
        \node[red] (N1) at (9,-0.1) {$\widetilde{\text{O3}}^-$};
        \draw[blue] (10,0.2)--(12,0.2);
        \node[blue] (N1) at (11,-0.1) {$\widetilde{\text{O3}}^+$};
    \end{tikzpicture}
    \caption{The D3-D5-NS5 brane system in Type IIB description the 3d $\mathcal{N}=4$ SCFT known as $T_\rho(USp^\prime(2n))$. The number of D3-branes in each interval is fixed by $\rho$. Here all numbers of branes are counted in the covering space.}
    \label{fig:ellbranesUSp}
\end{figure}

When $J$ is a classical Lie algebra, the theories $T_\rho(G^\vee)$ are, in fact, Lagrangian quivers. For each of the theories $T(G^\vee)$ appearing in Table \ref{tbl:part}, the linear quiver describing $T_\rho(G^\vee)$ is given in \cite{Cremonesi:2014uva}; while we do not repeat the general case here, we give one example of the $T_\rho(SU(n))$ theory. The quiver is
\begin{equation}
    T_{\rho}(SU(n)) : \quad \begin{gathered}\begin{tikzpicture}[scale=1.2,every node/.style={scale=1.2},font=\scriptsize]
    \node[gauge] (t0) at (0,0) {$N_1$};
    \node[gauge] (t1) at (1.5,0) {$N_2$};
    \node (t2) at (3,0) {$\cdots$};
    \node[gauge] (t3) at (4.5,0) {$N_{\ell-1}$};
    \node[flavorD] (t4) at (6,0) {$n$};
    \draw (t0)--(t1)--(t2)--(t3)--(t4);
\end{tikzpicture}\end{gathered} \quad,
\end{equation}
where the ranks of the $U(N_j)$ gauge factors, and the length $\ell$, are fixed in terms of the partition $\rho$. Specifically, if we consider the partition $\rho = [\rho_1, \cdots, \rho_{\ell}]$, where we see that the length $\ell$ is defined as the number of elements in the partition, then
\begin{equation}\label{eqn:NjU}
    N_{j} \,= \sum_{k=\ell-j+1}^\ell\rho_k \,.
\end{equation}
When $G^\vee$ is of type BCD, then $T_\rho(G^\vee)$ has a similar description as a linear quiver of alternating orthogonal and symplectic algebras.\footnote{In \cite{Beratto:2020wmn}, it was argued that in the $T_\rho(USp'(2n))$ theories the global structure of the orthogonal groups should be that of $SO(2N_i)$ in order to reproduce various properties expected for the dimensional reduction of the twisted $A_{2n}$ theories. In all the drawings of this paper we make the same assumption, though our analysis is insensitive to this issue. Indeed, for the 3d mirrors we mainly compute dimensions of branches of the moduli space, which do not depend on discrete gaugings, and the Coulomb branch Hilbert series of the 3d mirrors using the Hall--Littlewood formula of \cite{Cremonesi:2014kwa,Cremonesi:2014vla,Cremonesi:2014uva}, which is blind to the precise global structure of the gauge groups of the $T_\rho^\sigma(G)$ theories.} Each of these 3d $\mathcal{N}=4$ theories can also be realized via a D3-D5-NS5 brane configuration in Type IIB, which is schematically depicted for $T_\rho(SU(n))$ in Figure \ref{fig:ellbranes} and $T_\rho(USp'(2n))$ in Figure \ref{fig:ellbranesUSp}. We emphasize that the flavor symmetry of $T_\rho(SU(2n+1))$ is $SU(2n+1)$, not the naive $U(2n+1)$ that one might expect from Figure \ref{fig:ellbranes}, as the center of the putative $U(2n+1)$ acts trivially \cite{Gaiotto:2008ak}.

To illustrate the procedure, we now determine the 3d mirror for a particular three-punctured sphere which appears in the main text of this paper. Consider the class $\mathcal{S}$ theory of type $A_{2n}$ with punctures $2\times [2n]_t + [1^{2n+1}]$. The Lagrangian $T_\rho(G^\vee)$ quivers associated to each of the two different types of punctures appearing in this tinkertoy are
\begin{equation}
    \begin{aligned}
      &T_{[1^{2n+1}]}(SU(2n+1)) & &: \quad \begin{gathered}\begin{tikzpicture}[scale=1.2,every node/.style={scale=1.2},font=\scriptsize]
    \node[gauge] (t0) at (0,0) {$1$};
    \node[gauge] (t1) at (1.5,0) {$2$};
    \node (t2) at (3,0) {$\cdots$};
    \node[gauge] (t3) at (4.5,0) {$2n$};
    \node[flavorD] (t4) at (6,0) {$2n+1$};
    \draw (t0)--(t1)--(t2)--(t3)--(t4);
\end{tikzpicture}\end{gathered} \quad, \\
    &T_{[2n]}(USp'(2n)) & &: \quad \begin{gathered}\begin{tikzpicture}[scale=1.2,every node/.style={scale=1.2},font=\scriptsize]
    \node[gauge, red] (t0) at (0,0) {$1$};
    \node[flavor, blue] (t4) at (1.5,0) {$2n$};
    \draw (t0)--(t4);
\end{tikzpicture}\end{gathered} \quad.
    \end{aligned}
\end{equation}
Note that the $SO(1)$ gauge node appearing in $T_{[2n]}(USp'(2n))$ simply specifies that there exists a half-hypermultiplet in the fundamental representation of the $USp(2n)$ flavor algebra. The 3d mirror of the class $\mathcal{S}$ theory is then obtained by gauging the common $USp(2n)$ flavor symmetry of each of the $T_{[2n]}(USp'(2n))$ theories associated to the twisted punctures and the $T_{[1^{2n+1}]}(SU(2n+1))$ theory associated to the untwisted puncture. In the latter case, as the $T_{\rho}(SU(2n+1))$ theory has an $SU(2n+1)$ flavor symmetry, we gauge the $USp(2n)$ subgroup under the branching
\begin{equation}
    SU(2n+1) \rightarrow USp(2n) \times U(1) \,.
\end{equation}
As such, after gauging there is a leftover fundamental hypermultiplet, rotated by the commutant $U(1) = SO(2)$, attached to the $U(2n)$ gauge node that was adjacent to the $SU(2n+1)$ flavor node of the $T_{\rho}(SU(2n+1))$ theory. Specifically, the 3d mirror of type $A_{2n}$ with punctures $2\times[2n]_t + [1^{2n+1}]$ is
\begin{equation}
    \begin{gathered}
    \begin{tikzpicture}[scale=1.2,every node/.style={scale=1.2},font=\scriptsize]
    \node[gauge, blue] (cL) at (0,0) {$2n$};
    \node[flavor,red] (r1) at (0,1) {$2$};
    \node[gauge] (n1) at (1,0) {$2n$};
    \node[gauge] (n2) at (2.3,0) {$2n-1$};
    \node[empty] (n3) at (3.6,0) {$\cdots$};
    \node[gauge] (n4) at (4.6,0) {$2$};
    \node[gauge] (n5) at (5.5,0) {$1$};
    \node[flavor,red] (r2) at (1,-1) {$2$};
    \draw (cL)--(n1)--(n2)--(n3)--(n4)--(n5);
    \draw (cL)--(r1);
    \draw (n1)--(r2);
    \end{tikzpicture}
    \end{gathered}  \quad.
\end{equation}
It is similarly straightforward to determine the explicit quivers describing the 3d mirrors of any three-punctured sphere.

To determine the 3d mirror for a class $\mathcal{S}$ theory arising from an arbitrary $n$-punctured sphere, we proceed as follows.\footnote{We do not include the extension to higher genus Riemann surfaces in this appendix. Typically this involves adding extra adjoint-valued hypermultiplets, cf.~\cite{Benini:2010uu} for more details.} For an $n$-punctured sphere we can always consider an S-duality frame where the pair-of-pants decomposition is in terms of three-punctured spheres glued along \emph{maximal} punctures. These maximal punctures can be either untwisted or twisted maximal punctures. To determine the general 3d mirror we proceed inductively: consider gauging together an $(n-1)$-punctured sphere with a maximal puncture, and a three-punctured sphere with a maximal puncture, along the maximal punctures. Similarly to the three-punctured case just discussed, we assume that the 3d mirror of the $(n-1)$-punctured sphere also contains a copy of the $T(G^\vee)$ theory where part of the $G^\vee$ Higgs branch symmetry is gauged. On each sphere the $T(G^\vee)$ has an unbroken Coulomb branch symmetry, which is $G$. The 3d mirror of the $n$-punctured sphere is then obtained by gauging the diagonal Coulomb symmetry $G$ from the two maximal punctures which we are gluing. This involves introducing a twisted vector multiplet, rather than the more standard vector multiplet that was used to gauge the Higgs symmetries when constructing three-punctured spheres.\footnote{Note: despite the similarities in naming, twisted vector multiplets have no relation to twisted punctures.}

It remains to determine whether a Lagrangian quiver description of this 3d $\mathcal{N}=4$  theory exists after Coulomb-gauging, and what it consists of. The simplest case, which was discussed in detail in \cite{Benini:2010uu}, is when all of the $n$ punctures are untwisted. In this case, the Coulomb gauging simply collapses the quiver tails associated to the $T(J)$ theories, and one ends up with a star-shaped quiver where the central node is a gauge group $J$, obtained by gauging the common Higgs symmetry $J$ of the $T_{\rho_i}(J)$ theories associated to each of the $n$ punctures. This is due to the fact that the gauging of two $T(J)$ theories gives a theory with a quantum deformed moduli space of vacua where the $J\times J$ symmetry is spontaneously broken to its diagonal subgroup \cite{Benvenuti:2011ga,Nishioka:2011dq,Bottini:2021vms}.\footnote{This result translates at the level of the three-sphere partition function into an identity for the partition function of the gluing of two $T(J)$ theories, which is proportional to a delta-function identifying the mass parameters of the two $J$ symmetries. This can be shown for $J=SU(N)$ in the case of the $S^3$ partition function using the results of \cite{Benvenuti:2011ga,Nishioka:2011dq} and it was later extended in \cite{Bottini:2021vms} to the $S^3_b$ partition function, as well as to the $S^3\times S^1$ index of the 4d $\mathcal{N}=1$ parent of the $T(SU(N))$ theory called $E(USp(2N))$ \cite{Pasquetti:2019hxf,Hwang:2020wpd}. This is similar to what happens for the 4d $\mathcal{N}=1$ $SU(2)$ theory with four chiral multiplets which has a quantum deformed moduli space of vacua \cite{Seiberg:1994bz} and whose index is proportional to a delta-function as shown in \cite{Spiridonov:2014cxa}.} This theory has also additional Nambu--Goldstone modes in the adjoint representation of $J$. Since in the situation that we are considering of gluing two star-shaped quivers the $J\times J$ symmetry is gauged, this means that it is Higgsed to its diagonal subgroup and one combination of the two vectors multiplets in the adjoint representation of $J$ becomes massive by eating the aforementioned Nambu--Goldstone modes.

When twisted punctures are involved, the central node of the star-shaped quiver is $G^\vee$ instead of $J$, and there can be additional matter fields attached to the $G^\vee$ node from the Coulomb gauging. These arise because the gauging of two $T(J)$ theories still gives Nambu--Goldstone modes in the adjoint representation of $J$, but of these only those that are in the adjoint representation of the Higgsed gauge group $G^\vee$ get eaten, while the others remain as massless fields. For class $\mathcal{S}$ of type $D$ with $\mathbb{Z}_2$-twisted punctures, these additional multiplets have been observed to be vector-valued hypermultiplets of $G^\vee = SO(2n-1)$ in \cite{Benini:2010uu}.

In Section \ref{sec:3dmirror}, we conjecture, when $J = A_{2n}$, how the Coulomb gauging works when we gauge Coulomb branch symmetries of $T_\rho(SU(2n+1))$ associated to gluing a pair of untwisted punctures belonging to two distinct three-punctured spheres, both with two $\mathbb{Z}_2$-twisted punctures and one untwisted puncture. This then leads directly to Conjecture \ref{conj:3dmirror}, for the 3d mirror of a class $\mathcal{S}$ theory of type $A_\text{even}$ with arbitrary twisted and untwisted punctures.

\section{HL indices and HB Hilbert series for twisted \texorpdfstring{\boldmath{$A_2$}}{A2}}\label{app:summaryHLHS}

In this appendix, we provide a short summary table of the different Hall--Littlewood indices and Higgs branch Hilbert series of the class $\mathcal{S}$ theories of type $A_2$.
These quantities were computed throughout Section \ref{sec:Aeven}, and they are repeated here in Table \ref{tab:HLHSforA2} for convenience.

\begin{table}[H]
\centering
\renewcommand{\arraystretch}{1.3}
\resizebox{\textwidth}{!}{
    $\begin{array}{cccc}
        \toprule
        k & \text{Punctures} & \text{HL index} & \text{HS of the HB}  \\\midrule
        \multirow{14}{*}{2} & 2\times[1^3]+2\times[1^2]_t & \begin{array}{c}
            1+\tau^2(\textbf{adj})+\tau^4(\textbf{adj}^2+\chi_{\bf{8}}(a)+\chi_{\bf{8}}(b)+1)\\
            +\tau^6(\textbf{adj}^3+\chi_{\bf{27}}(a)+\chi_{\bf{10}}(a)+\chi_{\overline{\bf{10}}}(a)+\chi_{\bf{8}}(a)\\
            +\chi_{\bf{27}}(b)+\chi_{\bf{10}}(b)+\chi_{\overline{\bf{10}}}(b)+\chi_{\bf{8}}(b)+\chi_{\bf{3}}(c)+\chi_{\bf{3}}(d)\\
            +2\chi_{\bf{8}}(a)\chi_{\bf{8}}(b)+\chi_{\bf{8}}(a)\chi_{\bf{3}}(c)+\chi_{\bf{8}}(a)\chi_{\bf{3}}(d)\\
            +\chi_{\bf{8}}(b)\chi_{\bf{3}}(c)+\chi_{\bf{8}}(b)\chi_{\bf{3}}(d))+\mathcal{O}\left(\tau^8\right)
        \end{array} & \text{HL}=\text{HS} \\\cline{2-4}
         & 2\times[1^3] + 2\times[2]_t & \begin{array}{c}
            1+\tau^2(\textbf{adj})+\tau^4(\textbf{adj}^2+\textbf{adj}+1)+\tau^6(\textbf{adj}^3\\
            +\textbf{adj}^2+2\chi_{\bf{8}}(a)\chi_{\bf{8}}(b)+\textbf{adj}+\chi_{\bf{10}}(a)\\
            +\chi_{\bf{\overline{10}}}(a)+\chi_{\bf{10}}(b)+\chi_{\bf{\overline{10}}}(b))+\mathcal{O}(\tau^8)
        \end{array} & \text{HL}=\text{HS} \\\cline{2-4}\\[-12pt]
         & 4\times[2]_t & \dfrac{1+\tau^4-\tau^6}{1-\tau^4} & \dfrac{1-\tau^2+\tau^4}{(1-\tau^2)(1-\tau^4)}\\[-12pt] \\\cline{2-4}
         & 1\times[1^2]_t+3\times[2]_t & \begin{array}{c}
            1+\tau^2(\textbf{adj})+\tau^4(\textbf{adj}^2+1)+\tau^5\chi_{2}(a)\\
            +\tau^6(\textbf{adj}^3+\textbf{adj}-1)+\mathcal{O}\left(\tau^7\right)\end{array} &  \begin{array}{c}
           \dfrac{1-\tau^{12}}{(1-\tau^2)(1-a^{\pm2}\tau^2)(1-a^{\pm1}\tau^5)}\end{array} \\\cline{2-4}
         & 2\times[1^2]_t+2\times[2]_t & \begin{array}{c}
            1+\tau^2(\textbf{adj})+\tau^4(\textbf{adj}^2+1)+\tau^6(\textbf{adj}^3+\\
            \textbf{adj}+\chi_{2}(a)\chi_{2}(b)-1)+\mathcal{O}\left(\tau^8\right) 
        \end{array} &  \begin{array}{c}
            1+\tau^2(\textbf{adj})+\tau^4(\textbf{adj}^2+1)+\tau^6(\textbf{adj}^3+\\
            \textbf{adj}+\chi_{2}(a)\chi_{2}(b))+\mathcal{O}\left(\tau^8\right)
        \end{array}\\\cline{2-4}
         & \begin{array}{c}
            3\times[1^2]_t+1\times[2]_t
         \end{array} & \begin{array}{c}
            1+\tau^2(\textbf{adj})+\tau^4(\textbf{adj}^2+1)\\
            +\tau^6(\textbf{adj}^3+\textbf{adj}-1)+\mathcal{O}\left(\tau^7\right) 
        \end{array} & \begin{array}{c}
           1+\tau^2(\textbf{adj})+\tau^4(\textbf{adj}^2+1)\\
            +\tau^6(\textbf{adj}^3+\textbf{adj})+\mathcal{O}\left(\tau^7\right)
        \end{array} \\\cline{2-4}
         & \begin{array}{c}
            4\times[1^2]_t
         \end{array} & \begin{array}{c}
            1+\tau^2(\textbf{adj})+\tau^4(\textbf{adj}^2+1)\\
            +\tau^6(\textbf{adj}^3+\textbf{adj}-1)+\mathcal{O}\left(\tau^8\right) 
        \end{array} & \begin{array}{c}
            1+\tau^2(\textbf{adj})+\tau^4(\textbf{adj}^2+1)\\
            +\tau^6(\textbf{adj}^3+\textbf{adj})+\mathcal{O}\left(\tau^8\right)
        \end{array} \\\midrule\\[-12pt]
        \multirow{3}{*}{$\geq 3$} & 2k\times[2]_t & \dfrac{\tau^{6k-8}(1-\tau^2)^{k-2}}{1-\tau^{6k-8}}+\dfrac{(1-\tau^6)^{k-1}}{1-\tau^4} & \dfrac{1+\tau^{6k-6}}{(1-\tau^4)(1-\tau^{6k-8})} \\[-12pt] \\\cline{2-4}
         & n\hspace{-2pt}\times\hspace{-2pt}[1^2]_t+(2k\hspace{-2pt}-\hspace{-2pt}n)\hspace{-2pt}\times\hspace{-2pt}[2]_t & \begin{array}{c}
            1+\tau^2(\textbf{adj})+\tau^4(\textbf{adj}^2+1)\\
            +\tau^6(\textbf{adj}^3+\textbf{adj}-(k-1))+\mathcal{O}\left(\tau^8\right) 
        \end{array} & \begin{array}{c}
            1+\tau^2(\textbf{adj})+\tau^4(\textbf{adj}^2+1)\\
            +\tau^6(\textbf{adj}^3+\textbf{adj})+\mathcal{O}\left(\tau^8\right)\end{array}  \\ \bottomrule
    \end{array}$}
    \caption{For each class $\mathcal{S}$ theory of type $A_2$ on a sphere with $2k$ punctures, we list the Hall--Littlewood index and the Hilbert series for the Higgs branch. The $4\times[2]_t$ theory corresponds to the $\widehat{D}_4(SU(3))$ theory in \cite{Kang:2021lic}. As in the main text, $\textbf{adj}^n$ is defined as in equation \eqref{eq:adjchar}.
    }
    \label{tab:HLHSforA2}
\end{table}

\section{The Higgs branch Hilbert series of \texorpdfstring{\boldmath{$\widehat{D}_4(SU(3))$}}{D4hat(SU(3))}}
\label{app:HBHSD4hatSU3}

In this appendix we briefly explain how to use the quiver description in Figure \ref{fig:ex3b}, for the 3d reduction of the $\widehat{D}_4(SU(3))$ theory, to compute the Hilbert series of the Higgs branch with the aid of \texttt{Macaulay2} \cite{M2}. 

The first step is to consider the polynomial ring $\mathbb{C}^{24}$ whose variables are all the $\mathcal{N}=2$ chiral fields inside the $\mathcal{N}=4$ hypermultiplets of our theory. We denote those in the bifundamental representation of the central $SU(3)$ gauge node and the $i$th $U(1)$ gauge node by $Q^a_{(i)}$, $\tilde{Q}_{(i)a}$ for $i=1,2,3,4$, where $a=1,2,3$ is the $SU(3)$ color index. We then quotient this ring by the ideal generated by all the F-term equations obtained by varying the superpotential with respect to the $\mathcal{N}=2$ adjoint chiral multiplets inside of the $\mathcal{N}=4$ vector multiplets. Denoting by $A_{(i)}$ those of the $i$th $U(1)$ gauge group for $i=1,2,3,4$, which are singlets, and by $\Phi^a{}_b$ with $\Phi^3{}_3=-\Phi^1{}_1-\Phi^2{}_2$ those of the $SU(3)$ gauge group, the full superpotential, dictated by $\mathcal{N}=4$ supersymmetry, is
\begin{align}
    \mathcal{W}=\sum_{i=1}^4A_{(i)}Q^{a}_{(i)}\tilde{Q}_{(i)a}+\Phi^a{}_b\sum_{i=1}^4Q^{b}_{(i)}\tilde{Q}_{(i)a} \,.
\end{align}
The F-term relations given by the adjoint chiral fields are
\begin{align}
    &Q^{a}_{(i)}\tilde{Q}_{(i)a}=0, & &\, i=1,2,3,4 \,,\\
    &\sum_{i=1}^4Q^{a}_{(i)}\tilde{Q}_{(i)b}=0, & &\, a\neq b=1,2,3 \,, \\
    &\sum_{i=1}^4Q^{a}_{(i)}\tilde{Q}_{(i)a}-Q^{3}_{(i)}\tilde{Q}_{(i)3}=0, & &\, a=1,2 \,,
\end{align}
where the first line comes from the variations with respect to the fields $A_{(i)}$, while the second and third lines come from the variations with respect to the fields $\Phi^a{}_b$. Such a quotient ring, sometimes also called F-flat space $\mathcal{F}^\flat$, can be computed with \texttt{Macaulay2}, which can also compute the associated Hilbert series $g^{\mathcal{F}^\flat}(\tau;\mathbf{z};\mathbf{w})$ where $\mathbf{z}=(z_1,z_2,z_3)$ with $\prod_{i=1}^3z_i=1$ are the $SU(3)$ fugacities while $\mathbf{w}=(w_1,w_2,w_3,w_4)$ are the four $U(1)$ fugacities. Remarkably, \texttt{Macaulay2} is able to produce a closed form expression for this Hilbert series, which nevertheless is too long to report here. Finally, we integrate this against the Haar measure of the gauge group so as to project onto the gauge invariant operators. Even more remarkably, the result of this integration also has a closed form expression:
\begin{align}\label{eqn:app1HS}
\begin{split}
    \operatorname{HS}[\text{Figure } \ref{fig:ex3b}](\tau)&=\frac{1}{6}\oint\prod_{i=1}^4\frac{\mathrm{d}\,w_i}{2\pi iw_i}\prod_{a=1}^2\frac{\mathrm{d}\,z_a}{2\pi iz_a}\left(z_1-\frac{1}{z_2}\right)\left(\frac{1}{z_1}-\frac{z_1}{z_2}\right)\left(\frac{1}{z_1}-z_2\right)\\
    &\times\left(\frac{z_1}{z_2}-z_2\right)\left(z_1-\frac{z_2}{z_1}\right)\left(\frac{z_2}{z_1}-z_2\right)g^{\mathcal{F}^\flat}(\tau;\mathbf{z};\mathbf{w})\\
    &=\frac{1-\tau^2+\tau^4}{(1-\tau^2)(1-\tau^4)}\,,
\end{split}
\end{align}
where we used a parameterization of the $SU(3)$ fugacities where the character of the fundamental representation is given by
\begin{align}
    \chi_{\bf 3}^{SU(3)}(\mathbf{z})=z_1+\frac{1}{z_2}+\frac{z_2}{z_1} \,.
\end{align}
As we can see, the Hilbert series in equation \eqref{eqn:app1HS}, determined explicitly from the 3d reduction, matches with the Coulomb branch Hilbert series, in equation \eqref{eqn:D4hatmirmon} (or equivalently equation \eqref{eqn:D4hatmirHL}), of the proposed 3d mirror, which is depicted in Figure \ref{fig:ex3c}.

\bibliographystyle{sortedbutpretty}
\bibliography{references}
	
\end{document}